\documentclass{elsarticle}

\usepackage{hyperref}

\journal{Journal of Theoretical Biology}

\usepackage{amsmath} \usepackage{amsfonts,euscript,amssymb}
\usepackage{graphicx} \usepackage{subcaption} \usepackage[font=scriptsize]{caption}
\usepackage{epsf}
\usepackage{subcaption}
\usepackage{tabularx}
\usepackage{hyperref}
\hypersetup{colorlinks,citecolor={blue}}  
\usepackage{threeparttable}
\usepackage[strict]{changepage}
\usepackage{gensymb}

\usepackage{mathtools}
\usepackage{calc}   

\usepackage{color}

\definecolor{Navy}{rgb}{0.2,0.2,0.6}

\definecolor{Darkgreen}{rgb}{0.2,0.26,0.15}

\newcommand{\rs}[1]{\textcolor{black}{#1}}
\newcommand{\dt}[1]{\textcolor{black}{#1}}

\usepackage[]{todonotes} 

\usepackage{wasysym}

\newcommand{\eequ}{\end{equation}}
\newcommand{\bequ}{\begin{equation}}
\newcommand{\eequd}{\end{eqnarray*}}
\newcommand{\bequd}{\begin{eqnarray*}}

\def\A{\mathcal{A}}

\def\K{\mathcal{K}}

\def\Bila{\mathbf{B}}

\def\normal{\mathbf{n}}

\def\K{\mathcal{K}}

\def\R{\mathbb{R}}

\def\Bila{\mathbf{B}}

\newcommand {\nor} [1]{\parallel #1 \parallel}

\biboptions{authoryear}

\begin{document}

\begin{frontmatter}

\title{Multiscale dynamics of a heterotypic cancer cell population within a fibrous extracellular matrix}

\author{Robyn Shuttleworth}
\ead{r.shuttleworth@dundee.ac.uk}
\author{Dumitru Trucu}
\ead{trucu@maths@dundee.ac.uk}

\address{University of Dundee, Dundee, Scotland, DD1 4HN}

\begin{abstract}

Local cancer cell invasion is a complex process involving many cellular and tissue interactions and is an important prerequisite for metastatic spread, the main cause of cancer related deaths. Occurring over many different temporal and spatial scales, the first stage of local invasion is the secretion of matrix-degrading enzymes (MDEs) and the resulting degradation of the extra-cellular matrix (ECM). This process creates space in which the cells can invade and thus enlarge the tumour. As a tumour increases in malignancy, the cancer cells adopt the ability to mutate into secondary cell subpopulations giving rise to a heterogeneous tumour. This new cell subpopulation often carries higher invasive qualities and permits a quicker spread of the tumour. 

\dt{Building upon the recent multiscale modelling }framework \dt{for cancer invasion within a fibrous ECM} introduced in \cite{Shutt_2018}, \dt{in this paper we consider} the process of local invasion by a \dt{heterotypic} tumour \dt{consisting of two cancer cell populations mixed with a two-phase ECM.} \dt{To that end, we address the double feedback link between the tissue-scale cancer dynamics and the cell-scale molecular processes through the development of a two-part modelling framework} \dt{that crucially} incorporates the \dt{multiscale} dynamic redistribution of \dt{oriented} fibres occurring within a two-\dt{phase} extra-cellular matrix \dt{and combines this with the multiscale leading edge dynamics exploring key }matrix-degrading enzymes molecular processes \dt{along the tumour interface that drive the} movement of the \dt{cancer} boundary. The modelling framework will be accompanied by computational results that explore the effects of the underlying fibre network on the overall pattern of cancer invasion. 
\end{abstract}

\end{frontmatter}

\section{Introduction}
\dt{Regarded} as one of the hallmarks of cancer \citep{Hanahan2000}, \dt{cancer cell invasion of tissue} is a highly complex process that occurs over many spatial and temporal levels. The invasion of surrounding tissues is a key process for tumour progression and plays a crucial role during the formation of metastases. One of the first steps of tumour invasion is the production and secretion of matrix-degrading enzymes (MDEs) by the cancer cells. These enzymes degrade the surrounding extracellular matrix (ECM) \dt{enabling this way an immediate} \dt{spatial progression of the cancer cells population into} neighbouring tissues via \dt{both random motility and cell-adhesion mediated migration as well as} enhanced proliferation, \dt{and,} eventually, \dt{via a cascade of invasion events (including angiogenesis)} leading to the spread \dt{at} distant sites in the body (metastasis). \dt{In this context, as the} prognosis for patients is \dt{still} poor, with limited treatment options \dt{(}such as chemotherapy and radiation\dt{)} \dt{and} metastases \dt{being currently attributed} to $90\%$ of cancer related deaths \citep{Sporn1996}, a deeper understanding of the processes that occur during local invasion is key \dt{for the improvement and future development of treatment strategies}. 

The local invasion of cancer is, in the first instance, stimulated by the secretion of MDEs. There are several classes of MDEs, such as matrix metalloproteinases (MMPs) and the urokinase-type plasminogen activator (uPA), produced by cancer cells and these enzymes degrade and reshape the structure of the ECM. Through full or partial degradation of the surrounding matrix the cells create free space in which they can invade, further advancing tumour progression. MMPs are substrate specific and can be either membrane bound (MT-MMPs), or can move freely around the cells, each working on different components of the ECM ensuring a \dt{significant} breakdown of the matrix \citep{Brinck2002,Parsons1997}. 

The invasive abilities of cancer can be strengthened in many ways, namely, enhanced proliferation, increased migrative capability and adaptive cellular adhesion properties. Cell-cell and cell-matrix adhesion are both key players during cancer invasion and play important roles in tumour progression \citep{Behrens1989,Kawanishi_1995,Todd_2016,Berrier_2007}. Any perturbations to either factor will contribute to a change in tumour morphology and the direction of migration of the tumour. A decrease in cell-cell adhesion allows the cells to detach from the primary tumour and invade further into the matrix, \dt{and coupling this} with cell-matrix adhesion increases \dt{leads to a notable escalation of} tumours' invasive capabilities \citep{CAVALLARO200139}. 

In healthy cells, cell-cell adhesion is mediated by a family of specific molecules on the cell surface known as cell adhesion molecules (CAMs). Adhesion is dependent on the cell-cell signalling pathways that are formed due to the interactions between the distribution of calcium-sensing receptors and $\text{Ca}^{2+}$ ions within the ECM. Essential for cellular adhesion is the family of transmembrane glycoproteins cadherins. These are calcium dependent adhesion molecules that interact with intra-cellular proteins, known as catenins. In particular, the subfamily, E-cadherins, are responsible for binding with these catenins, typically $\beta$-catenin, forming an E-cadherin/catenin complex. Any alteration to the function of $\beta$-catenin will result in the loss of ability of the E-cadherin to initiate cell-cell adhesion \citep{Wijnhoven2000}. The direct correlation between this calcium-based cell signalling mechanism and the regulation of E-cadherin and $\beta$-catenin was first discovered in colon carcinoma \citep{Bhagavathula2007}. 

On the other hand, cell-matrix adhesion is regulated by the subfamily of calcium independent CAMs, known as integrins, that enable the cells to bind to different components of the ECM. Integrins link the cytoskeleton inside a cell to the ECM outside and this is aided by their ability to attach to a wide variety of ligands \citep{Humphries2006}. Additionally, integrins can bind to actin proteins within the cytoskeleton, aiding in cell migration by creating a leading and trailing edge to the cell, resulting in persistent migration in one direction \citep{Delon_actin,Moissoglu2006}. 

All of these properties of invasive tumour cells have one thing in common, they all require or interact with the ECM. The extracellular matrix is comprised from a variety of proteins \dt{including} collagen and elastin, glycoproteins such as fibronectin and laminins as well as a large array of other molecules, MDEs, fibroblasts, etc. It is \dt{however the} complex network of \dt{ECM} fibres, \dt{such as collagen and fibronectin fibres,} that provides not only support for cells, but it acts as a platform through which cells can communicate. This feature is of particular use to cancer cells, which use the matrix as a means of invasion. Through a combination of degradation and cell-matrix adhesion, the \dt{ECM} is manipulated and exploited to further \dt{advance} their \dt{spatial} progression. A main \dt{fibrous} component \dt{that} provides \dt{ECM with a scaffolding structure and integrity} is collagen, \dt{which is} the most abundant protein in the human body, with collagen type I, II and III making up around $90\%$ of the overall collagen present \citep{Lodish_22.3}. 

Moving away from the \dt{scaffolding} structure of the ECM, there are \dt{other fibrous ECM} components that provide functional qualities within the matrix. One such component is the glycoprotein fibronectin. Fibronectin contributes to cell migration, growth and proliferation, ensuring the normal functionality of healthy cells. It also plays a crucial role in cell adhesion, having the ability to anchor cells to collagen and other components of the ECM. 

Biological experiments have revealed that increased stromal collagen density promotes tumour formation and results in tumours exhibiting a more invasive phenotype \citep{Provenzano_2008}. In addition, it has been demonstrated that local invasion is further accelerated by collagen reorganisation \citep{Provenzano_06}, and this behaviour is significantly increased in regions of high collagen density. The realignment of fibronectin fibrils has also been associated with increased local tumour invasion \citep{Erdogan3799} enabling a smooth invasion of the cancer cells. On the other hand, collagen type I has been shown to down-regulate E-cadherin gene expression in pancreatic cancer cell lines which leads to a reduction in cell-cell adhesion and increased proliferation and cell migration \citep{Menke_2001}. \dt{This motivated us to explore the migratory behaviour of the overall cancer cell population by assuming in this work the presence of a secondary cancer cell sub-population} exhibiting a decrease in cell-cell adhesion and an increase in migratory activity arising in places of high matrix density.

\dt{The past two decades or so have witnessed a vast} of interest in the modelling of cancer invasion, see, for example, \citep{Perumpanani_et_al_1998,Anderson_et_al_2000,Chaplain2006a,Hillen2006,Chauviere_2007,szymanska_08,Andasari_et_al_2011,Ramis-Conde_et_al_2000,Chaplain_et_al_2011,Scianna_Presiosi_2012}. With biological experiments advancing, there is an increasing need for more extensive modelling of the processes involved in cancer invasion. Biological and mathematical models of both \textit{in vivo} and \textit{in vitro} experiments have given us a deeper insight into many processes involved during tumour invasion. Great focus has been placed on modelling the effects of cell-cell and cell-matrix adhesion \citep{Painter2010,Armstrong_et_al_2006,Anderson2005,Turner2002,Gerisch2008,Domschke_et_al_2014,Bitsouni_2017}. On the other hand, \dt{recent works such as} \citep{Chauviere_2007,Painter2008,hillen_10,Schluter_et_al_2012,hillen_painter_winkler_2013,Engwer2015} have \dt{highlighted} the vital importance that \dt{the composition of the ECM has on} the overall invasion of cancer. Finally, the multi-scale nature of cancer invasion has received special attention over the past decade \citep{Ramis-Conde2008,Dumitru_et_al_2013,Peng2016,shutt_chapter}, \dt{with significant advancements towards two-scale approaches appropriately linking  the spatio-temporal dynamics occurring at different scales being proposed in \cite{Dumitru_et_al_2013,Shutt_2018}}. 

In this paper we extend the model developed in \citep{Shutt_2018} to investigate the invasiveness of a heterogeneous tumour in a multi-phase \dt{fibrous tissue} environment by introducing a secondary cancer cell subpopulation. This \dt{will} examine the invasive abilities of a \dt{malignant tumour with two cancer cell sub-populations} under the presence of a two-component ECM consisting of \dt{both an oriented fibre phase distribution \dt{(such as collagen and fibronectin)}, and a soluble phase, that accounts for all other ECM non-fibrous components (such as Ca$^{2+}$ ions, laminin, and other soluble matrix constituents). To that end, in the context of multiscale moving boundary approach introduced in \cite{Dumitru_et_al_2013} (capturing the key influence of the underlying tumour invasive edge two-scale MDEs proteolytic activity), we will consider here the dynamics of the two cancer cell sub-populations progressing within the surrounding fibrous ECM and explore its multiscale evolution and interaction with the ECM fibre phase in the presence of both homogeneous and heterogeneous non-fibre ECM soluble phases.} 

The paper \dt{is} structured as follows. \dt{I}n Section 2 we will \dt{detail the distribution of oriented macroscopic (tissue-scale) ECM fibres vector field induced by the microscale (cell-scale) mass distribution of micro-fibres alongside the multiscale interaction that arises between the fibre ECM phase and the two cancer cell populations (both in terms of dynamic micro-scale fibres rearrangements by the cancer cells and the fibre impact within the macro-scale cancer evolution).} In Section 3 we will present the numerical approaches used and initial conditions for the computations and in Section 4 we present the simulation results. Finally, we will conclude with Section 5 where we discuss the implications of our results as well as future work.

\section{The mathematical model}

Here we will build upon the key aspects of the two-part multiscale model introduced and developed in \citep{Shutt_2018} that utilises the two-scale moving boundary framework first proposed in \citep{Dumitru_et_al_2013} and introduces a two-component ECM in which the fibre phase plays a central role in the two-scale dynamic redistribution of microscopic fibres. \dt{As we will briefly detail below, this complex dynamics will be captured by two} interconnected multiscale systems \dt{that} share the same macro-scale cancer dynamics at the tissue-scale, whilst having their own distinct micro-scale dynamics occurring at cell-scale that are linked to the macro-dynamics through two double feedback loops, as illustrated in Fig. \ref{two-part_model}. 

During the invasion process a tumour can become increasingly malignant, whereby the \dt{primary} cancer cell population acquires the ability to mutate, \dt{giving rise to} a secondary subpopulation of cancer cells \dt{that exhibits} more aggressive invasion qualities, \dt{including:} faster \dt{random motility}, increased proliferation, and changes in cell-cell and cell-matrix adhesion properties enabling an acceleration of local cancer invasion. \dt{Thus, the presence of this secondary cancer cell subpopulation has implications for cancer dynamics at both macro-  and micro- scales, and to address all these, we will devote the following sections to develop and adapt the modelling approach introduced initially in \citep{Shutt_2018} to the new context created by the two cancer cell sub-populations.}

Using here the same terminology as in \citep{Shutt_2018}, let us denote the support of the invading tumour region by $\Omega(t)$, and assume this evolves in the maximal reference \dt{tissue} cube $Y \in \R^{N}$ with $N=2,3$, centred at the origin of the space. At any spatio-temporal node $(x,t) \in \Omega(t) \times [0,T]$ we consider the tumour to be a mixture of cancer cells $c_{n}(x,t)$, $n=1,2$, with their combined vector denoted $\textbf{c}(x,t)=[c_{1}(x,t),c_{2}(x,t)]$, integrated within a multiphase distribution of ECM, $v(x,t)$, whose components will be defined in the next section.

\subsection{Multiscale fibre structure and their dynamic contribution in tumour progression}
\begin{figure}[h]
\centering
\includegraphics[scale=0.4]{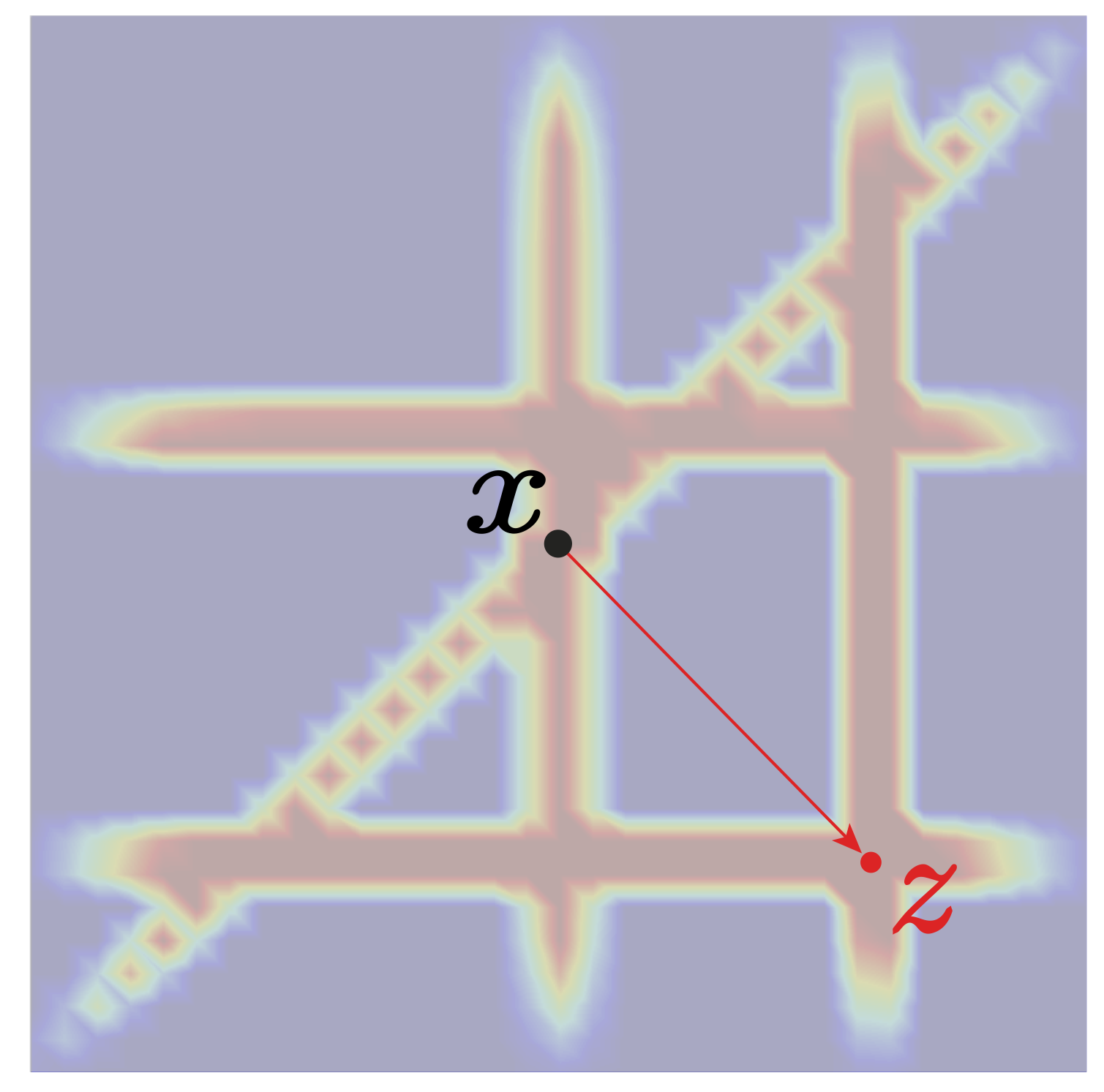}
\caption{\emph{A 2D contour plot of the micro-fibres distribution on the micro-domain $\delta Y(x)$, centred at $x$, with the barycentral position vector $\protect\overrightarrow{x \,z}:=z-x $} pointing towards an arbitrary micro-location $z\in \delta Y(x)$ illustrated by the red arrow.}
\label{dyadic_cubes}
\end{figure}

Adopting the approach \dt{and terminology} introduced in \citep{Shutt_2018}, \dt{to capture} the dynamics of \dt{the two sub-populations of cancer cells} in the presence of a two component heterogeneous extra-cellular matrix (consisting of a fibre and non-fibre phase), \dt{we proceed as follows}. \dt{As derived in \citep{Shutt_2018} and briefly outlined below, at any macroscale point $x\in \Omega(t)$, the ECM-fibre phase is represented through a macro-scale vector field $\theta_{_{f}}(x,t)$. This vector field captures and represents at tissue-scale not only the amount of fibres distributed at $(x,t)$ but also their naturally arising macroscopic fibres orientation induced by the revolving barycentral orientation $\theta_{_{f,\delta Y(x)}}(x,t)$ generated by the microscopic mass distribution of microfibres $f(\cdot,t)$ within the micro-domain $\delta Y(x) := \delta Y + x$. An example of a mass distribution of ECM micro-fibres $f(z,t)$, $z\in\delta Y(x)$, is illustrated in Fig. \ref{dyadic_cubes} below and is defined in \ref{microfibres}. }     

\dt{In brief, while referring the reader to its full derivation presented in \citet{Shutt_2018}, the naturally generated revolving barycentral orientation $\theta_{_{f,\delta Y(x)}}(x,t)$ associated with $\delta Y(x)$ is given by the \emph{Bochner-mean-value} of the position vectors function $\delta Y(x)\ni z\mapsto z-x\in\R^{N}$ with respect to the density measure $f(x,t)\lambda(\cdot)$, where $\lambda(\cdot)$ is the usual Lebesgue measure (see \citet{yosida1980}), and so this is expressed mathematically as:
\bequ
\theta_{_{f,\delta Y(x)}}(x,t)=\frac{\int\limits_{\delta Y(x)}f(z,t)(z-x)dz}{\int\limits_{\delta Y(x)} f(z,t)  dz}. 
\eequ}

Following on, at any spatio-temporal node $(x,t)$, the macroscopic fibre \dt{oriented} is defined as 
\bequ\label{fiborien}
\theta_{_{f}}(x,t)=\frac{1}{\lambda (\delta Y(x))} \int_{\delta Y(x)} f(z,t) \ dz \cdot \frac{\theta_{_{f,\delta Y(x)}} (x,t)}{||\theta_{_{f,\delta Y(x)}} (x,t)||}
\eequ
where $\lambda(\cdot)$ is the usual Lebesgue measure. The macroscopic mean-value fibre representation at any $(x,t)$ is then given by the Euclidean magnitude of $\theta_{_{f}}(x,t)$, namely,
\bequ\label{fibmean}
F(x,t):=||\theta_{_{f}}(x,t)||_{2}.
\eequ
Finally, \dt{concerning the \emph{non-fibre soluble phase} of the ECM, we \dt{consider that this include} all the other non-fibre components of the ECM, i.e., elastin, laminins, fibroblasts, etc.. Thus, denoting the spatial distribution of the non-fibre ECM phase distribution $l(x,t)$, the total ECM distributed at any spatio-temporal node $(x,t)$ is therefore given by $v(x,t)=l(x,t)+F(x,t)$.}

\subsection{Macro-scale dynamics}
To \dt{explore the dynamics of a heterotypic cancer cell population with two sub-populations consisting of a primary and a mutated cell population, within the modelling framework}, let's denote the \dt{\emph{global macro-scale tumour vector distribution}} by 
\[
\textbf{u}(x,t)=[\textbf{c}(x,t)^{T},F(x,t),l(x,t)]^{T},
\]
with the tumour's \emph{volume fraction of occupied space} being given as 
\[
\rho(\textbf{u}(x,t))=\vartheta_{v}(l(x,t)+F(x,t)) + \vartheta_{c}\bar{c}(x,t),
\]
where $\vartheta_{v}$ and $\vartheta_{c}$ represent the fraction of physical space occupied by the entire ECM (\dt{including} both the fibre and non-fibre phase) and \dt{$\bar{c}(x,t)$ denotes the combined cell populations $c_{1}$ and $c_{2}$, this being given by
\[
\bar{c}(x,t):=c_{1}(x,t)+c_{2}(x,t).
\]}
Therefore, \dt{adopting the same approach as in \citet{Andasari_et_al_2011} and \citet{shutt_chapter}}, focusing first on the cancer cell population, the dynamics of the two cell \dt{sub-}populations are much \dt{similar}. \dt{Under the presence of} a logistic proliferation law, per unit time, the spatial movement of the primary tumour cells $c_{1}(x,t)$ is described by a combination of local Brownian movement (approximated here by diffusion) and cell-adhesion. They experience a loss of cells through mutation towards a second, more aggressive population $c_{2}(x,t)$. \dt{Similar to population $c_{1}(x,t)$, o}nce mutations have begun, per unit time, \dt{sub-}population $c_{2}(x,t)$ \dt{exercises} spatial movement through a local Brownian movement and \dt{a perturbed} cell-adhesion under the presence of logistic proliferation law. Hence, mathematically, the dynamics of these cell populations can be represented as 
\begin{align} 
\frac{\partial c_1}{\partial t} &= \nabla \cdot [D_{1} \nabla c_1 - c_1 \mathcal{A}_{1}(x,t,\textbf{u}(t,\cdot),\theta_{f}(\cdot,t))] +\mu_{1}c_1(1-\rho(\textbf{u})) - M_{c}(\textbf{u},t)c_1, \label{eq:c1} \\ 
\frac{\partial c_2}{\partial t} &= \nabla \cdot [D_{2} \nabla c_2 - c_2 \mathcal{A}_{2}(x,t,\textbf{u}(t,\cdot),\theta_{f}(\cdot,t))] +\mu_{2}c_2(1-\rho(\textbf{u})) + M_{c}(\textbf{u},t)c_1,
 \label{eq:c2}
\end{align}
where: $D_n$ and $\mu_{n}$, $n=1,2$ are the non-negative diffusion and proliferation coefficients of sub\dt{-}populations $c_{1}(t,x)$ and $c_{2}(t,x)$ respectively, $M_{c}$ describes the conversion from population $c_{1}(t,x)$ to $c_{2}(x,t)$ and finally, the non-local flux $\A_{n}(x,t,\textbf{u}(t,\cdot),\theta_{f}(\cdot,t))$ accounts for the cellular adhesion processes which directly influence the spatial movement of the tumour \dt{cell population $c_{n}$}, $n=1,2$. Embracing the \dt{modelling} concept proposed in \citet{Shutt_2018}, the cells will not only interact with other cells, i.e., cell-cell adhesion, but also with the surrounding multi-phase ECM, that in this instance constitutes of both cell-fibre and cell-ECM-non-fibre adhesion. Hence, within a sensing radius $R$ at time $t$, the non-local adhesive flux can be expressed as:
\bequ
\begin{split}
\A_{n}(x,t,\textbf{u}(\cdot, t), \theta_{_{f}}(\cdot, t))=\frac{1}{R} \int_{B(0,R)} \mathcal{K}(\nor{\!y\!}_{2}) & \big(n(y) (\textbf{S}_{_{cc}} \textbf{c}(x+y,t) + \textbf{S}_{_{cl}} l(x+y,t)) \\
&+ \hat{n}(y) \ \textbf{S}_{_{cF}} F(x+y,t) \big)(1-\rho(\textbf{u}))^{+}
\end{split}
\label{adhesionterm}
\eequ
Whilst we consider the adhesive activities of the cells to become less influential as the distance $r$ from $x$ increases, \dt{and account for this} through the radial kernel $\K(\cdot)$ defined in \citet{Shutt_2018} as 
\bequ
\K(r)=\frac{2 \pi R^2}{3}\left(1-\frac{r}{2R}\right),
\eequ
we \dt{explore the strength of the adhesion bonds created between} the cancer cells \dt{distributed at $x$ and the cells or the non-fibre phase of the ECM distributed at $y$} in the direction of the unit normal
\bequ
\normal(y):=
\left\{
\begin{array}{l}
y/||y||_2 \quad\text{if}   \quad y \in B(0,R) \setminus \{ (0,0) \}, \\
\left(0,0\right) \quad  \textrm{otherwise}
\end{array}
\right.
\eequ
\dt{and let us denote the cell-cell and cell-non-fibre-ECM} adhesive strengths by $\textbf{S}_{_{cc}}$ and $\textbf{S}_{_{cl}}$, respectively. \dt{Furthermore,} proceeding as in \citep{Shutt_2018}, we consider $\textbf{S}_{_{cl}}$ to be constant whilst the coefficient representing cell-cell adhesion $\textbf{S}_{_{cc}}$ is \dt{monotonically} dependent on the level of extracellular Ca$^{2+}$ ions enabling strong adhesive bonds between cells \citep{Hofer2000,Gu2014}. Hence we assume that cell-cell adhesion is dependent on the density of the underlying non-fibre ECM phase, ranging from $0$ to \dt{a} Ca$^{2+}$-saturation level \dt{denoted} $S_{_{max}}$, and is taken as

\bequ
\textbf{S}_{_{cc}}(x,t):=\textbf{S}_{_{max}} e^{\left(1-\frac{1}{1-(1-l(x,t))^2}\right)}.
\label{scc}
\eequ
The final term in \eqref{adhesionterm} describes the adhesive behaviour between the cancer cells and the fibres distributed on $\Bila(x,R)$. Within this term we account for the role of the fibres in two ways. \dt{On one hand, we account for the influence of the} macroscopic \dt{fibres magnitude} $F(x+y,t)$ \dt{has over the adhesion that the cells distributed at the spatial location $x$ exercise with adhesion strength $\textbf{S}_{_{cF}}$ upon the fibres distributed at $x+y$. On the other hand, their full macroscopic orientation $\theta_{f}(x,t)$ (induced by their micro-scale mass distribution of microfibres),  biases the cell-matrix adhesion in the direction of the resultant vector}
\bequ
\hat{n}(y):=
\left\{
\begin{array}{l}
\frac{y+\theta_f(x+y)}{||y+\theta_f(x+y)||_2} \quad\text{if} \  (y+\theta_f(x+y))  \neq (0,0) \\[0.2cm]
(0,0) \in \mathbb{R}^2 \quad \quad  \text{otherwise}.
\end{array}
\right.
\label{nhat}
\eequ

The last \dt{important aspect that} we consider here \dt{concerns mutations that enable cells from primary tumour cell subpopulation $c_{1}$ to undergo genetic conversions  and become secondary cancer cells $c_{2}$, process that is explored here through the mutation term $M_{c}(\textbf{u},t)$, which captures the direct correlation between the presence of significant ECM levels, as detailed in the following. Indeed, as}  cellular adhesion is controlled by the binding of the calcium dependent adhesion molecules, \emph{E-cadherins}, with the intra-cellular proteins, \emph{catenins}\dt{, it is significant to note that c}ollagens present in the ECM (in particular collagen type I) down-regulate the gene expression of E-cadherins, resulting this way in a loss of cell-cell adhesion \citep{Menke_2001}. This loss of adhesion is accompanied by an increase in proliferation and migratory activity, \dt{and so, to reflect this behaviour, the mutation term $M_{c}(\textbf{u},t)$} is dependent on the underlying ECM density \dt{levels and is taken to be}  
\bequ
M_{c}(x,t):=
\left\{
\begin{array}{l}
\frac{\text{exp}\left(\frac{-1}{\kappa^{2}-(1-v(x,t))^{2}}\right)}{\text{exp}\left(\frac{1}{\kappa^{2}}\right)}  \cdot H(t-t_m) \quad \text{if} \ 1-\kappa<v(x,t)<1, \\
0 \quad  \quad \quad \textrm{otherwise},
\end{array}
\right.
\label{mutation}
\eequ
where \dt{$\kappa$ is a certain level of ECM beyond which mutations can occur and} $H(\cdot)$ is the usual Heaviside function, \dt{with} $t_m$ \dt{being} the time at which mutations begin. 

Finally, we describe the dynamics of the ECM, considering the individual constituents of the matrix, namely the fibre and non-fibre component. Both constituents of the matrix are simply degraded by the cancer cells, and so their macroscopic dynamics can be mathematically written as
\begin{align}
\frac{d F}{d t}&=-\gamma_{1} c F \label{fibdeg}\\
\frac{d l}{d t}&=-\gamma_{2} c l + \omega (1-\rho(\textbf{u}))\label{ldeg}
\end{align}
where $\gamma_{1}$ and $\gamma_{2}$ are the degradation rates of the fibre and non-fibre components respectively, and $\omega$ describes the rate of non-fibre ECM remodelling. Furthermore, the matrix remodelling, \dt{which is important both in the development and in progression of cancer, contributing to processes such as metastasis and tumour cell invasion \citep{cox_2011}}, is controlled \dt{here} by the volume filling factor $(1-\rho(\textbf{u}))$.

\subsection{Microscopic fibre rearrangement}
\dt{As explored and modelled in \citet{Shutt_2018}, during their invasion,} the cancer cells push the surrounding fibres \dt{in accordance with the emerging cell-flux direction and rearrange their micro-fibre mass distribution managing this way to reorient the macro-scale fibres. Indeed,} in addition to their macroscopic degradation described in \eqref{fibdeg}, the fibres go through a microscopic rearrangement process induced by the macro-dynamics of the cancer cells. \dt{Specifically}, at time $t$ and at any spatial location $x \in Y$, the cancer cells will realign the micro-fibres through a microscopic rearrangement process in each micro-domain $\delta Y(x)$ \dt{that is induced} by the combined macro-scale spatial flux of both cancer cell \dt{sub-}populations 
\[
\mathcal{F}(x,t):=\mathcal{F}_{1}(x,t)+\mathcal{F}_{2}(x,t)
\]
where
\begin{align*}
\mathcal{F}_{1}(x,t):=D_1 \nabla c_{1}(x,t)-c_{1}(x,t)\mathcal{A}_{1}(x,t,\textbf{u}(\cdot,t),\theta_{_{f}}(\cdot,t)), \\
\mathcal{F}_{2}(x,t):=D_2 \nabla c_{2}(x,t)-c_{2}(x,t)\mathcal{A}_{2}(x,t,\textbf{u}(\cdot,t),\theta_{_{f}}(\cdot,t)).
\end{align*}
The combined flux \dt{acts upon the micro-scale distribution $f(z,t)$, $\forall z\in \delta Y(x)$ in accordance to the magnitude that the total mass of cancer cells has relative to the combined mass of cells and fibres at $(x,t)$, which is given by the weight} 
\[
\omega(x,t)=\frac{\bar{c}(x,t)}{\bar{c}(x,t)+F(x,t)}.
\]
\dt{At the same time, the} spatial flux of cancer cells $\mathcal{F}(x,t)$ is balanced \dt{in a weighted manner by the orientation $\theta_{_{f}}(x,t)$ of the existing distribution of fibres at $(x,t)$ that is appropriately magnified by a weight that accounts for the magnitude of fibres versus the combine mass of cells and fibres at $(x,t)$ and is given by $(1-\omega(x,t))$. As a consequence, the microscale distribution of micro-fibres $f(z,t)$, $\forall z\in \delta Y(x)$ is therefore acted upon uniformly by the  resultant force given by the following macro-scale vector-valued function} 
\bequ
r(\delta Y(x),t):=\omega(x,t)\mathcal{F}(x,t)+(1-\omega(x,t))\theta_{_{f}}(x,t).
\eequ
\dt{As detailed in \citet{Shutt_2018}, under the uniform incidence of the resultant force $r(\delta Y(x),t)$ upon the mass of micro-fibres distributed at any $z\in \delta Y(x)$, an} \emph{on-the-fly spatial} microscopic rearrangement of \dt{this micro-}fibres \dt{mass takes place. Specifically, under the influence of $r(\delta Y(x),t)$, an appropriate level of micro-fibres mass $f(z,t)$ will undergo a spatial transport towards a new position  
\[
z^{*}:=z+\nu_{_{\delta Y(x)}}(z,t)
\]
where the relocation direction and magnitude is given by
\begin{equation}
\nu_{_{\delta Y(x)}}(z,t)=\left(x_\text{dir}(z) + r(\delta Y(x), t)\right) \cdot \frac{f(z,t)(f_{\text{max}}-f(z,t))}{f^{*}+||r(\delta Y(x)) - x_\text{dir}(z)||_{2}} \cdot \chi_{_{\{f(\cdot,t)>0\}}}(z)
\label{eq:fibnu}
\end{equation}
Here we have $x_{\text{dir}}(z)=\overrightarrow{ x \, z}$ representing the barycentric position vector pointing to $z$ in $\delta Y(x)$, which enables us also  the quantification of the \emph{position defect} of $z$ with respect to $r(\delta Y(x), t)$ namely
\[
||r(\delta Y(x)) - x_\text{dir}(z)||_{2}.
\]
which affects the spatial relocation of micro-fibres mass. Furthermore, $f_{\text{max}}$ represents a level of micro-fibres that can be distributed at $z\in\delta Y(x)$, with the micro-fibres mass relocation in the direction of $\left(x_\text{dir}(z) + r(\delta Y(x), t)\right)$ being enabled provided that their level is below $f_{\text{max}}$. Finally, alongside the \emph{position defect} of $z$ with respect to $r(\delta Y(x), t)$, another aspect that affects the micro-fibres relocation is the level of micro-fibres distributed at location $z$, which is accounted for in \eqref{eq:fibnu} through the micro-fibres saturation fraction 
\[
f^{*}=\frac{f(z,t)}{f_{\text{max}}}.
\]
Thus, a micro-fibres mass transport from $z$ to the location $z^{*}$ is exercised provided that micro-fibres level at $z$ are not at their maximum level $f_{\text{max}}$, while lower levels of micro-fibres saturations at $z$ together with a better position alignment given by a smaller position defect lead to a relocation of the micro-fibres mass in direction $\left(x_\text{dir}(z) + r(\delta Y(x), t)\right)$ at a greater distance, resulting in reaching a position $z^{*}$ that is further away from $z$. Finally, this micro-fibres transport is also regulated by the capacity available} at the new position $z^{*}$, \dt{which is explored here} through \dt{a} movement probability
\[
p_{move}:=\max\big(0,\frac{f_\text{max}-f(z^{*},t)}{f_\text{max}}\big)
\]
\dt{that} enables only an amount of $p_{move}f(z,t)$ of micro-fibres to be transported to position $z^{*}$, while the rest of $(1-p_{move})f(z,t)$ remain at $z$.

\subsection{The multiscale moving boundary approach}
Let us now briefly revisit the novel multiscale moving boundary framework initially introduced in \citep{Dumitru_et_al_2013} and later evolved to consider two cancer cell subpopulations in \citep{shutt_chapter}, \dt{which explore the cell-scale proteolytic activity of MDEs along the invasive edge of the tumour that is non-locally induced by the tissue-scale cancer cell population dynamics and that degrades the peritumoural ECM determining this way the direction and associated displacement for tumour boundary progression.} As detailed in \citep{Dumitru_et_al_2013}, the link between the tumour macro-dynamics \eqref{eq:c1}, \eqref{eq:c2}, \eqref{fibdeg} and \eqref{ldeg} and the proteolytic enzyme micro-dynamics is captured via a double feedback loop that is realised by a \emph{top-down} link describing the source of MDEs induced at cell-scale by the spatial distribution of cancer cells at tissue-scale, and a \emph{bottom-up} link describing the translation of the resulting boundary relocation to the tissue-scale. 

\paragraph{Top-down link} As previously discussed, cancer invasion is a multiscale process in which the MDEs secreted by tumour cells \dt{from} the outer \dt{proliferating ring undergo a cell-scale spatial transport in the neighbourhood of the invasive} edge of the tumour \dt{and are} responsible for the degradation of the surrounding ECM. It is this breakdown of peritumoural ECM that enables the tumour opportunities to expand and proceed with its local invasion. Continuing with the terminology of the framework introduced in \citet{Dumitru_et_al_2013}, during a time interval $[t_{0},t_{0}+\Delta t]$, the MDEs micro-dynamics is explored on the invasive leading edge of the tumour $\partial \Omega(t)$ enclosed by a complete cover of $\epsilon$-size half-way shifted overlapping micro-domains $\{\epsilon Y\}_{\epsilon Y \in P(t)}$. \dt{Furthermore, the specific topological requirements detailed in full in \citet{Dumitru_et_al_2013} that enable the construction of the covering bundle of microdomains $\{\epsilon Y\}_{\epsilon Y \in P(t)}$, allow us to} capture \dt{the cell-scale MDEs activity in a neighbourhood of $\partial \Omega(t)$, exploring this molecular dynamics in both the overlapping inner regions $\epsilon Y\cap \Omega(t)$ and the peritumoural outside regions $\epsilon Y\cap \setminus \Omega(t)$ where the MDEs get transported and degrade the ECM}. 

\begin{figure}[h]
\centering
\includegraphics[scale=0.4]{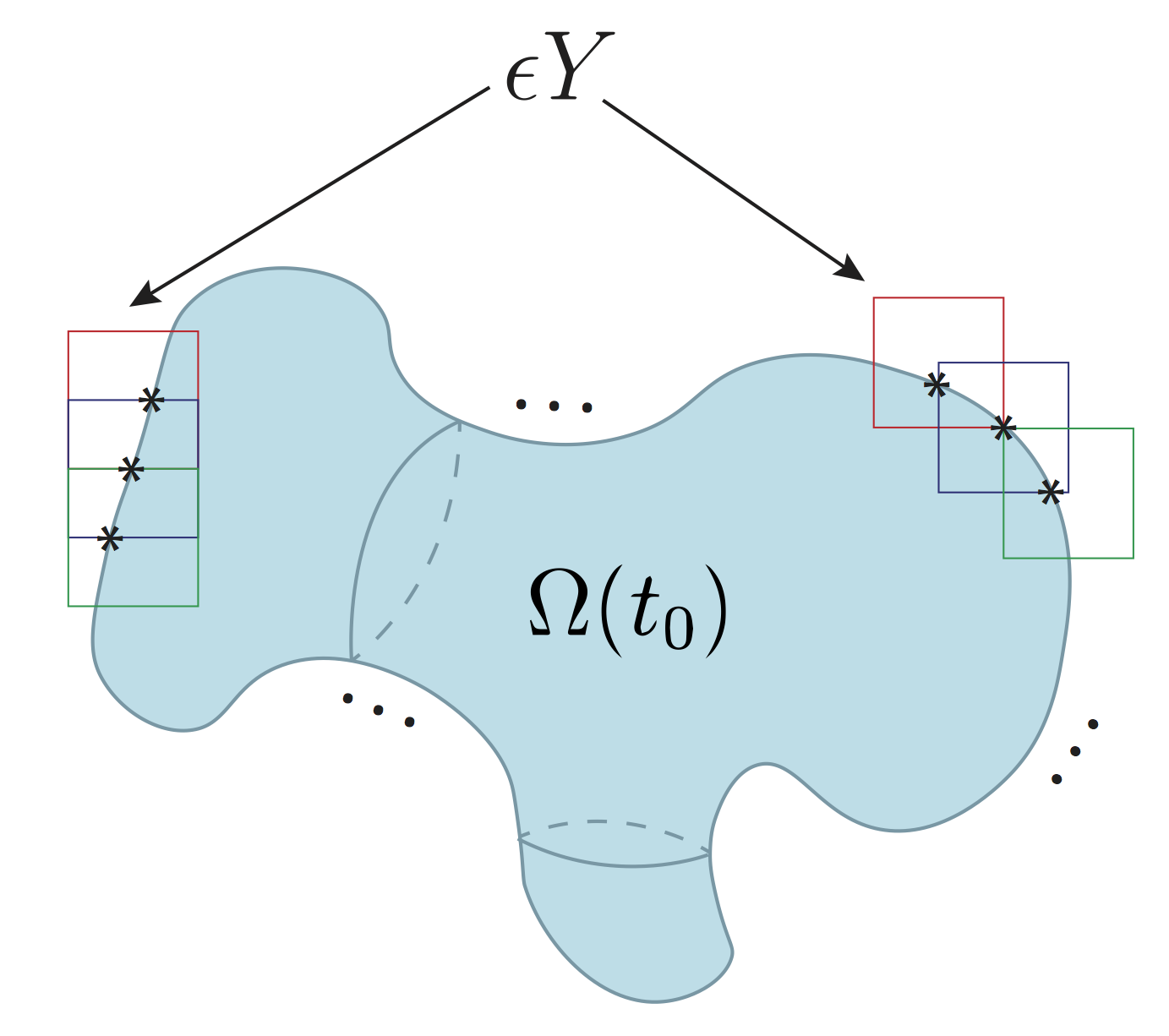}
\caption{\emph{Schematic of the bundle of $\epsilon Y$ micro-cubes covering boundary of the tumour $\partial \Omega(t_{0})$, including the half-way shifted overlapping $\epsilon Y$ cubes. Dots illustrate continuation of boundary coverage by $\epsilon Y$ cubes.}}
\label{epsilon_cubes_covering}
\end{figure}
At each $z\in\epsilon Y \cap \Omega(t)$ a source of MDEs is induced as \dt{a collective contribution of both cancer cell sub-populations that arrive during their dynamics in the outer proliferating rim within a maximal} distance $\gamma>0$ with respect to $z$ \dt{(given by the maximal thickness of the outer proliferating rim)}, and so this can be mathematically \dt{given by} 
\begin{align}
\begin{split}
1. \quad &g_{\epsilon Y}(y,\tau) = \frac{\int\limits_{\textbf{B}(z,\gamma)\cap\Omega(t_0)} \alpha_{1} c_{1} (x,t_0 + \tau) + \alpha_{2} c_{2} (x,t_0 + \tau) \ dx}{\lambda (\textbf{B}(y,\gamma)\cap\Omega(t_0))}, \quad  y \in \epsilon Y \cap \Omega(t_0), \\[0.7cm]
2. \quad &g_{\epsilon Y}(y,\tau) = 0,\quad  y \in \epsilon Y \setminus \big( \Omega(t_0)+\{ y \in Y|  \ ||z||_2 < \gamma\}), 
\end{split}
\label{eq:sourceMDEs}
\end{align}
where $\Bila(y,\gamma):=\{\xi\in Y\,|\, \nor{y-\xi}_{_{\infty}}\leq \gamma\}$; $\alpha_{i}$, $i=1,2$, are the MDE secretion rates for the two cancer cell sub\dt{-}populations, $\lambda(\cdot)$ is the standard Lebesgue measure on $\R^{2}$; \dt{and $\gamma$ is a small distance between the zero source level from outside $\Omega(t_{0})$ and the non-zero source levels on $\epsilon Y \cap \Omega(t_0)$ where via \emph{Urysohn Lemma} \citep{yosida1980} we ensure a continuous transition. Once secreted by the cancer cells, the MDEs molecular distribution denoted here by $m(y,\tau)$ exercise a cell-scale cross interface spatio-temporal transport which in the simplest context of a generic MDEs is considered as being given by a local diffusion. Thus, given the topological properties of the covering bundle of  overlapping micro-domains $\{\epsilon Y\}_{\epsilon Y \in P(t)}$ detailed in \citet{Dumitru_et_al_2013}, this enable us to decompose this cross-interface diffusion process (into a corresponding bundle of micro-processes) and to explore this per each individual $\epsilon Y$, where this micro-dynamics is given as }
\bequ
\frac{\partial m}{\partial \tau}=D_{m} \Delta m + g_{\epsilon Y}(z,\tau), \quad z \in \epsilon Y, \ \tau \in [0, \Delta t].
\label{MDE_micro}
\eequ

\begin{figure}[ht]
\centering
\includegraphics[scale=0.2]{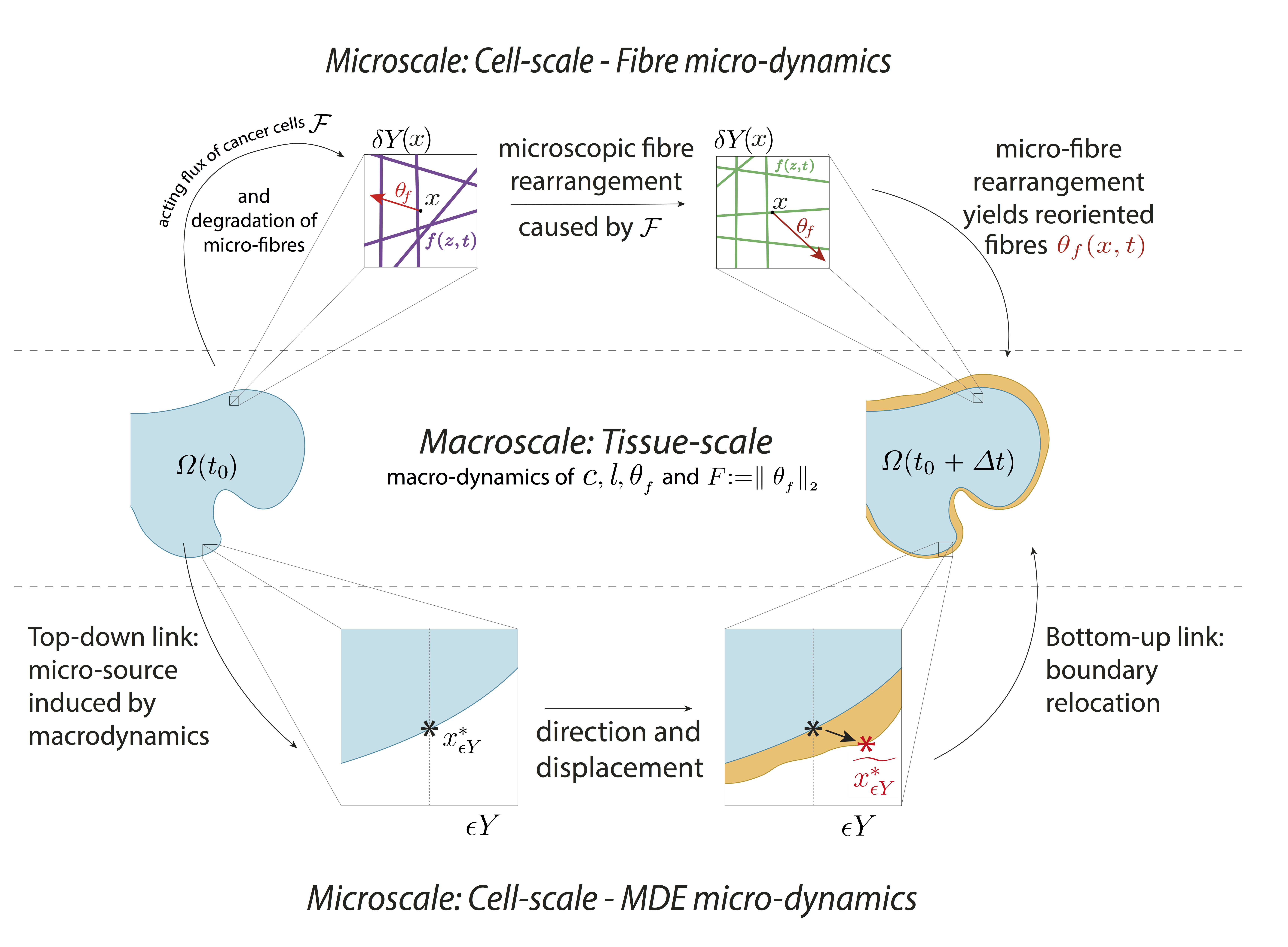}
\caption{\emph{Schematic summary of the two-part multiscale model}}
\label{two-part_model}
\end{figure}

\paragraph{Bottom-up link} During the micro-dynamics, the MDEs present in the peritumoural region interact with the ECM distribution captured within each $\epsilon Y$ micro-domain. The local degradation of ECM is dependent on the advancing spatial distribution of MDEs in $\epsilon Y \backslash \Omega(t)$ secreted by the cancer cells. This pattern of degradation gives rise to a movement direction, $\eta_{\epsilon Y}$, and a displacement magnitude, $\xi_{\epsilon Y}$ (detailed in \citet{Dumitru_et_al_2013}), that ultimately dictate the movement of the boundary midpoint $x_{\epsilon Y}^{*}$ to a new position $\widetilde{x^{*}_{\epsilon Y}}$. This process is the catalyst behind the expansion of the macroscopic tumour boundary. \dt{Thus, }the \emph{bottom-up} link of the model between the molecular activities of MDEs on the tumour invasive edge and the macroscopic boundary $\Omega(t_{0})$, is realised through the macro-scale boundary movement effectuated by the micro-dynamics of the proteolytic enzymes, resulting in an expanded tumour domain, $\Omega(t_{0}+\Delta t),$ on which the invasion process continues. 

\subsection{Summary of model}

As stated previously, the two-part multiscale model is comprised of two interconnected multiscale systems that share the same macro-scale whilst having their own distinct micro-scale dynamics, \dt{as schematically summarised in Figure \ref{two-part_model}}. The macro-scale dynamics governs the spatial distribution of cancer cells and both the fibre and non-fibre components of the ECM. The first multiscale system controls the dynamic redistribution of micro-fibres, weighted according to the cancer cell and macroscopic fibre distributions and \dt{triggers a micro-scale spatial rearrangement of micro-fibres} via the \dt{combined} spatial flux of the \dt{two} cancer cell \dt{sub-populations}. The second multiscale system reacts to a source of MDEs induced on the boundary by the spatial distribution of cancer cells. The microscopic distribution of MDEs instigates degradation of the peritumoural ECM and it is from this pattern of degradation that the position of the tumour boundary is changed. The microscopic change in the boundary is translated back to the tissue-level and the macro-dynamics continue \dt{leading to a} newly expanded tumour region, \dt{where the cancer invasion process continues its dynamics.}

\section{Numerical approaches and initial conditions for computations}

\dt{Building on the multiscale moving boundary computational framework introduced in \citet{Dumitru_et_al_2013}  combined with its very recent novel extension introduced in \citet{Shutt_2018} to address the multiscale fibres dynamics in the bulk of the tumour, we developed a new computational approach to explore the complex multiscale evolution of a heterotypic tumour with two cancer cell sub-populations within a fibrous environment}.  We consider a uniform spatial mesh of size $h=0.03125$ to solve the macroscale computations on the expanding tumour domain, whilst using an off-grid approach for the calculation of the macroscopic adhesion term that decomposes the cell-sensing region to approximate the adhesive flux at each spatio-temporal node. To complete the tumour macro-dynamics we use a novel predictor-corrector scheme developed in \citet{Shutt_2018} that accounts for the complexity of the cancer dynamics. \dt{Further, to obtain the microscopic boundary relocation, we explore the \emph{top-down} and \emph{bottom-up} links as well as solve the micro-dynamics via } finite element approximation \dt{as detailed in} \citet{Dumitru_et_al_2013}. \dt{The simulations of the model equations were all performed on MATLAB.}

\begin{figure}
\centering
\includegraphics[scale=0.3]{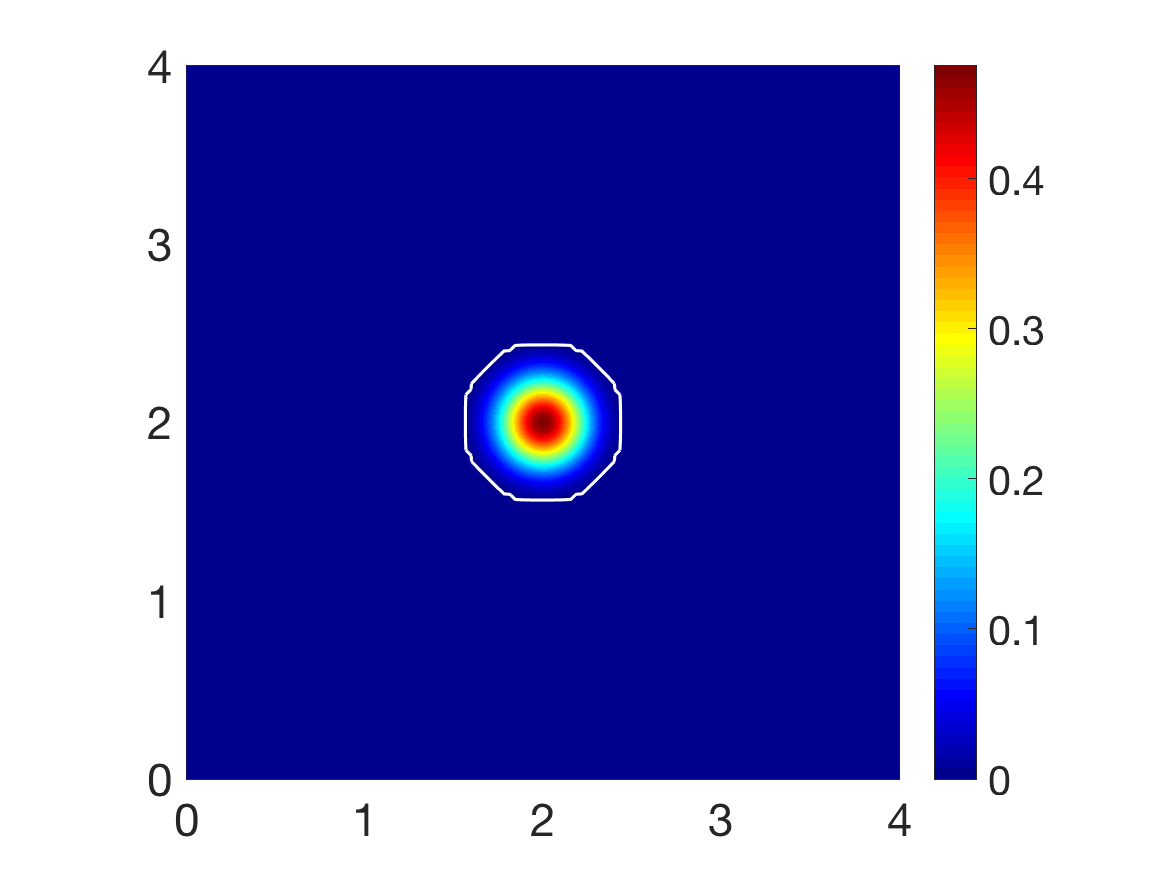}
\caption{\emph{Initial condition for cancer cell population $c_{1}$.}}
\label{fig:canceric}
\end{figure}

\begin{figure}
\centering
\begin{subfigure}{0.5\textwidth}
  \includegraphics[width=\linewidth]{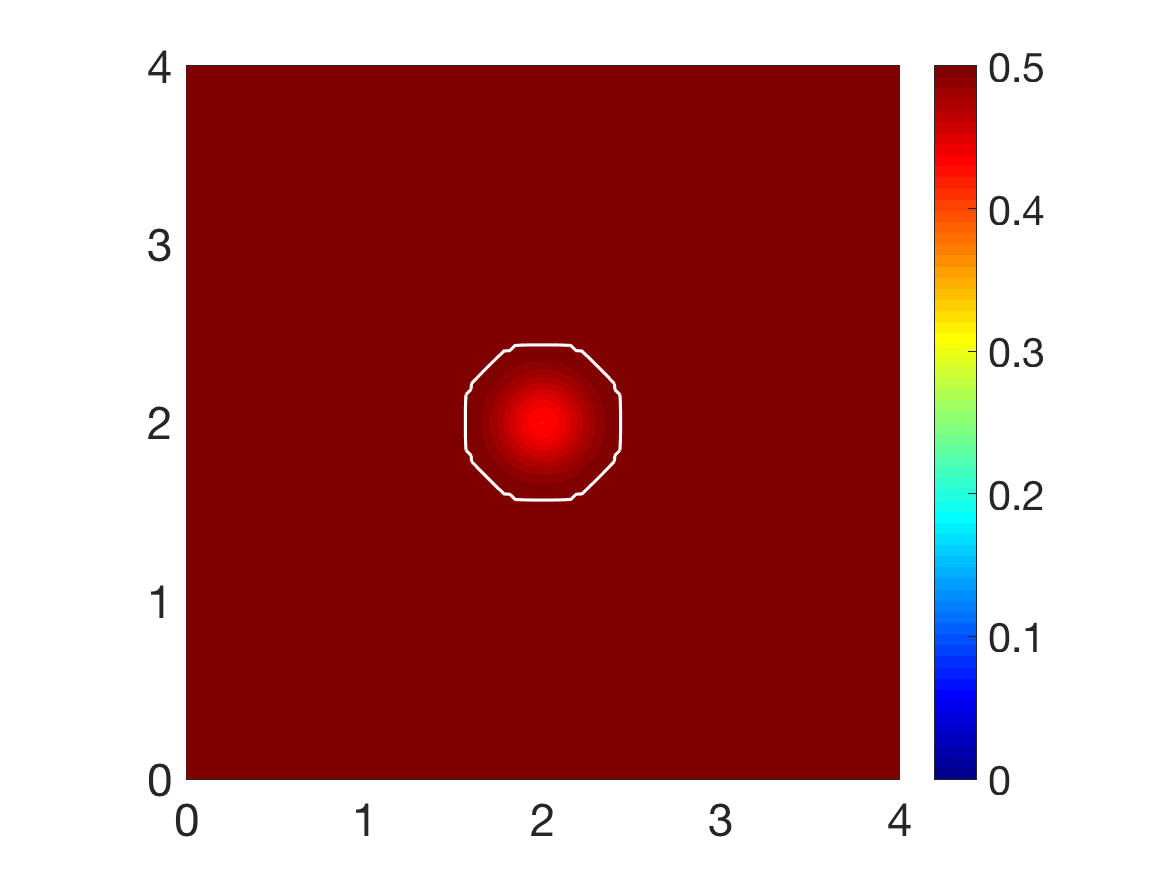}
  \caption{\emph{\rs{Homogeneous non-fibre ECM}}}
\end{subfigure}\hfil 
\begin{subfigure}{0.5\textwidth}
  \includegraphics[width=\linewidth]{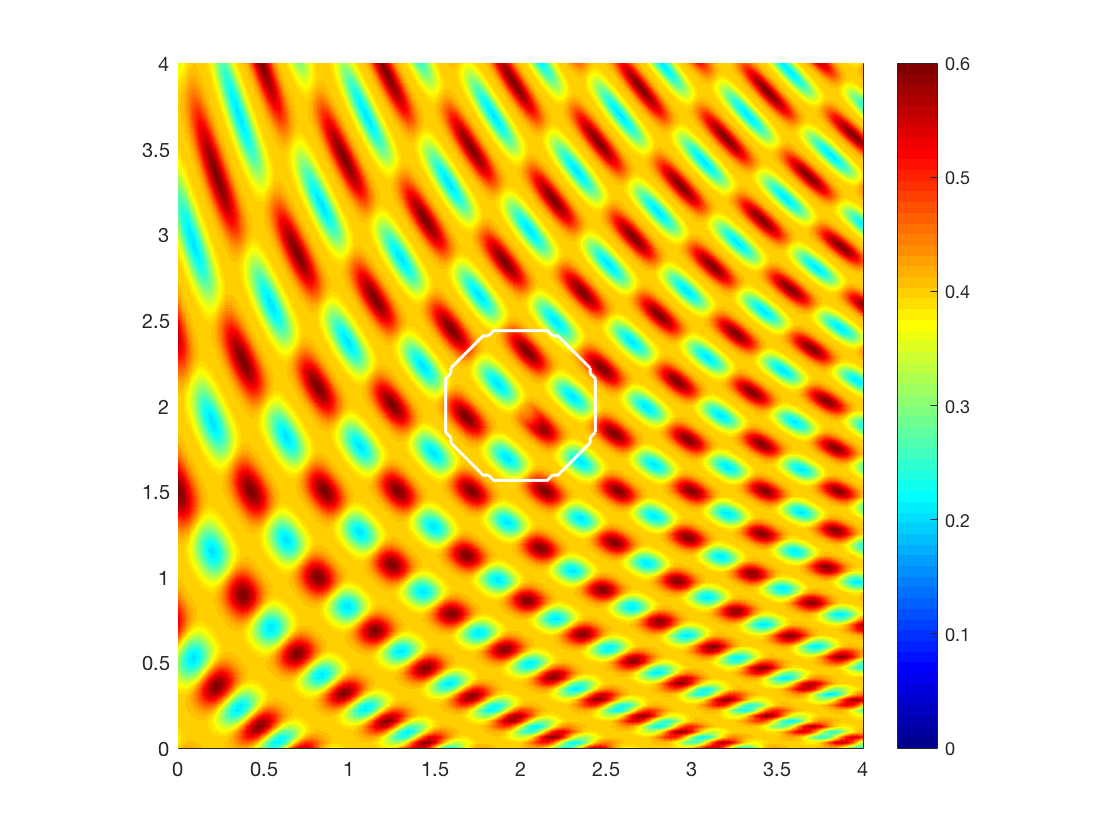}
  \caption{\emph{\rs{Heterogeneous non-fibre ECM}}}
\end{subfigure}\hfil 
\caption{\rs{\emph{Initial conditions \dt{for non-fibre ECM phase: (a)} homogeneous (a); and (b) heterogeneous}}}
\label{fig:non-fibecmic}
\end{figure}

In the following sections we consider the initial cancer cell population $c_{1}$ to occupy the region $\Omega(0)=\Bila((2,2),0.5)$ positioned in the centre of $Y$, Figure \ref{fig:canceric}, whilst the distribution of cell population $c_{2}$ is initially zero, i.e., 
\begin{align}
c_{1}(x,0)&=0.5\left(\text{exp}\left(-\frac{||x-(2,2)||^2_2}{0.03}\right)-\text{exp}(-28.125)\right)\left(\chi_{_{\Bila((2,2),0.5-\gamma)}} \ast \psi_{\gamma}\right), \label{cancer1ic} \\
c_{2}(x,0)&=0, \label{cancer2ic}
\end{align}
where $\psi_{\gamma}$ is the standard mollifier detailed in \ref{mollifier} that acts within a radius $\gamma <<\frac{\Delta x}{3}$ from $\partial \Bila((2,2),0.5-\gamma)$ to smooth out the characteristic function $\chi_{_{\Bila((2,2),0.5-\gamma)}}$. \dt{Furthermore, for the non-fibre ECM phase, }\rs{we consider both a homogeneous and heterogeneous initial conditions,  with the homogeneous initial conditions illustrated in Figure \ref{fig:non-fibecmic}(a) and given by
\bequ
l(x, 0) = min\{0.5, 1 - c_{1}(x,0)\},
\label{eq:homomatrix}
\eequ
and heterogeneous initial condition shown in Figure \ref{fig:non-fibecmic}(b) and given by 
\begin{equation}
l(x,0)=\text{min}\left\{ h(x_1,x_2), 1- c(x,0)\right\},
\label{eq:matrix_IC}
\end{equation}
where 
\begin{align*}
h(x_1,x_2)&=\frac{1}{2}+\frac{1}{4}\text{sin}(\zeta x_1 x_2)^3 \cdot \text{sin}(\zeta \frac{x_2}{x_1}),  \\
(x_1,x_2)&= \frac{1}{3}(x+1.5) \ \in [0,1]^2 \ \text{for} \ x \in D, \quad \zeta = 7\pi.
\end{align*}}
these being previous used also in \citet{Shutt_2018}.

\begin{figure}
    \centering 
\begin{subfigure}{0.5\textwidth}
  \includegraphics[width=\linewidth]{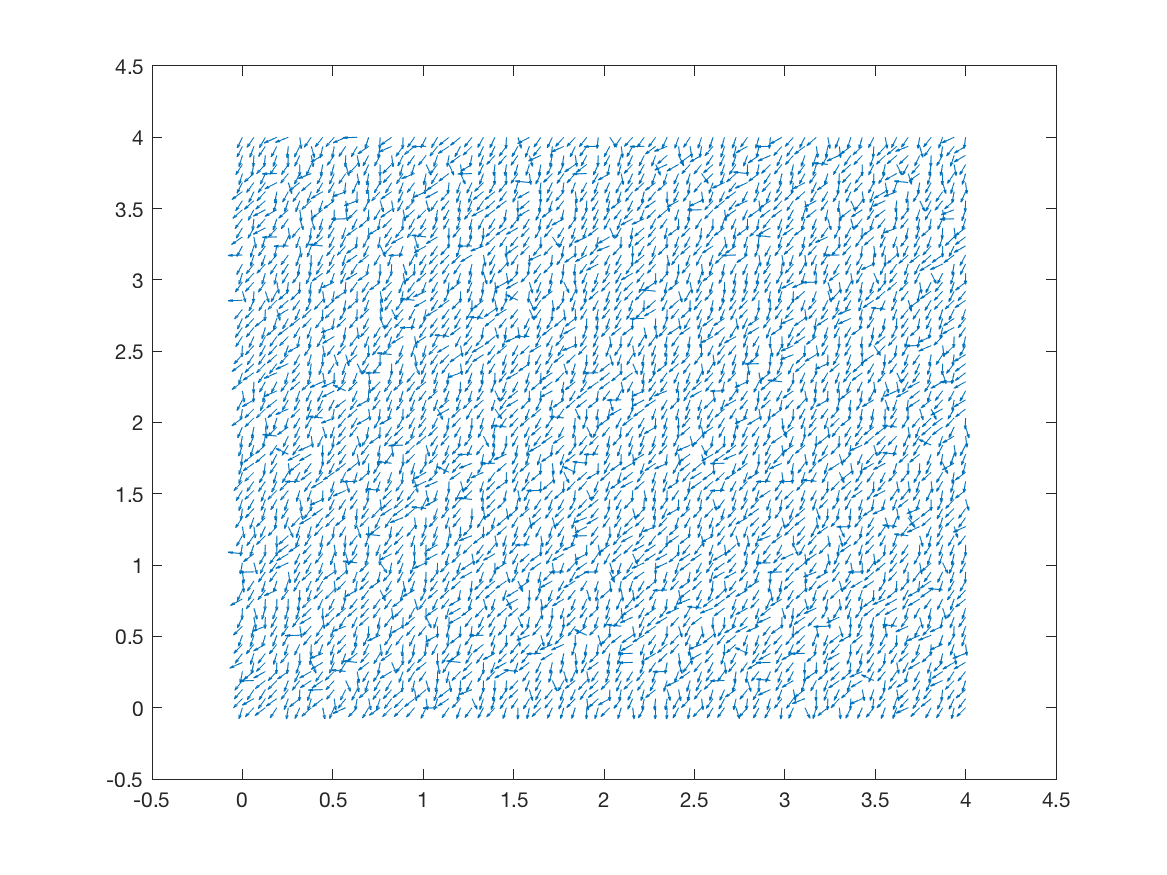}
  \caption{\emph{\dt{$\theta_{_{f}}(x,0)$ $-$ homogeneous magnitude}}}
\end{subfigure}\hfil 
\begin{subfigure}{0.5\textwidth}
  \includegraphics[width=\linewidth]{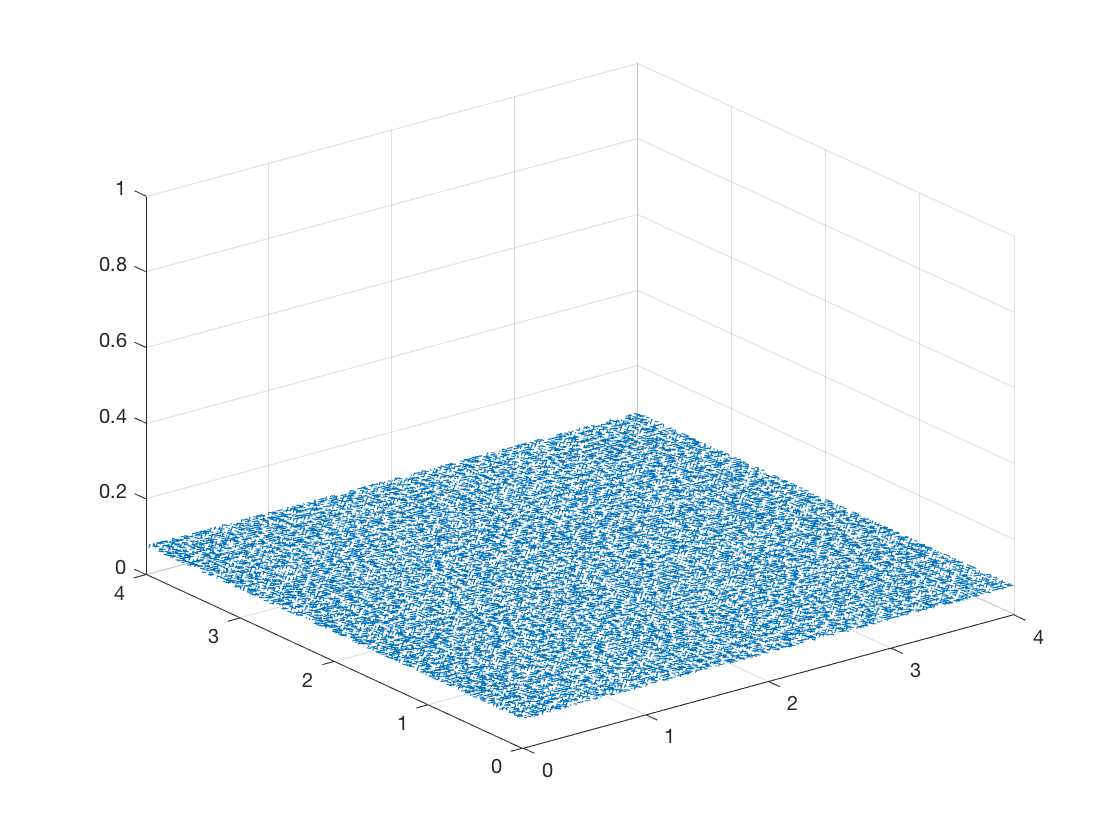}
  \caption{\emph{\dt{$\theta_{_{f}}(x,0)$ $-$ homogeneous magnitude in a 3D plot}}}
\end{subfigure}\hfil 

\medskip
\begin{subfigure}{0.5\textwidth}
  \includegraphics[width=\linewidth]{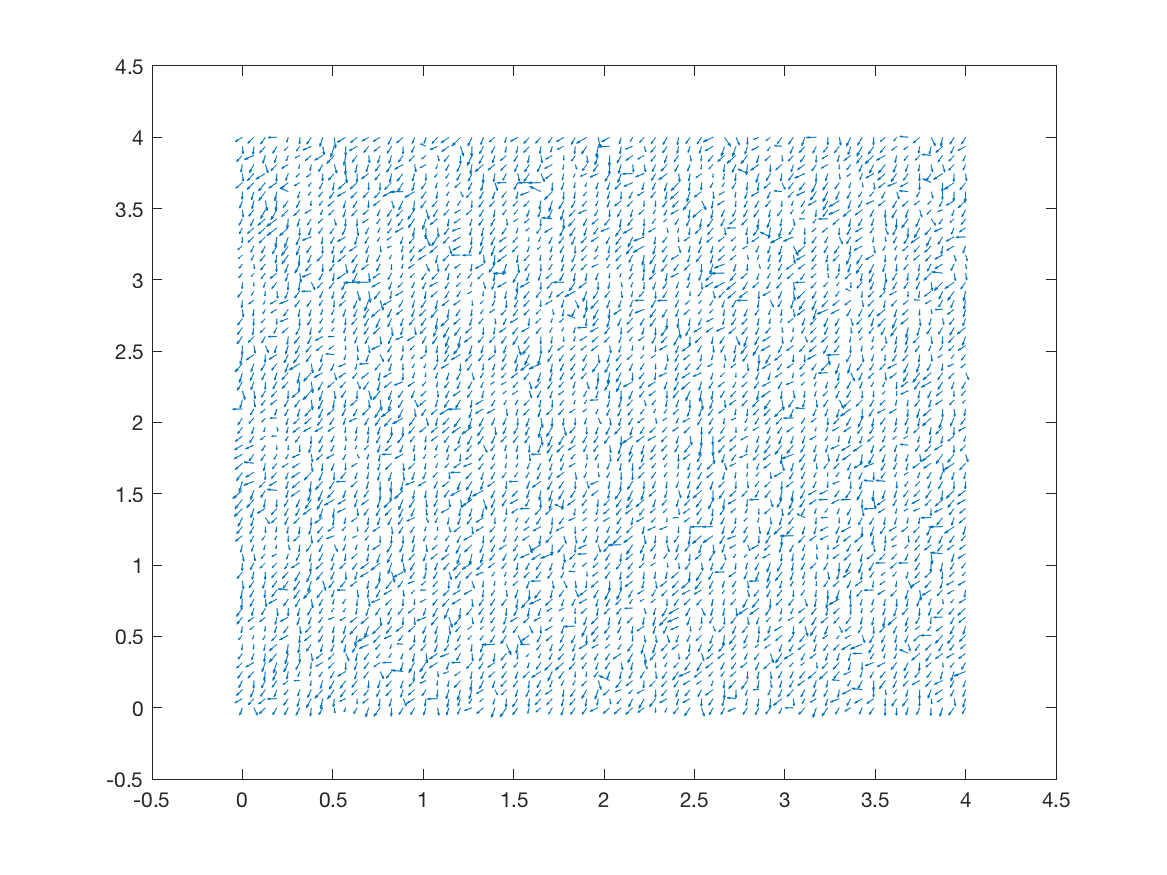}
  \caption{\emph{\dt{$\theta_{_{f}}(x,0)$ $-$ heterogeneous magnitude}}}
\end{subfigure}\hfil 
\begin{subfigure}{0.5\textwidth}
  \includegraphics[width=\linewidth]{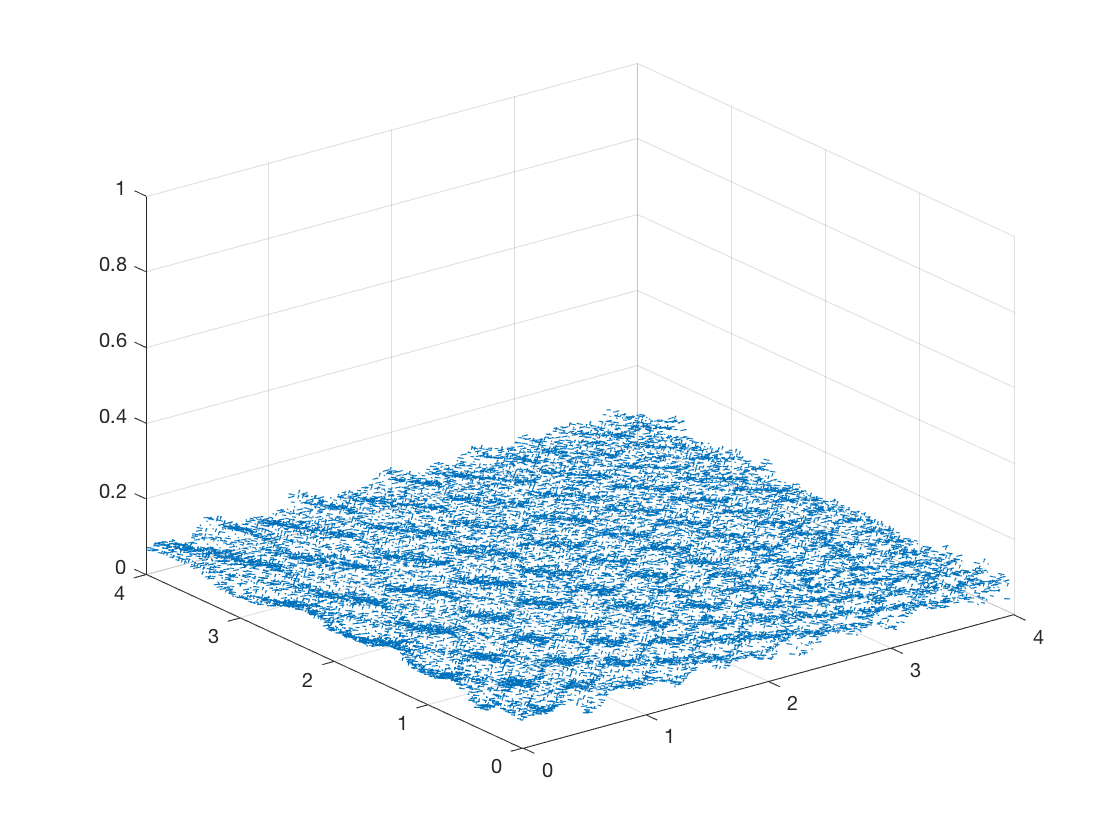}
  \caption{\emph{\dt{$\theta_{_{f}}(x,0)$ $-$ heterogeneous magnitude in a 3D plot}}}
\end{subfigure}\hfil 

\caption{\emph{\dt{Initial conditions for ECM fibres phase: (a) shows oriented fibres of homogeneous magnitude while (b) shows their corresponding 3D plot; (c) shows oriented fibres of heterogeneous magnitude while (d) shows their corresponding 3D plot.}}}
\label{fig:ichetero}
\end{figure}

For the initial distribution of the ECM fibre phase, we assume a distribution of five pre-assigned \dt{micro-}fibre distributions \dt{of the five different micro-fibre patterns generated along the family of path-networks $\{P^{1}_{i}\}_{i=1..5}$  and assigned randomly onto $\delta Y(x):=x+\delta Y$ as detailed in \ref{microfibres}).} \rs{To represent a homogeneous distribution of fibres, we calibrate the maximal height of the microfibres to be uniform across all micro-domains, resulting in a macroscopic fibre distribution $F(\cdot,t)$ that represents a percentage $p$ of the mean density of the non-fibrous homogeneous ECM phase \eqref{eq:homomatrix}, illustrated in Figure \ref{fig:ichetero}(a)-(b). On the other hand, a heterogeneous distribution of fibres is achieved by allowing the maximal height of the microfibres to be determined by a heterogeneous pattern, for example, the initial condition for a heterogeneous ECM non-fibre phase \eqref{eq:matrix_IC}, Figure \ref{fig:ichetero}(c)-(d). We set the maximal height of the micro-fibres in each $\delta Y(x)$, centred at $x$, to correspond to the distribution $l(x,0)$ for all $x \in Y$, resulting in the macroscopic fibre \dt{magnitude spatial} distribution $F(\cdot,t)$ to represent again a percentage $p$ of the heterogeneous non-fibre ECM phase.}

\section{Computational simulation results}
\subsection{Homogeneous non-fibre ECM component} 

\begin{figure}
    \centering 
\begin{subfigure}{0.5\textwidth}
  \includegraphics[width=\linewidth]{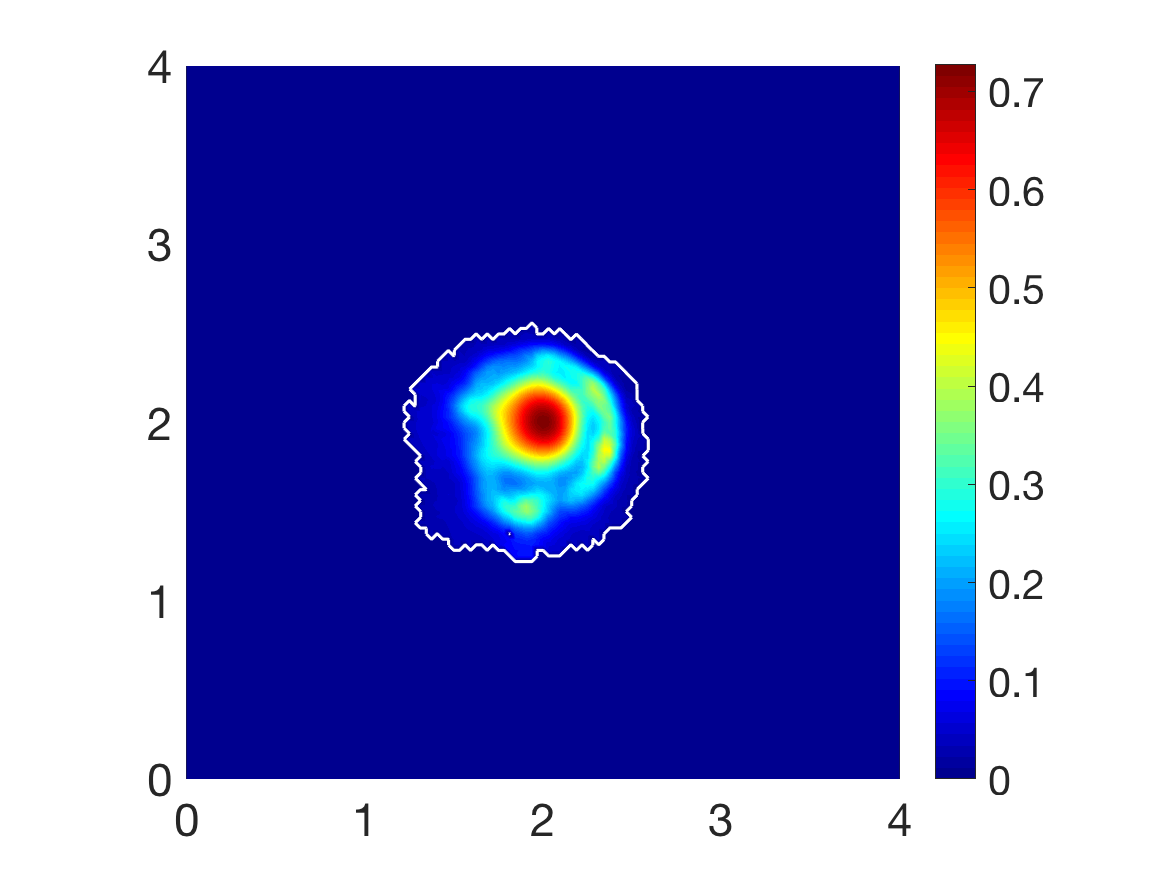}
  \caption{\emph{Cell population 1}}
\end{subfigure}\hfil 
\begin{subfigure}{0.5\textwidth}
  \includegraphics[width=\linewidth]{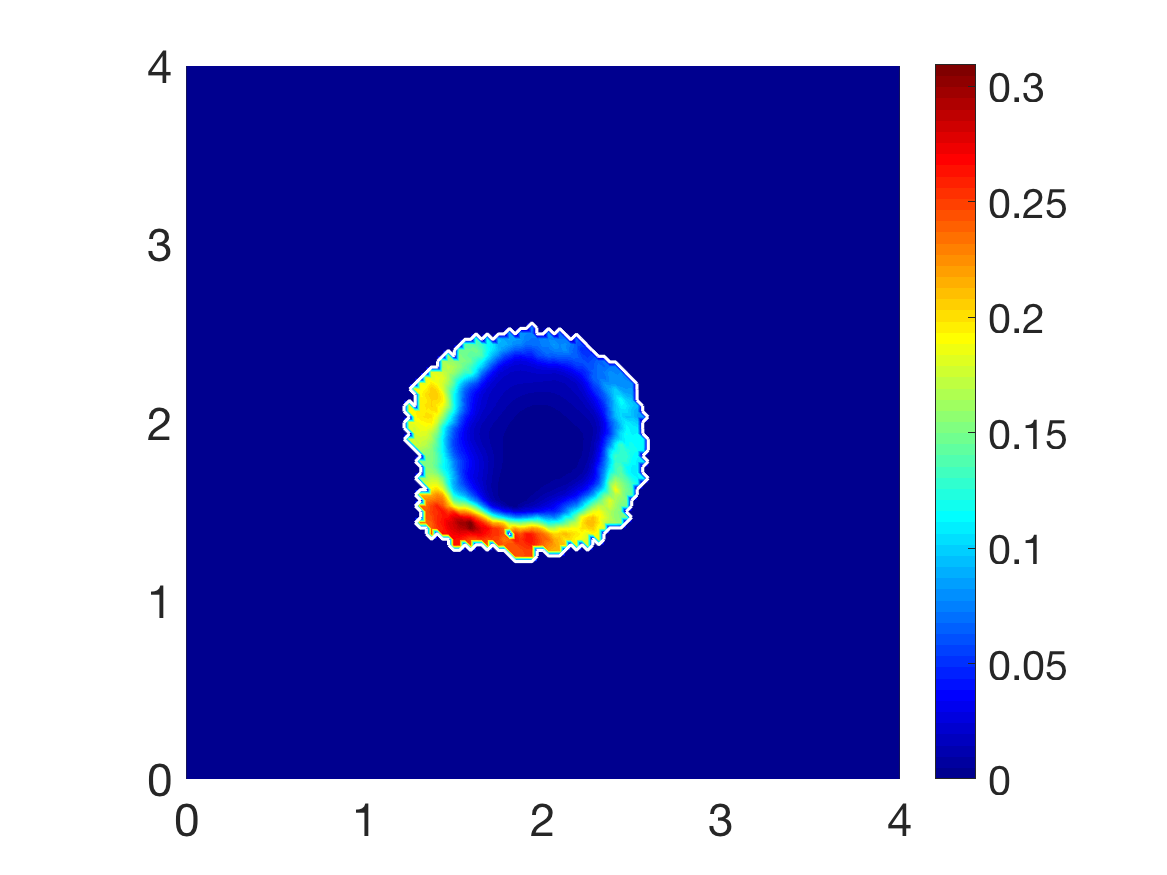}
  \caption{\emph{Cell population 2}}
\end{subfigure}\hfil 

\medskip
\begin{subfigure}{0.5\textwidth}
  \includegraphics[width=\linewidth]{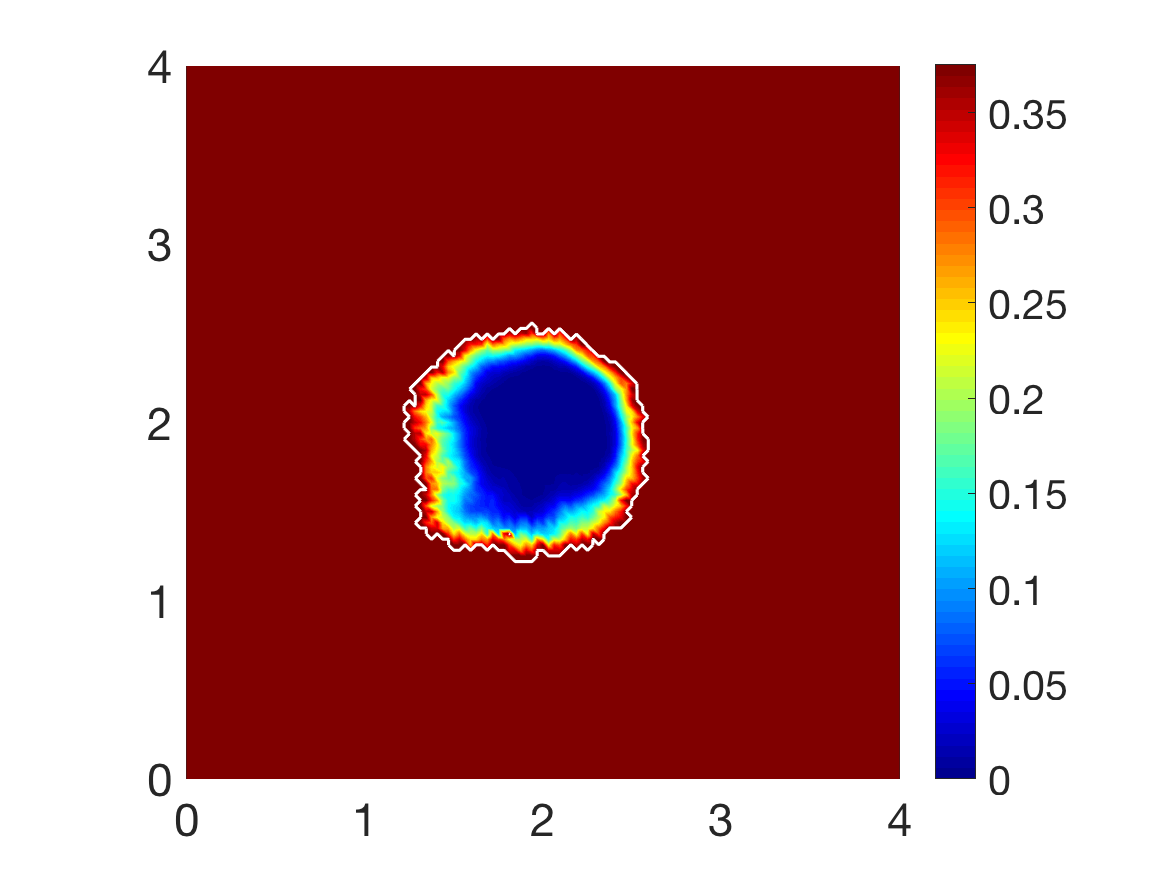}
  \caption{\emph{Non-fibres ECM phase}}
\end{subfigure}\hfil 
\begin{subfigure}{0.5\textwidth}
  \includegraphics[width=\linewidth]{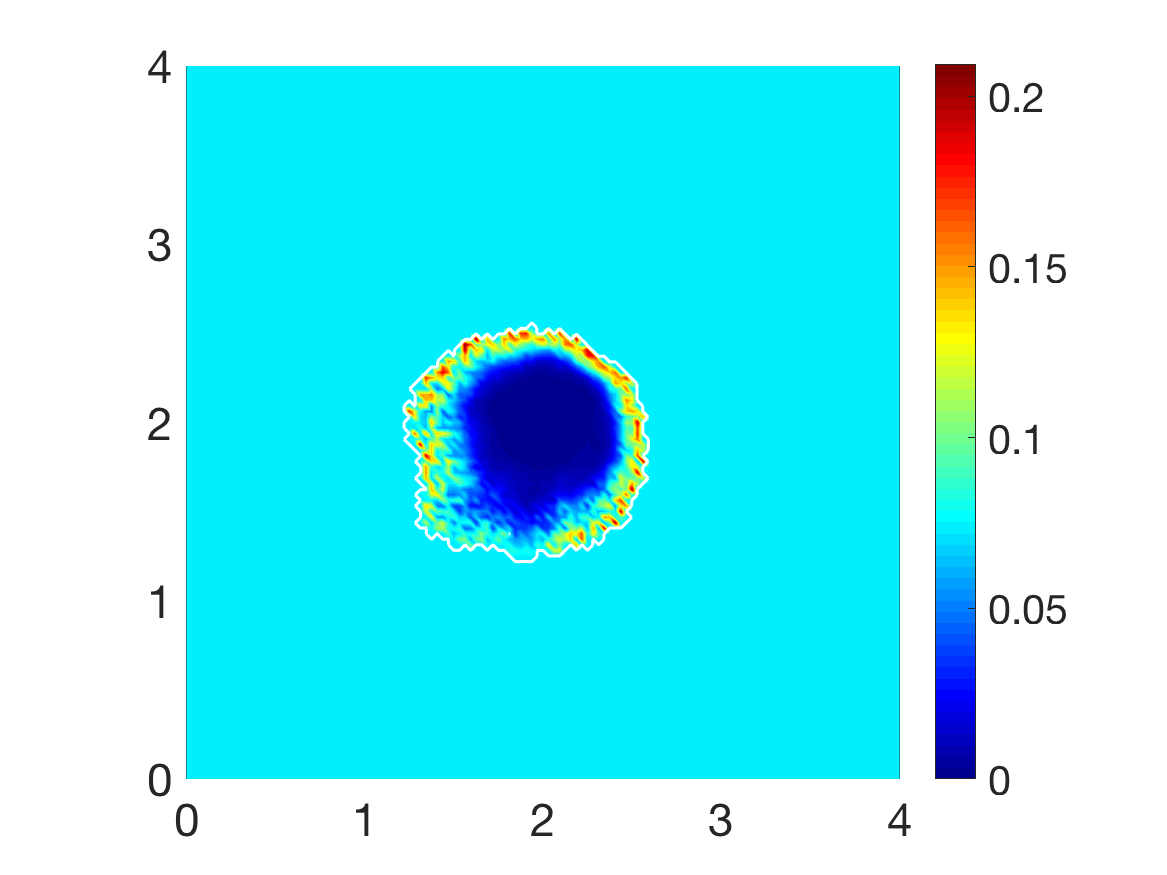}
  \caption{\emph{Macro-scale ECM fibres magnitude}}
\end{subfigure}\hfil 

\medskip
\begin{subfigure}{0.5\textwidth}
  \includegraphics[width=\linewidth]{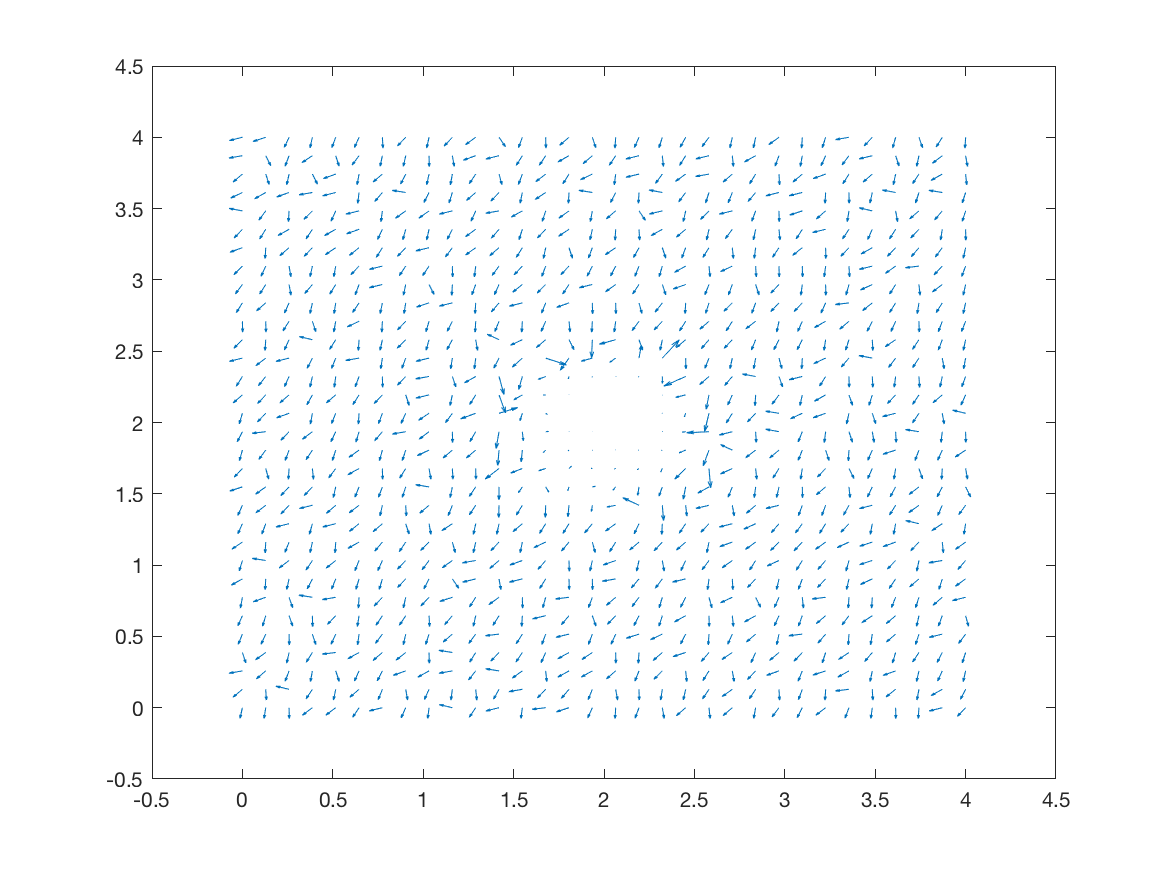}
  \caption{\emph{Oriented macro-scale ECM fibres}}
\end{subfigure}\hfil 
\begin{subfigure}{0.5\textwidth}
  \includegraphics[width=\linewidth]{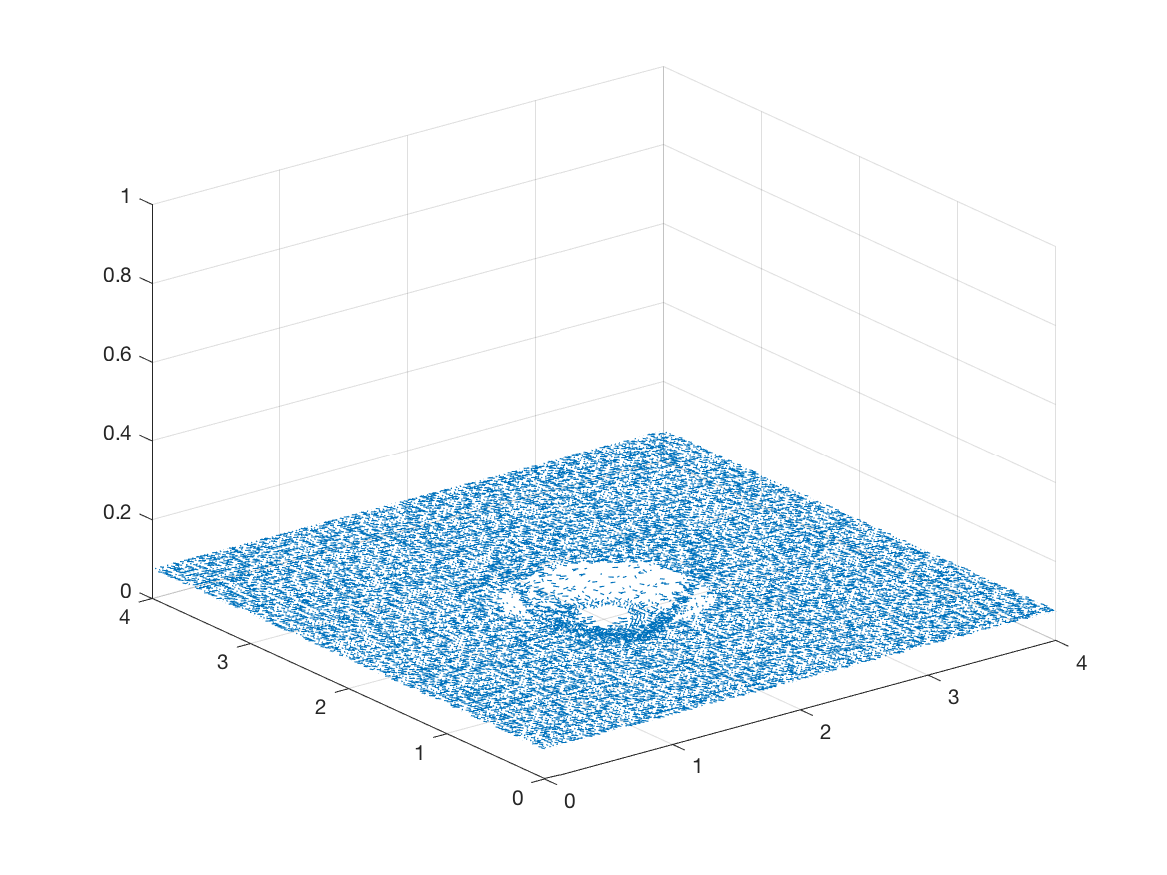}
  \caption{\emph{Oriented macro-scale ECM fibres in a 3D plot}}
\end{subfigure}\hfil 

\caption{Simulations at stage $75\Delta t$ with a homogeneous distribution of non-fibres and a random initial $15\%$ homogeneous distribution of fibres.}
\label{fig:homol_homof_15_75_cancer}
\end{figure}

In the first instance, the initial condition for the non-fibre ECM component will be taken as the homogeneous distribution, namely as $l(x,0)=min\{0.5,1-c_{1}(x,0)\}$. The initial condition for both cancer cell populations is given in \eqref{cancer1ic}, \eqref{cancer2ic}, and the fibres take an initial homogeneous macroscopic distribution of $15\%$ the non-fibre phase, with the combined ECM density $v(x,0)=l(x,0)+F(x,0)$. Using the parameter set $\Sigma$ from \ref{paramSection}, we show the computational results at the final time stage for the evolution of both individual cancer cell populations in subfigures \ref{fig:homol_homof_15_75_cancer}(a), \ref{fig:homol_homof_15_75_cancer}(b), the non-fibre ECM phase and fibre ECM phase in subfigures \ref{fig:homol_homof_15_75_cancer}(c) and \ref{fig:homol_homof_15_75_cancer}(d) respectively, presented with the vector field of oriented fibres coarsened four times, \ref{fig:homol_homof_15_75_cancer}(e), and a 3D representation of fibres \ref{fig:homol_homof_15_75_cancer}(f).

Figure \ref{fig:homol_homof_15_75_cancer} displays computations at final stage $75\Delta t$, during which time mutations from population $c_1$ to $c_2$ have begun, starting at stage $5\Delta t$. Previously it was observed that in the presence of a homogeneous ECM, a tumour was limited to a symmetric pattern of invasion \citep{shutt_chapter}, however with the addition of an initially homogeneous fibre distribution we witness a different pattern of invasion in \ref{fig:homol_homof_15_75_cancer}(c). The primary cell population $c_1$, \ref{fig:homol_homof_15_75_cancer}(a), has been manipulated and reshaped by the underlying fibre network. Although the fibre phase of the ECM begins as a macroscopically homogeneous distribution, once fibre rearrangement occurs, higher density regions of fibres are formed, \ref{fig:homol_homof_15_75_cancer}(d), and these areas become increasing influential on the direction of invasion. Since mutations have begun, we see the emergence of cells in population $c_2$, \ref{fig:homol_homof_15_75_cancer}(b). The pattern of cells in population $c_2$ is correspondent to the fibre orientations displayed in \ref{fig:homol_homof_15_75_cancer}(e), where the general direction of fibres are aiming \dt{roughly} towards the origin $(0,0)$. Thus, the distribution of cells present in Figure \ref{fig:homol_homof_15_75_cancer}(b) are more prominent in this direction. This behaviour is attributable to population $c_2$ retaining a higher affinity for cell-fibre adhesion than population $c_1$. Throughout the simulations, the ECM is degraded by the cancer cells, most notable by the large low density region in \ref{fig:homol_homof_15_75_cancer}(c), with the highest levels of degradation relative to the regions of highest cell distribution. 

It is clear from the computational results presented in Figure \ref{fig:homol_homof_15_75_cancer} that the tumour boundary has expanded in the general macroscopic fibre direction and strayed away from a symmetric route of invasion. Hence we conclude from these computations and comparisons with previous work, \citep{shutt_chapter}, that the underlying fibre network strongly influences the direction of invasion.

\begin{figure}
    \centering 
\begin{subfigure}{0.5\textwidth}
  \includegraphics[width=\linewidth]{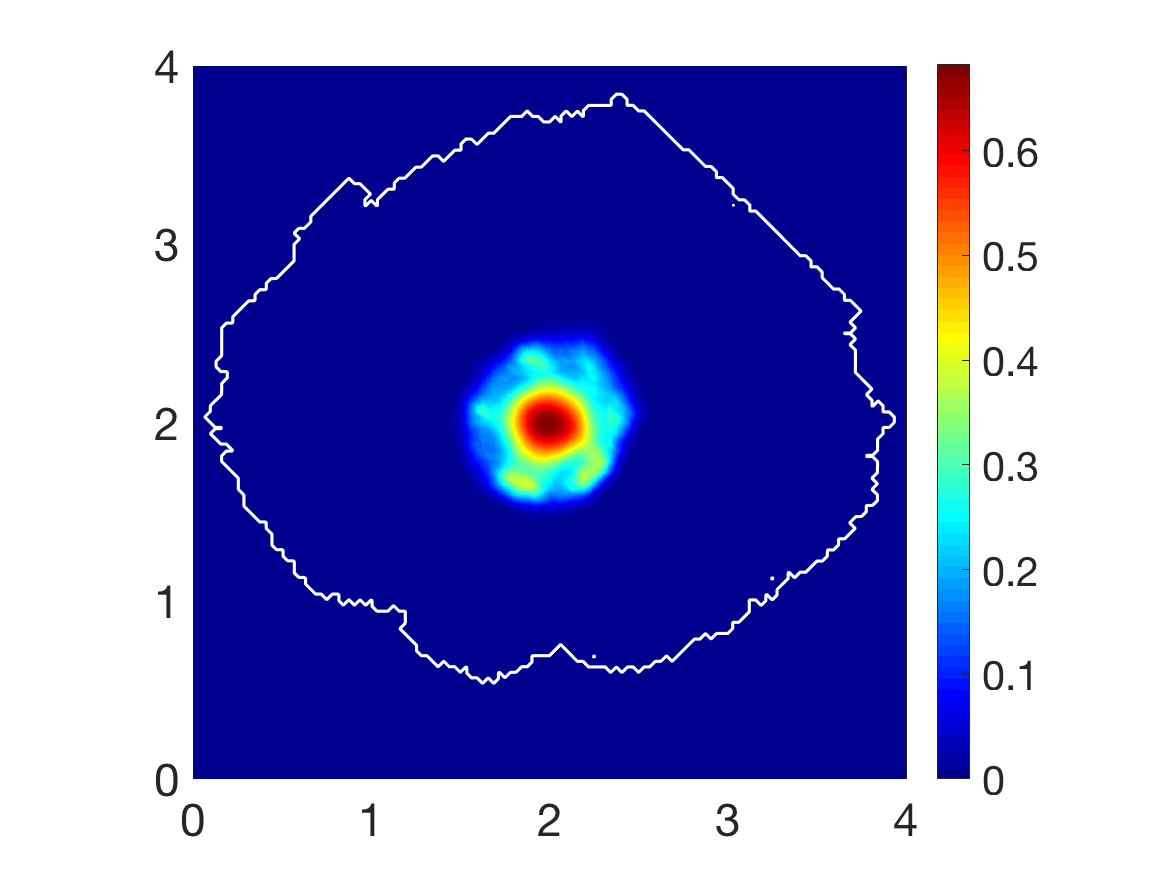}
  \caption{\emph{Cell population 1}}
\end{subfigure}\hfil 
\begin{subfigure}{0.5\textwidth}
  \includegraphics[width=\linewidth]{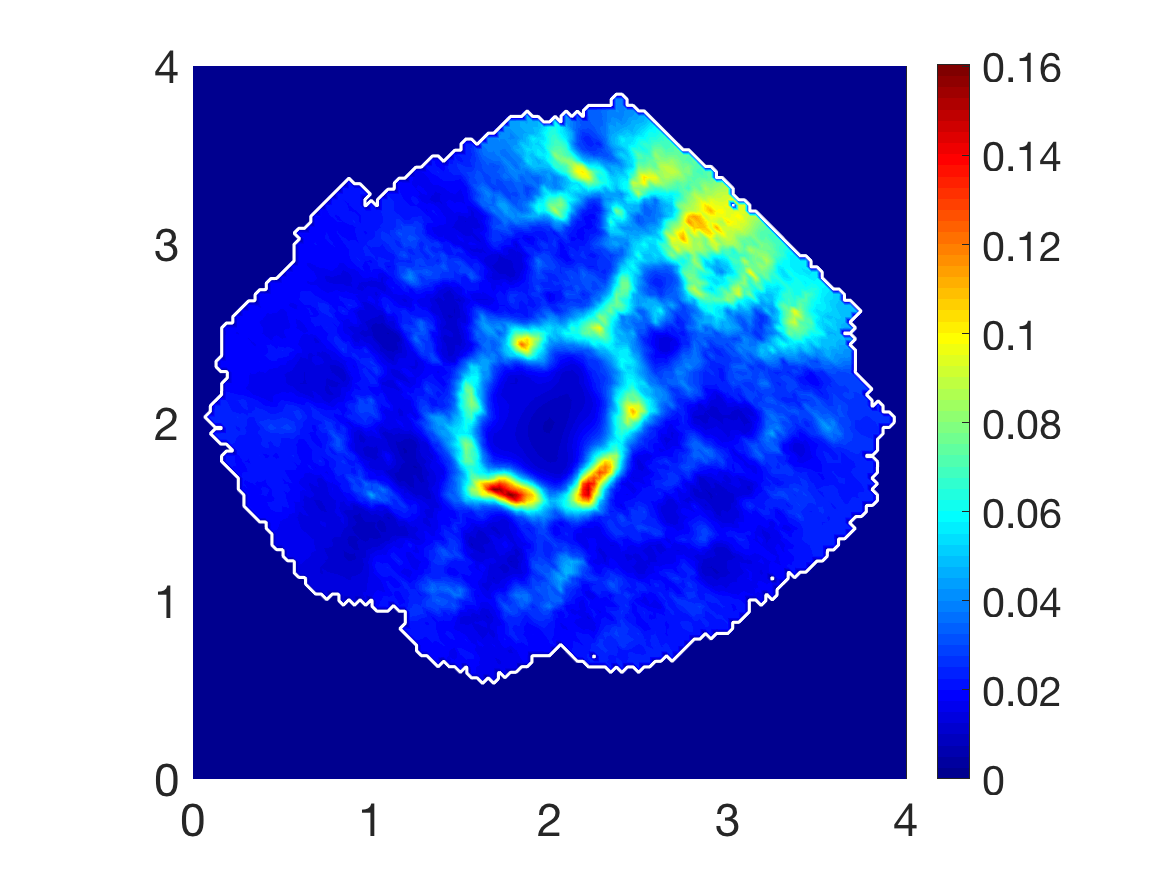}
  \caption{\emph{Cell population 2}}
\end{subfigure}\hfil 

\medskip
\begin{subfigure}{0.5\textwidth}
  \includegraphics[width=\linewidth]{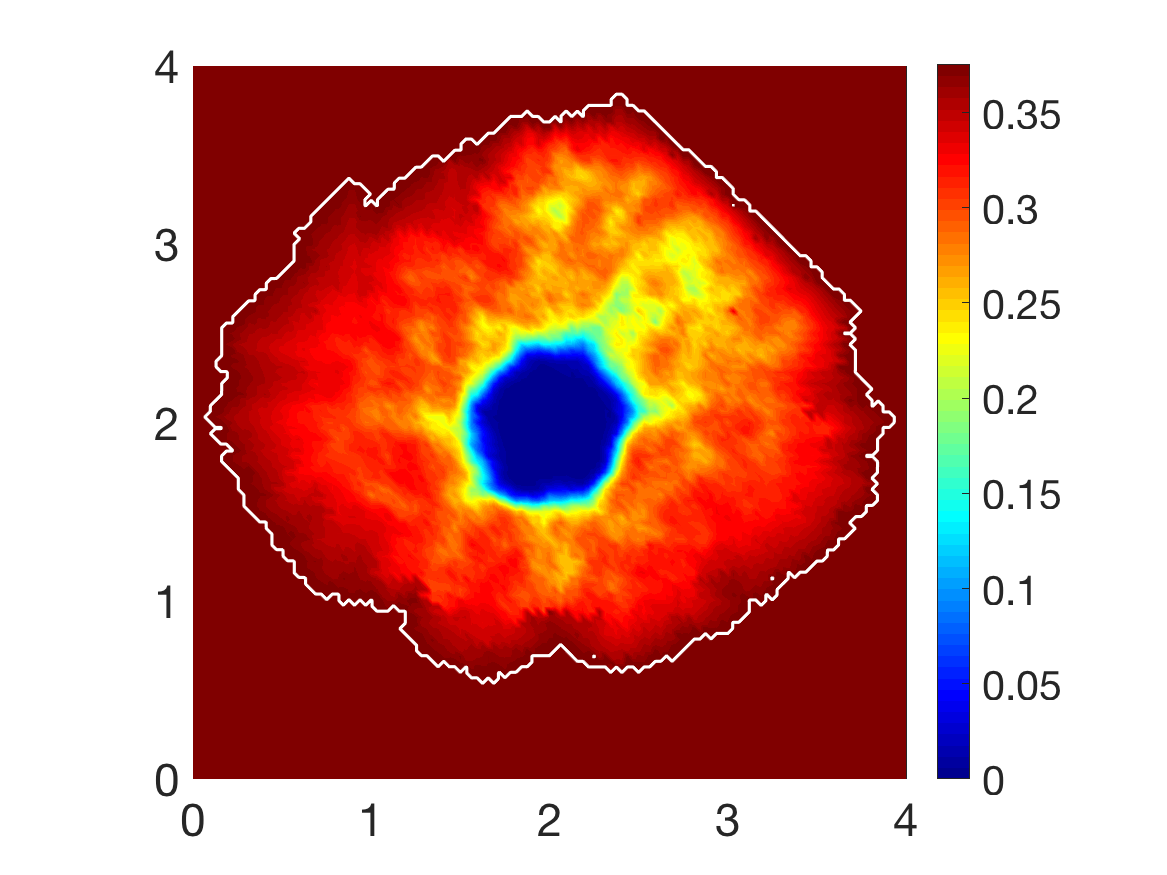}
  \caption{\emph{Non-fibres ECM phase}}
\end{subfigure}\hfil 
\begin{subfigure}{0.5\textwidth}
  \includegraphics[width=\linewidth]{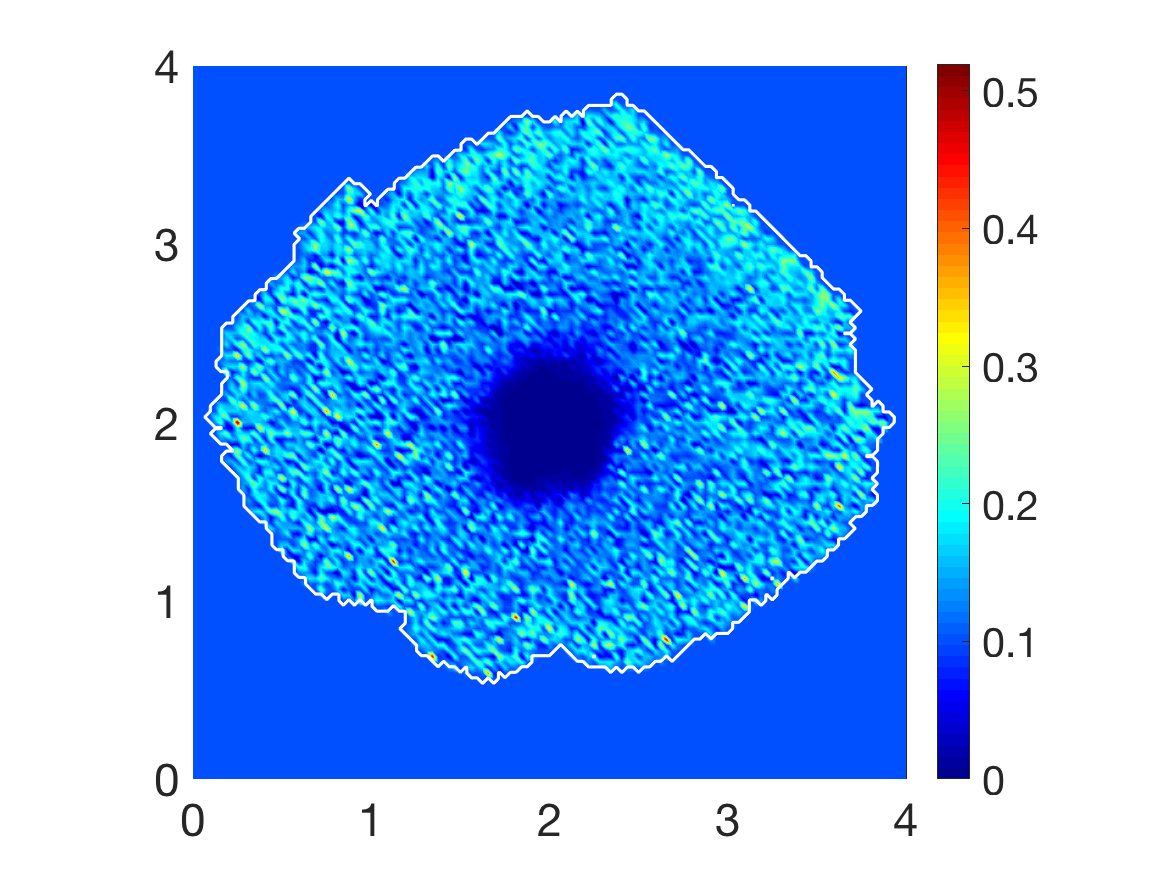}
  \caption{\emph{Macro-scale ECM fibres magnitude}}
\end{subfigure}\hfil 

\medskip
\begin{subfigure}{0.5\textwidth}
  \includegraphics[width=\linewidth]{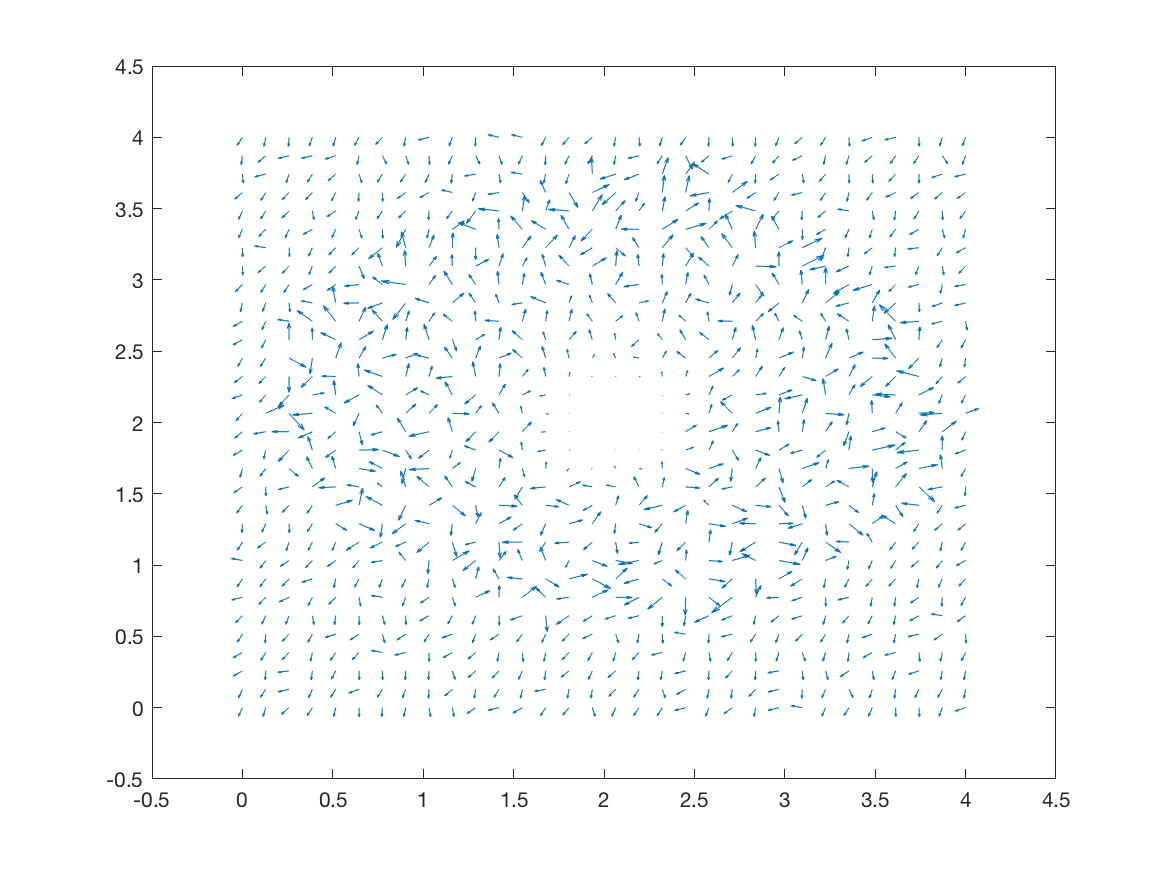}
  \caption{\emph{Oriented macro-scale ECM fibres}}
\end{subfigure}\hfil 
\begin{subfigure}{0.5\textwidth}
  \includegraphics[width=\linewidth]{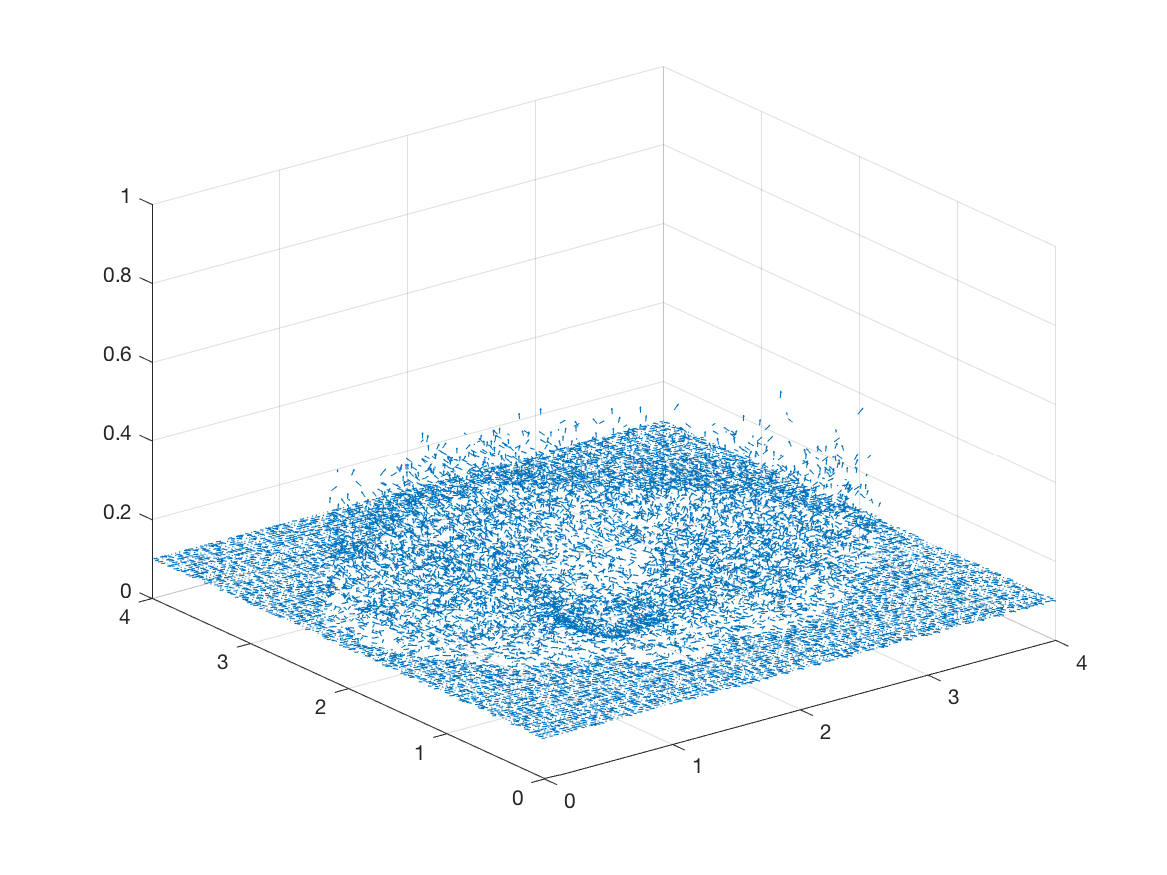}
  \caption{\emph{Oriented macro-scale ECM fibres in a 3D plot}}
\end{subfigure}\hfil 

\caption{Simulations at stage $70\Delta t$ with a homogeneous distribution of non-fibres and a random initial $20\%$ homogeneous distribution of fibres.}
\label{fig:homol_homof_20_70_cancer}
\end{figure}

To examine the effects of an underlying fibre distribution, we increase the initial homogeneous fibre distribution to $20\%$ of the non-fibre ECM phase. Continuing with the same initial conditions for the non-fibre ECM phase and cancer distributions, we present the computations at final stage $70 \Delta t$, Figure \ref{fig:homol_homof_20_70_cancer}. Comparing with Figure \ref{fig:homol_homof_15_75_cancer}, cells of population $c_{1}$ are exhibiting limited migration within the tumour boundary due to their high self-adhesion rate, \ref{fig:homol_homof_20_70_cancer}(a), whilst cells of population $c_{2}$, \ref{fig:homol_homof_20_70_cancer}(b), have migrated in a direction consistent with the macroscopic fibre orientations in \ref{fig:homol_homof_20_70_cancer}(e). The mutated cell population $c_{2}$ congregates around the cells of population $c_{1}$, this time showing more regions of high distribution, suggesting the increased fibre distribution enables faster migration of cells in population $c_{2}$. The entire ECM (both the fibre and non-fibre phase) presented in subfigures \ref{fig:homol_homof_20_70_cancer}(c)-(d) has been degraded in the centre of the domain where the cancer cells are most dense.

In line with the previous simulations, the micro-fibres have been pushed outwards towards the boundary, however this time the macroscopic representation of the fibres appears as a region of very low fibre density surrounded by a mottled pattern of fibre densities, Figure \ref{fig:homol_homof_20_70_cancer}(d). This behaviour occurs because of two reasons. Firstly, the tumour boundary is progressing faster than the fibres are being rearranged, and because the rearrangement of fibres is restricted to neighbouring domains only, the fibre distributions will never be found to build up explicitly at the rapidly expanding tumour boundary. Secondly, as the degradation of fibres is kept within the bounds of the tumour and dependent on the cancer cell distribution, out-with the dense main body of cancer cells, we see little degradation of the fibres. The increased size of spread of the tumour indicates that the ratio of the underlying fibre network to the non-fibre ECM phase plays a key role in the success of local tumour invasion.

\begin{figure}
    \centering 
\begin{subfigure}{0.5\textwidth}
  \includegraphics[width=\linewidth]{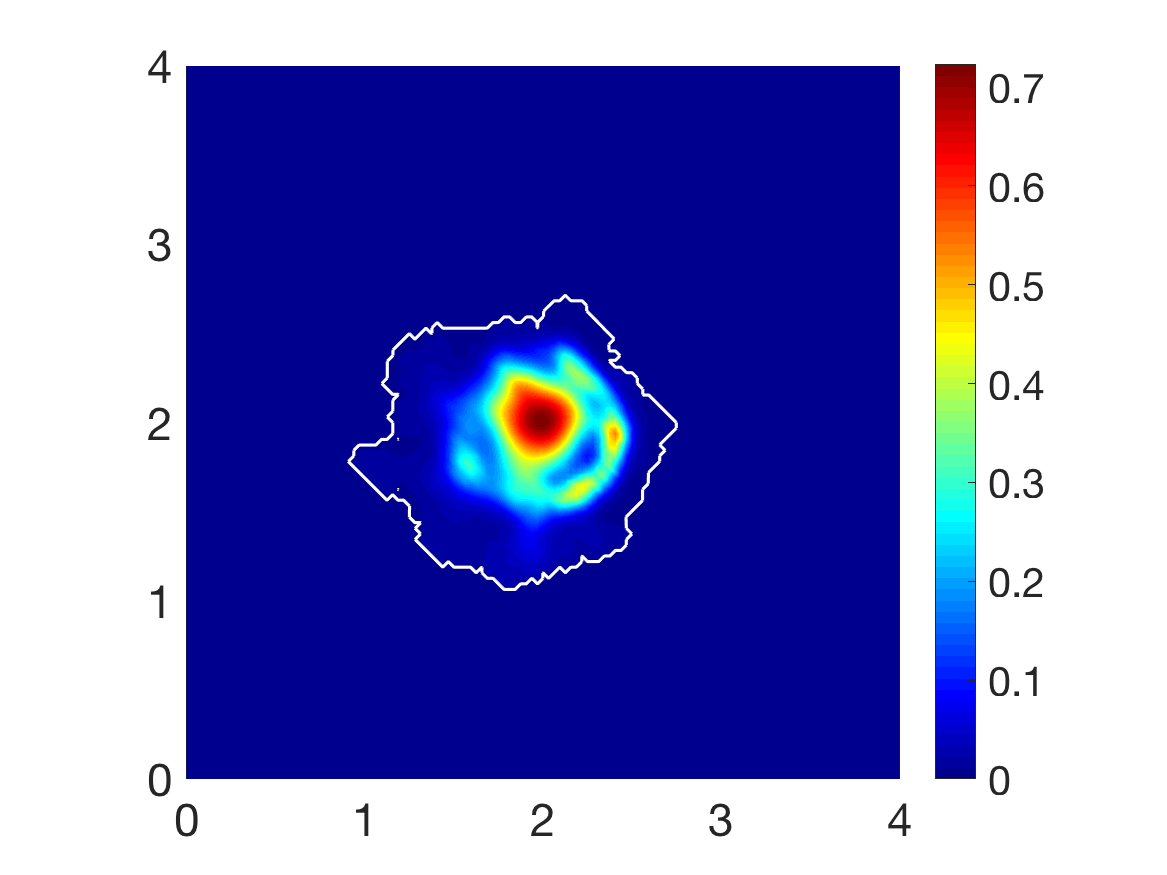}
  \caption{\emph{Cell population 1}}
\end{subfigure}\hfil 
\begin{subfigure}{0.5\textwidth}
  \includegraphics[width=\linewidth]{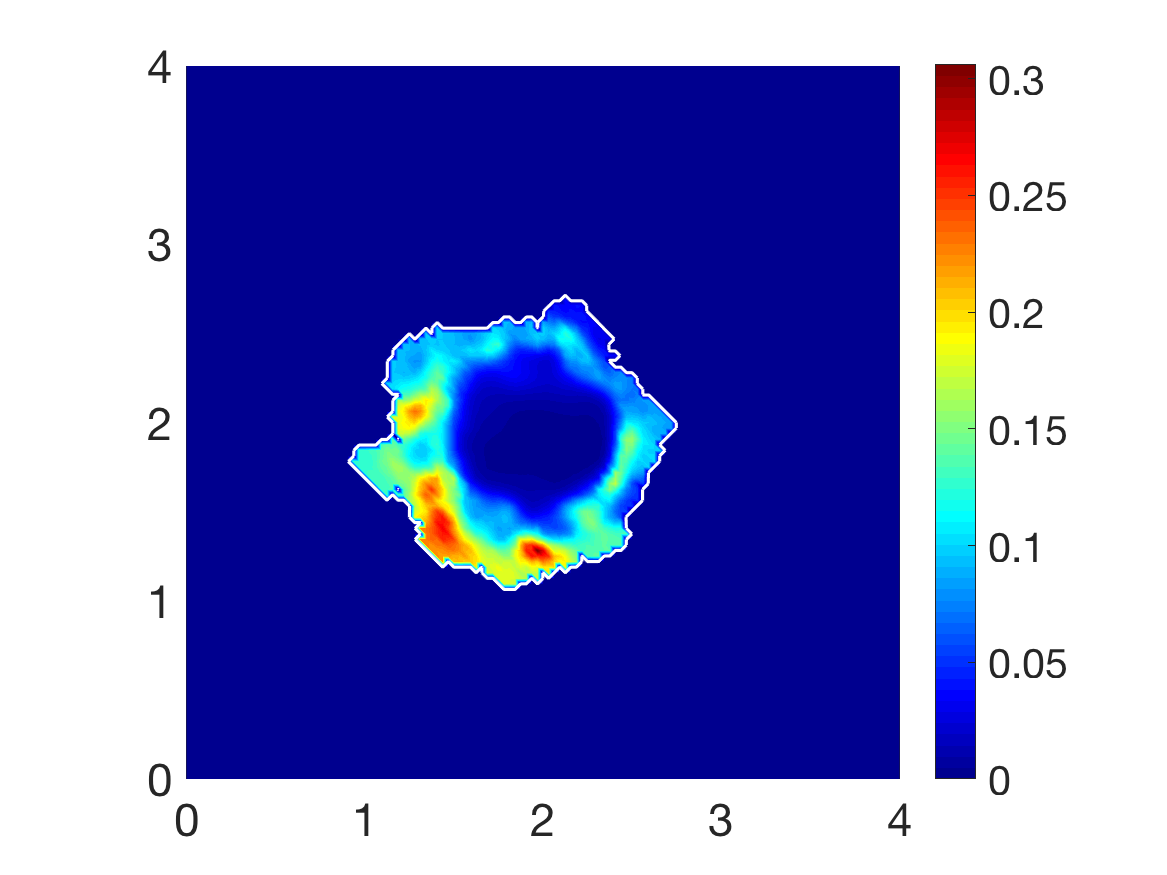}
  \caption{\emph{Cell population 2}}
\end{subfigure}\hfil 

\medskip
\begin{subfigure}{0.5\textwidth}
  \includegraphics[width=\linewidth]{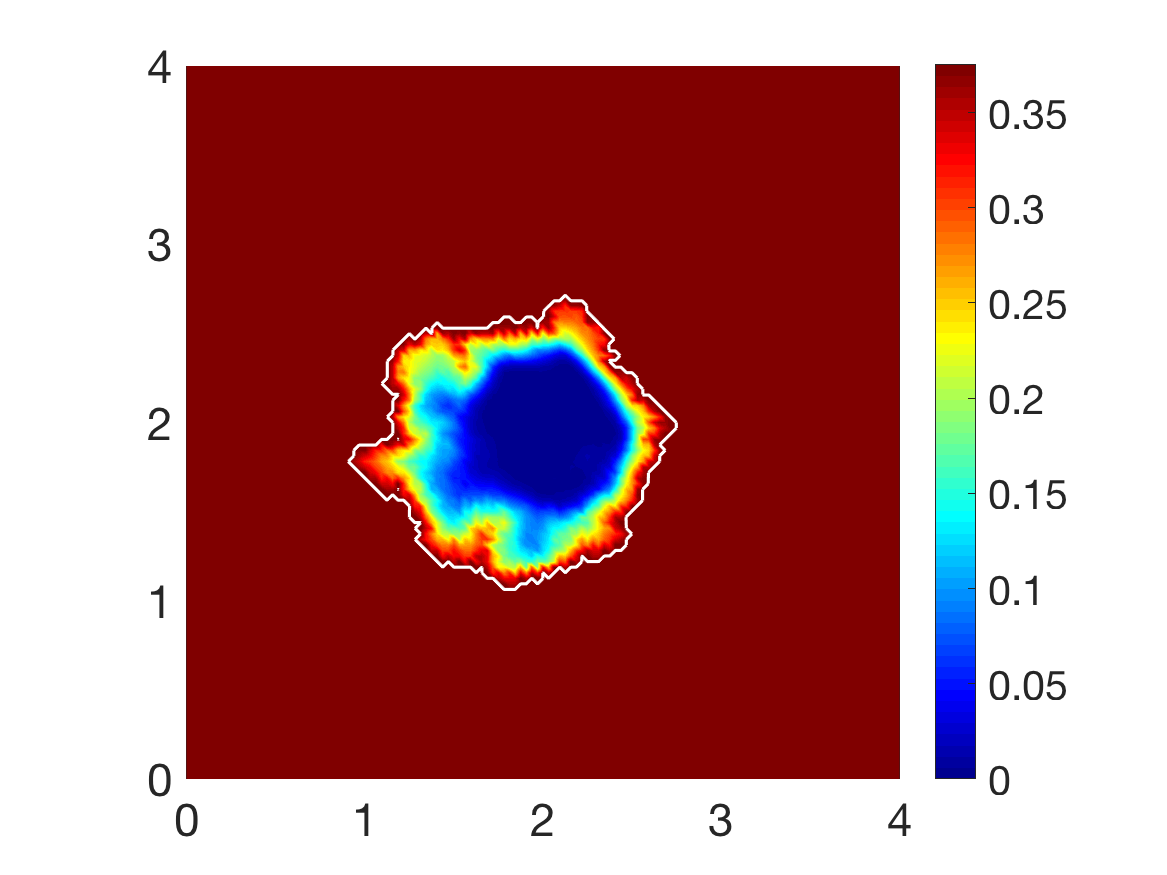}
  \caption{\emph{Non-fibres ECM phase}}
\end{subfigure}\hfil 
\begin{subfigure}{0.5\textwidth}
  \includegraphics[width=\linewidth]{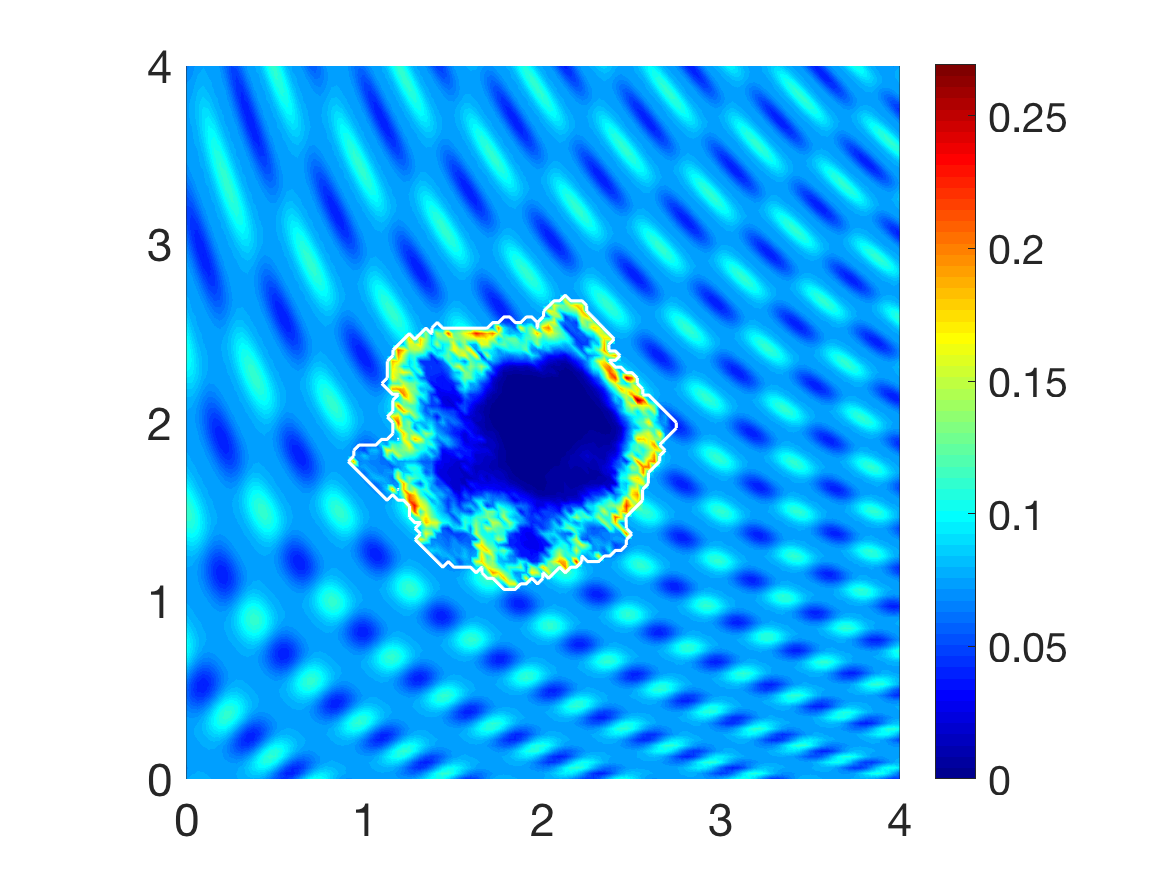}
  \caption{\emph{Macro-scale ECM fibres magnitude}}
\end{subfigure}\hfil 

\medskip
\begin{subfigure}{0.5\textwidth}
  \includegraphics[width=\linewidth]{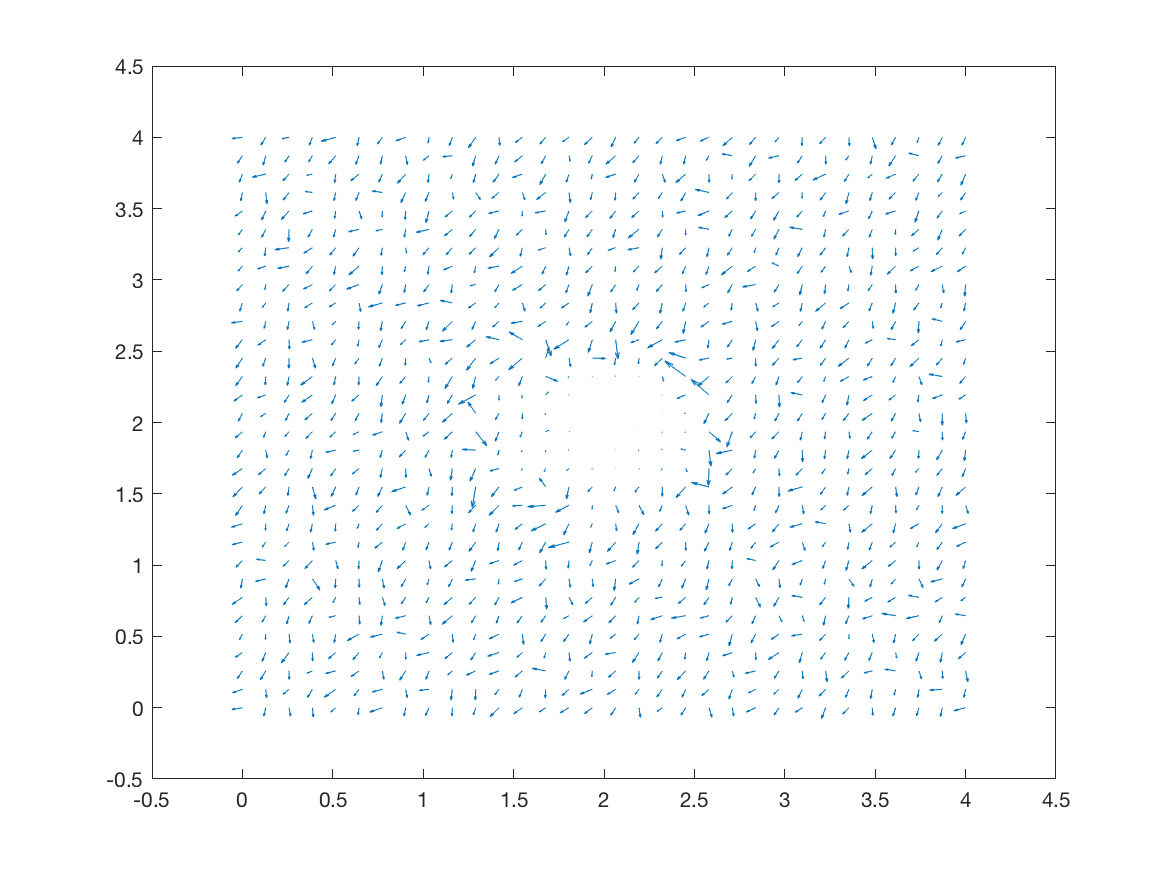}
  \caption{\emph{Oriented macro-scale ECM fibres}}
\end{subfigure}\hfil 
\begin{subfigure}{0.5\textwidth}
  \includegraphics[width=\linewidth]{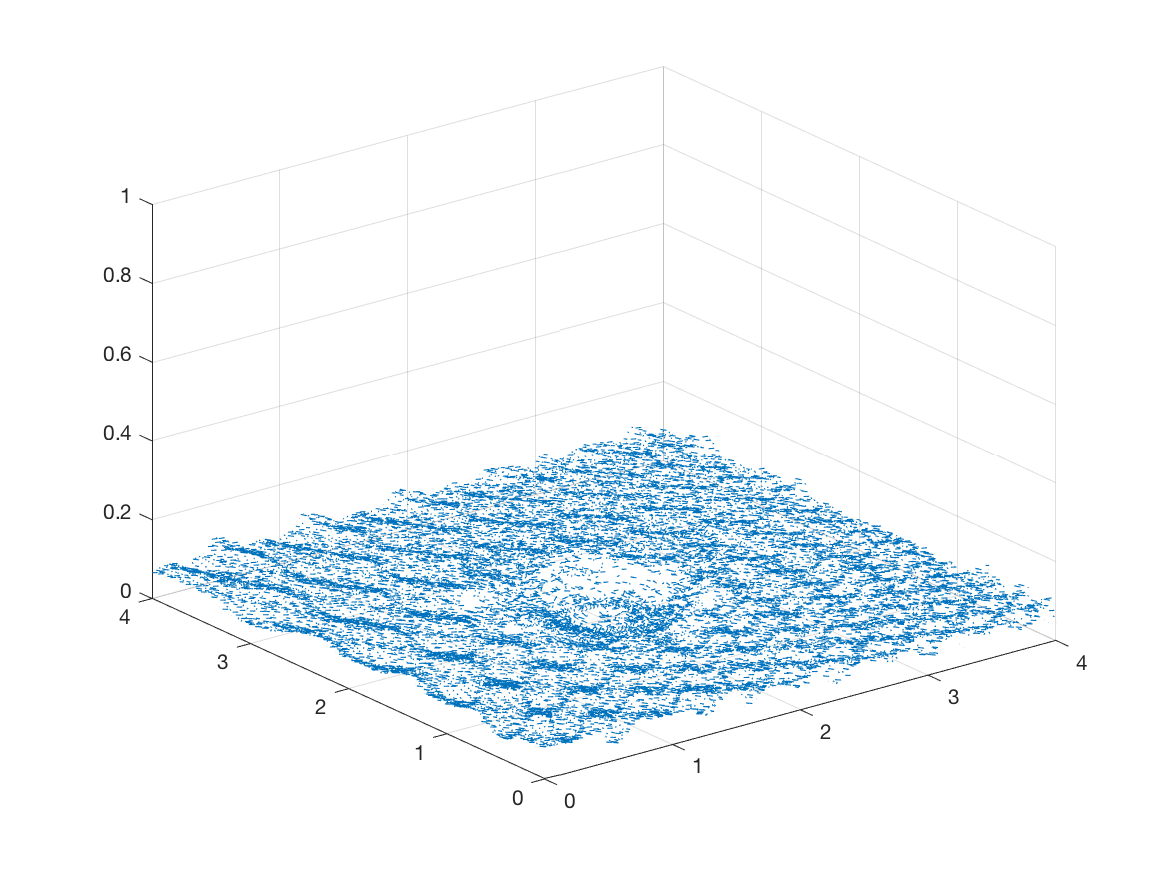}
  \caption{\emph{Oriented macro-scale ECM fibres in a 3D plot}}
\end{subfigure}\hfil 

\caption{Simulations at stage $75\Delta t$ with a homogeneous distribution of non-fibres and a random initial $15\%$ heterogeneous distribution of fibres.}
\label{fig:homoland15randfib_75_cancer}
\end{figure}

We can investigate a similar scenario to that presented previously whereby we now explore tumour invasion on an initially heterogeneous fibre network. The initial condition for the macroscopic fibre distributions will be taken as $15\%$ of the heterogeneous distribution defined in \eqref{eq:matrix_IC} whilst the non-fibre ECM phase will be kept as a homogeneous distribution. Figure \ref{fig:homoland15randfib_75_cancer} displays computations at final stage $75\Delta t$. Comparing to Figure \ref{fig:homol_homof_15_75_cancer}, the boundary of the tumour has formed lobules, specifically towards low density regions of the fibre network, most notable in subfigure \ref{fig:homoland15randfib_75_cancer}(d). A different pattern of mutations from population $c_1$ to $c_2$ is adopted due to the heterogeneity of the fibre network and thus high distribution regions of cells are formed in \ref{fig:homoland15randfib_75_cancer}(b) that differ to that when in the presence of a homogeneous fibre phase.

The macroscopic mean value of fibres has had a substantial impact on the overall invasion of cancer. The boundary of the tumour has grown many lobules, first reaching to the lower density areas of fibres in \ref{fig:homoland15randfib_75_cancer}(d) before engulfing the higher density regions as invasion progresses. The high distributions of cells in population $c_2$ are reminiscent of Figure \ref{fig:homol_homof_15_75_cancer}(b) in that they are building according to the general orientation of the fibre network in \ref{fig:homoland15randfib_75_cancer}(e). Under the presence of a heterogenous fibre network, tumour invasion is halted at regions of high macroscopic fibre density, whilst in a homogeneous environment there are no barriers during invasion, evidenced in Figure \ref{fig:homol_homof_15_75_cancer}. Here the tumour experiences a lobular expansion as it invades the surrounding ECM whilst the high macroscopic fibre distributions coupled with the higher cell-fibre adhesion rate that population $c_{2}$ carries allows the cells more migratory freedom within the tumour boundary.

\begin{figure}
    \centering 
\begin{subfigure}{0.5\textwidth}
  \includegraphics[width=\linewidth]{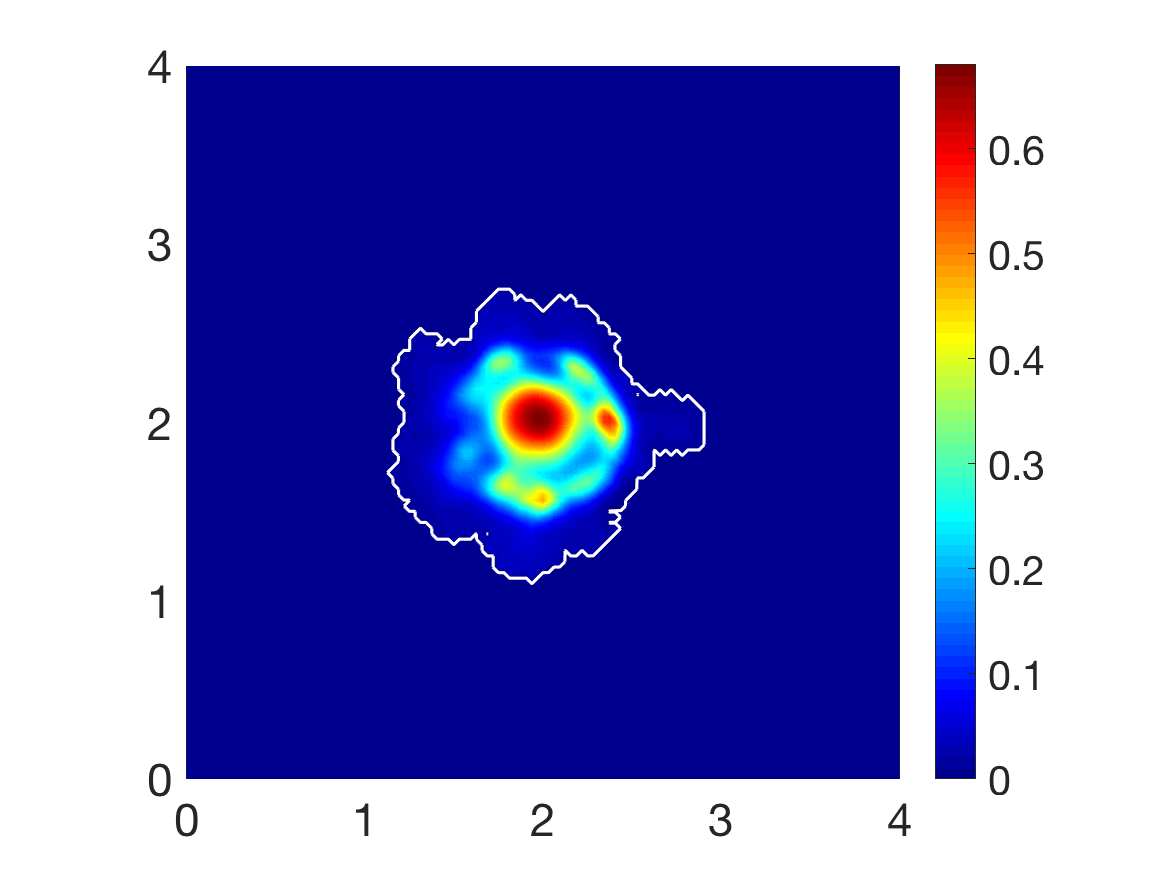}
  \caption{\emph{Cell population 1}}
\end{subfigure}\hfil 
\begin{subfigure}{0.5\textwidth}
  \includegraphics[width=\linewidth]{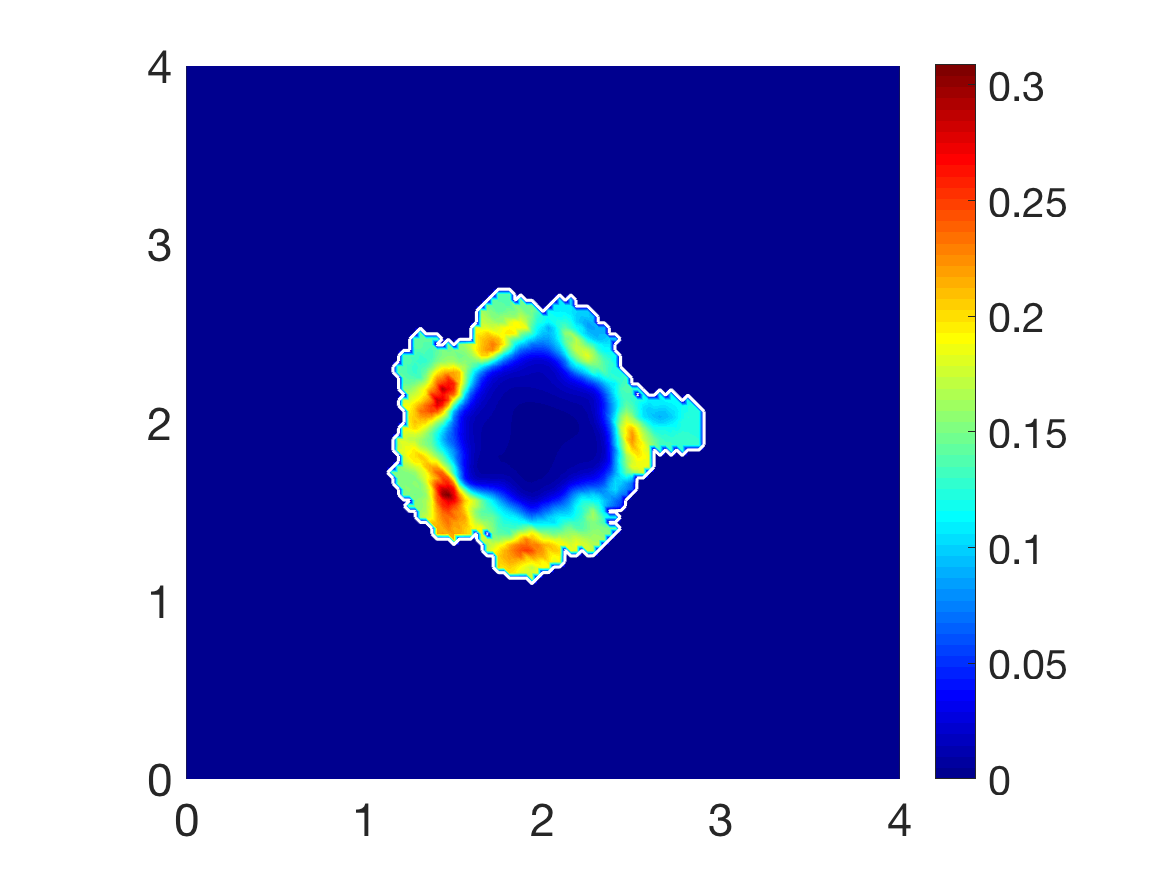}
  \caption{\emph{Cell population 2}}
\end{subfigure}\hfil 

\medskip
\begin{subfigure}{0.5\textwidth}
  \includegraphics[width=\linewidth]{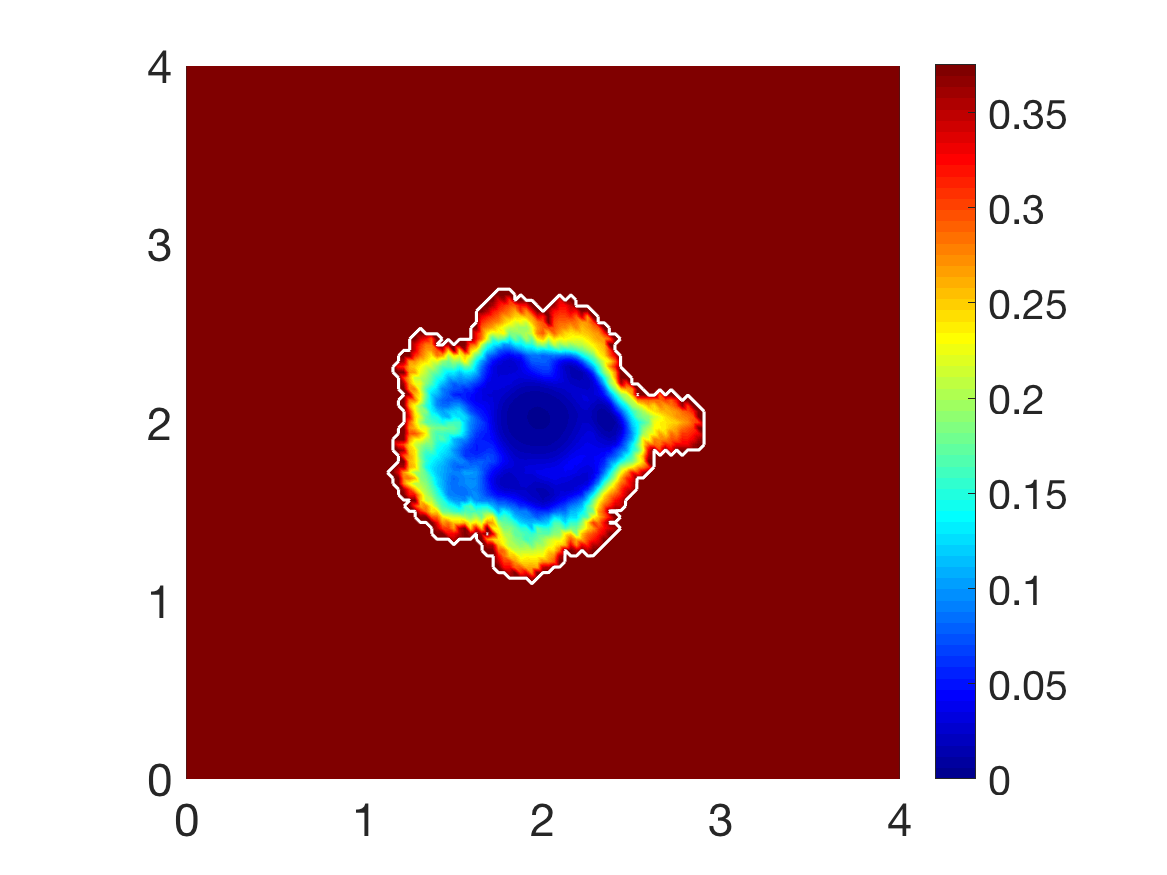}
  \caption{\emph{Non-fibres ECM phase}}
\end{subfigure}\hfil 
\begin{subfigure}{0.5\textwidth}
  \includegraphics[width=\linewidth]{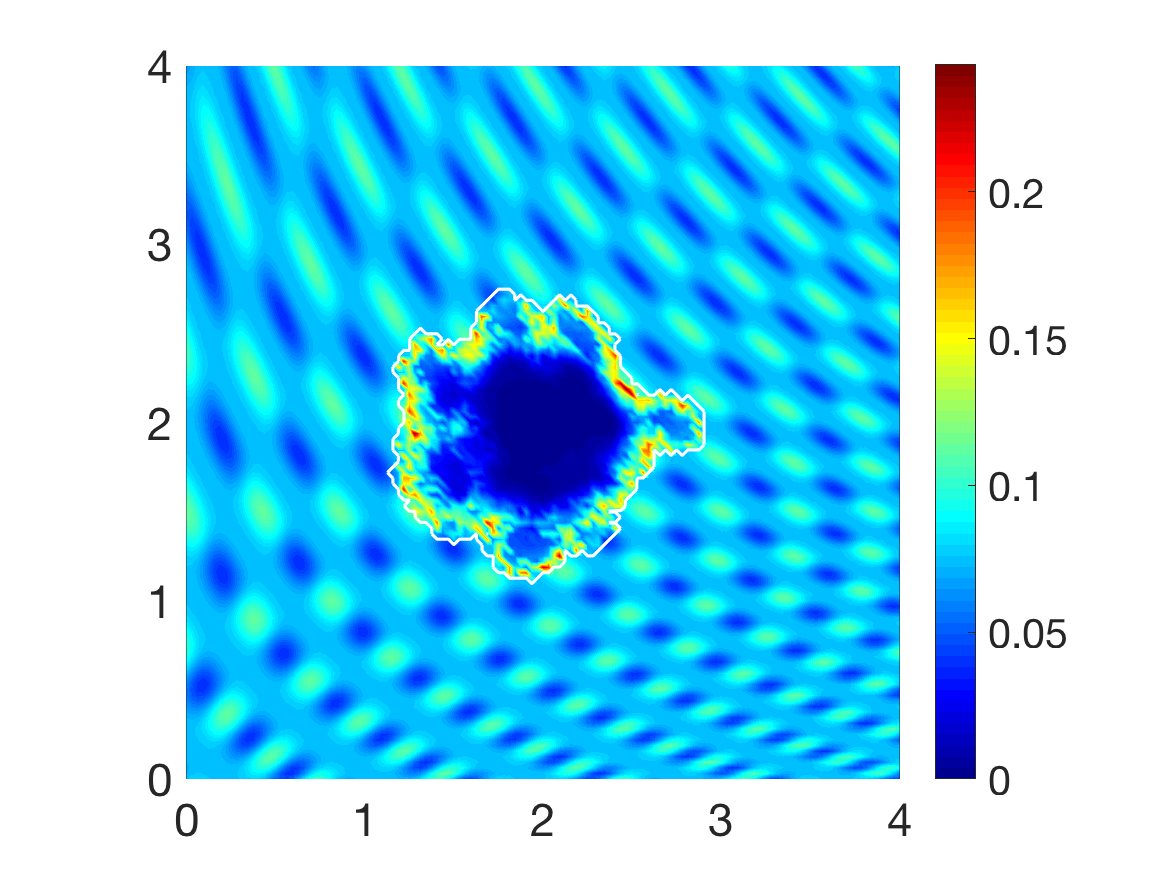}
  \caption{\emph{Macro-scale ECM fibres magnitude}}
\end{subfigure}\hfil 

\medskip
\begin{subfigure}{0.5\textwidth}
  \includegraphics[width=\linewidth]{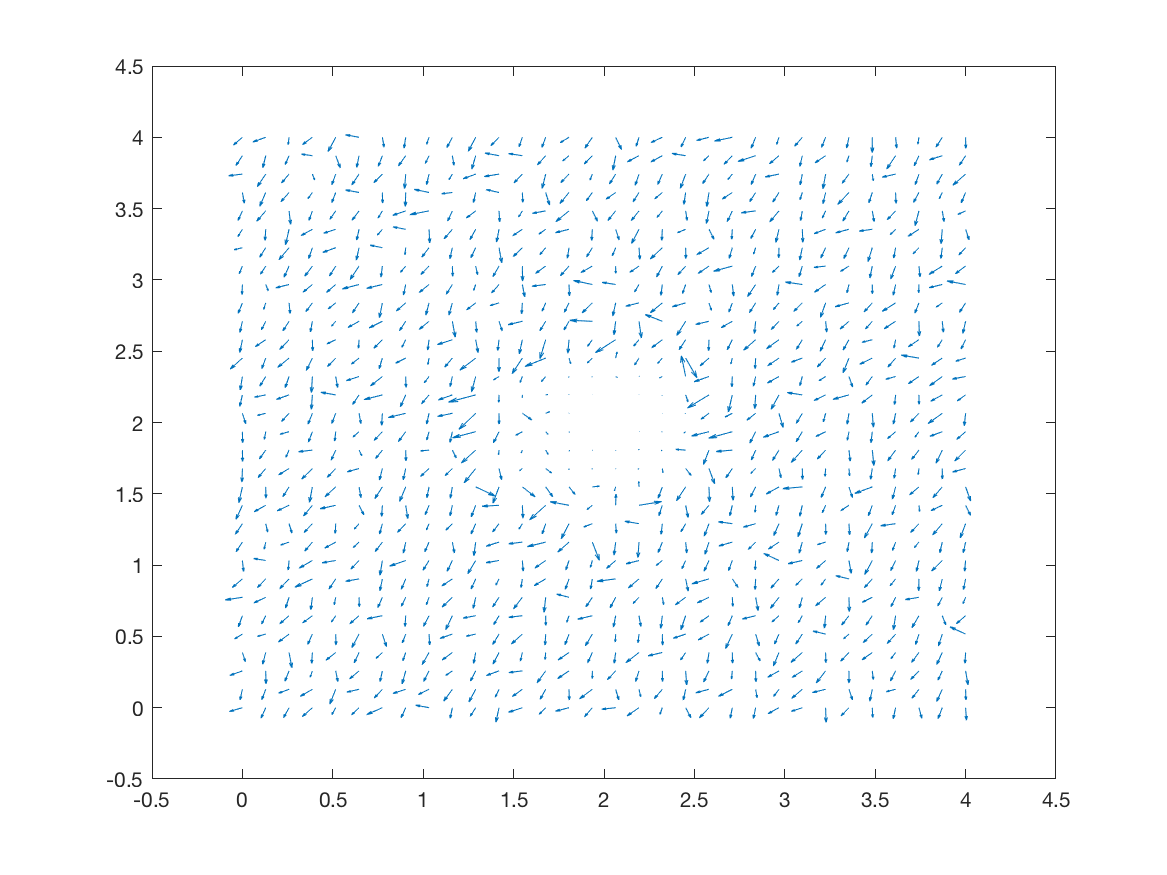}
  \caption{\emph{Oriented macro-scale ECM fibres}}
\end{subfigure}\hfil 
\begin{subfigure}{0.5\textwidth}
  \includegraphics[width=\linewidth]{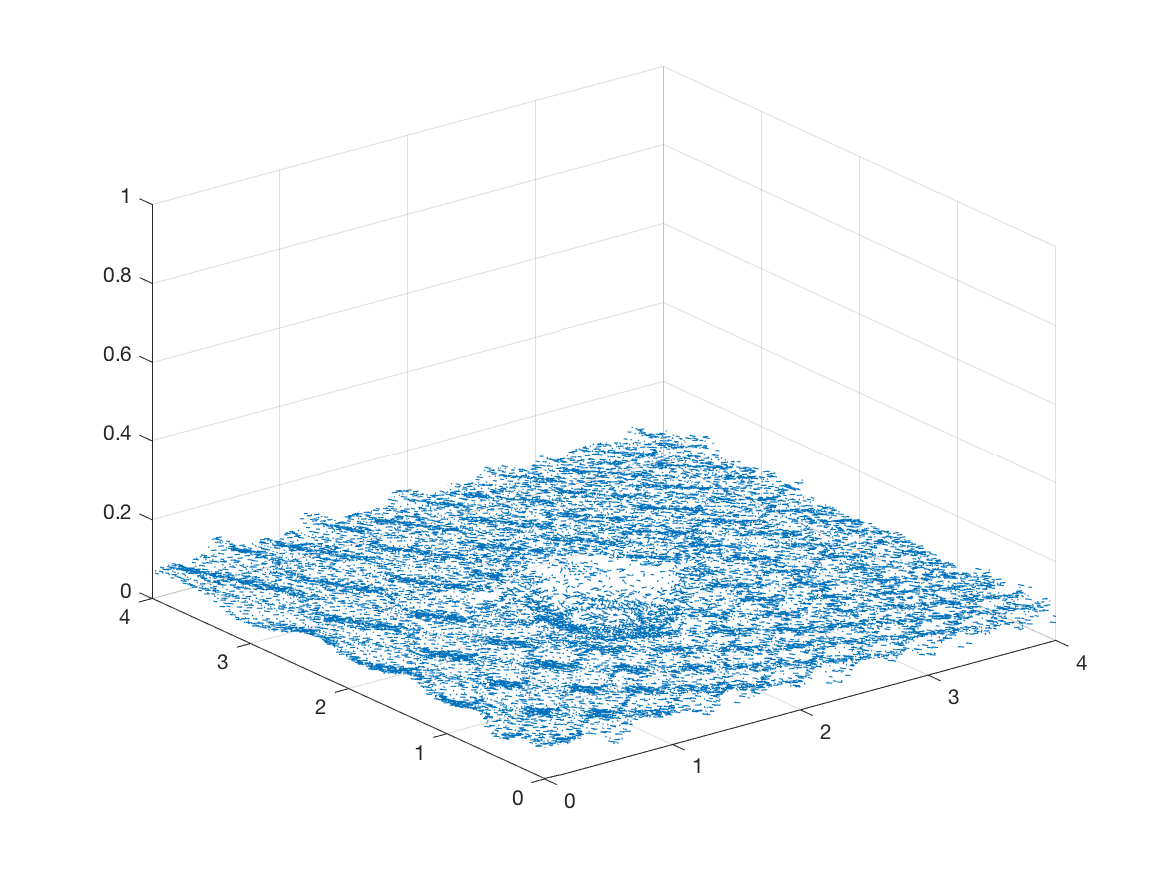}
  \caption{\emph{Oriented macro-scale ECM fibres in a 3D plot}}
\end{subfigure}\hfil 

\caption{Simulations at stage $75\Delta t$ with a homogeneous distribution of the non-fibres with remodelling and a random initial $15\%$ heterogeneous distribution of fibres.}
\label{fig:homolheterof_remod_75_cancer}
\end{figure}

We consider the same initial conditions, but we now introduce matrix remodelling to the non-fibre ECM phase, namely, increasing $\omega$ in \eqref{ldeg} from $0$ to $0.02$. Initially the computational results are very similar to that of the previous section, however, as tumour invasion progresses to stage $75 \Delta t$, Figure \ref{fig:homolheterof_remod_75_cancer}, both the boundary of the tumour and the main body of cells looks very different. Although the density of tumour cells is not as high as in Figure \ref{fig:homoland15randfib_75_cancer}, there is an increase in cell distribution within the boundary of the tumour, particularly in cell population $c_{2}$, \ref{fig:homolheterof_remod_75_cancer}(b), with the entire boundary of the tumour containing a visible amount of cells. This behaviour is as a consequence of matrix remodelling only occurring under the presence of cancer cells, hence, the increased density of matrix within the boundary of the tumour gives more opportunity for cell adhesion and opens more pathways in which the cells can invade. Due to the heterogeneity of the fibre ECM phase, the tumour again exhibits a lobular route of invasion in \ref{fig:homolheterof_remod_75_cancer}(d), first enveloping the low density regions where space is available for the cells to freely invade and consequently overrunning the high density regions in their path.

\subsection{Heterogeneous non-fibre ECM component}

\begin{figure}
    \centering 
\begin{subfigure}{0.5\textwidth}
  \includegraphics[width=\linewidth]{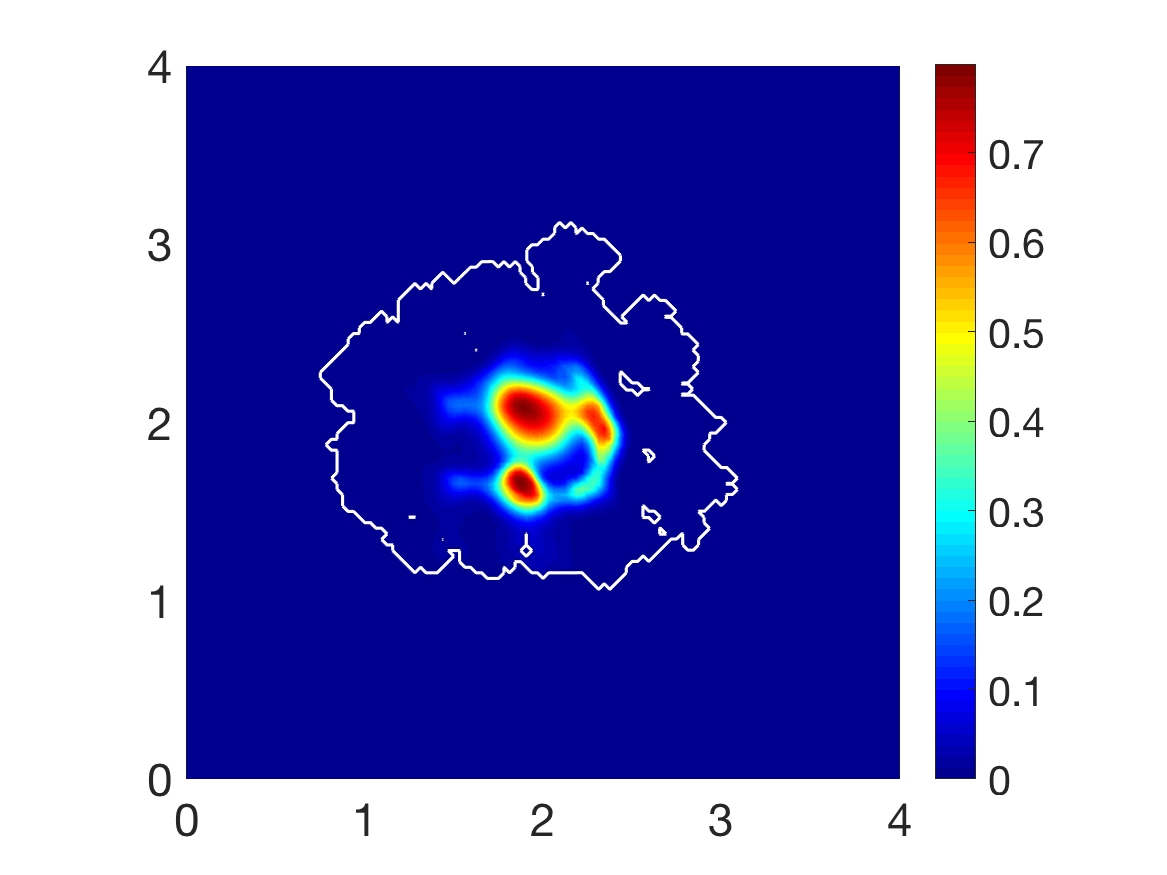}
  \caption{\emph{Cell population 1}}
\end{subfigure}\hfil 
\begin{subfigure}{0.5\textwidth}
  \includegraphics[width=\linewidth]{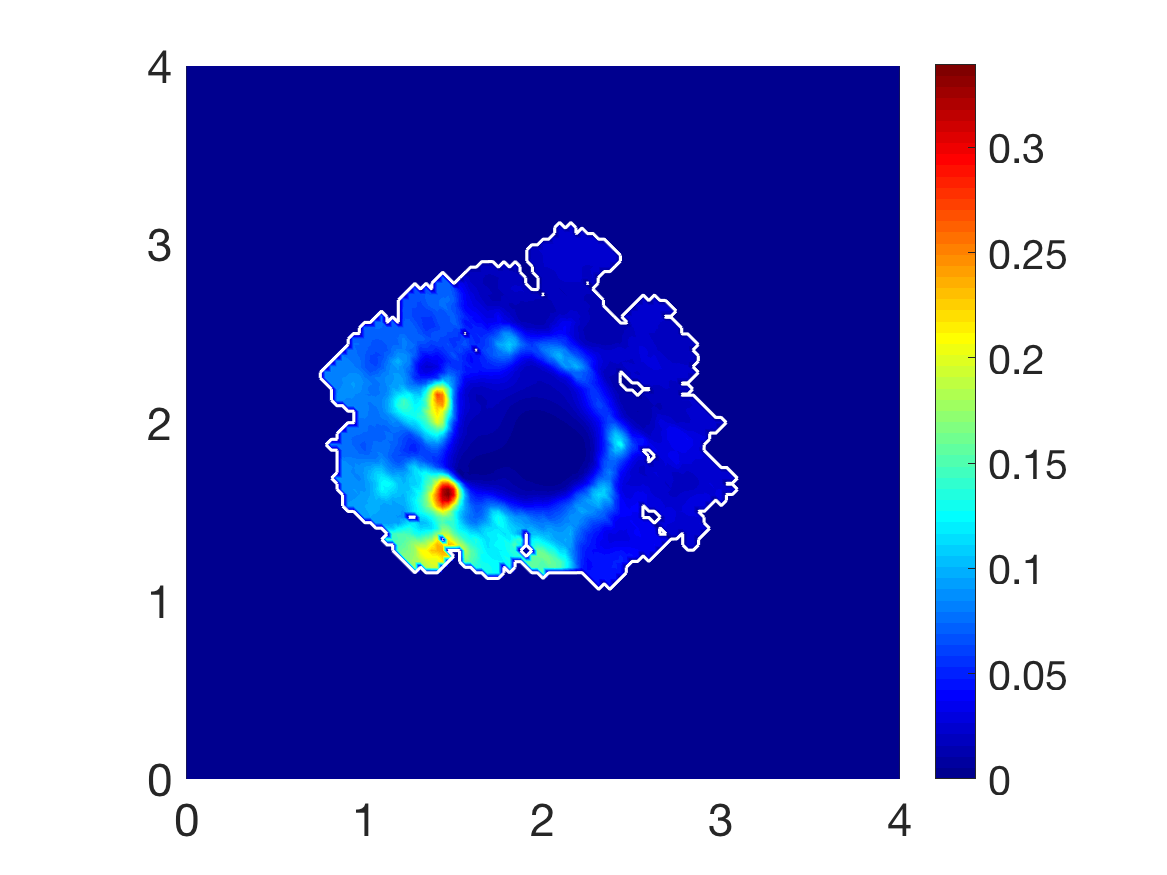}
  \caption{\emph{Cell population 2}}
\end{subfigure}\hfil 

\medskip
\begin{subfigure}{0.5\textwidth}
  \includegraphics[width=\linewidth]{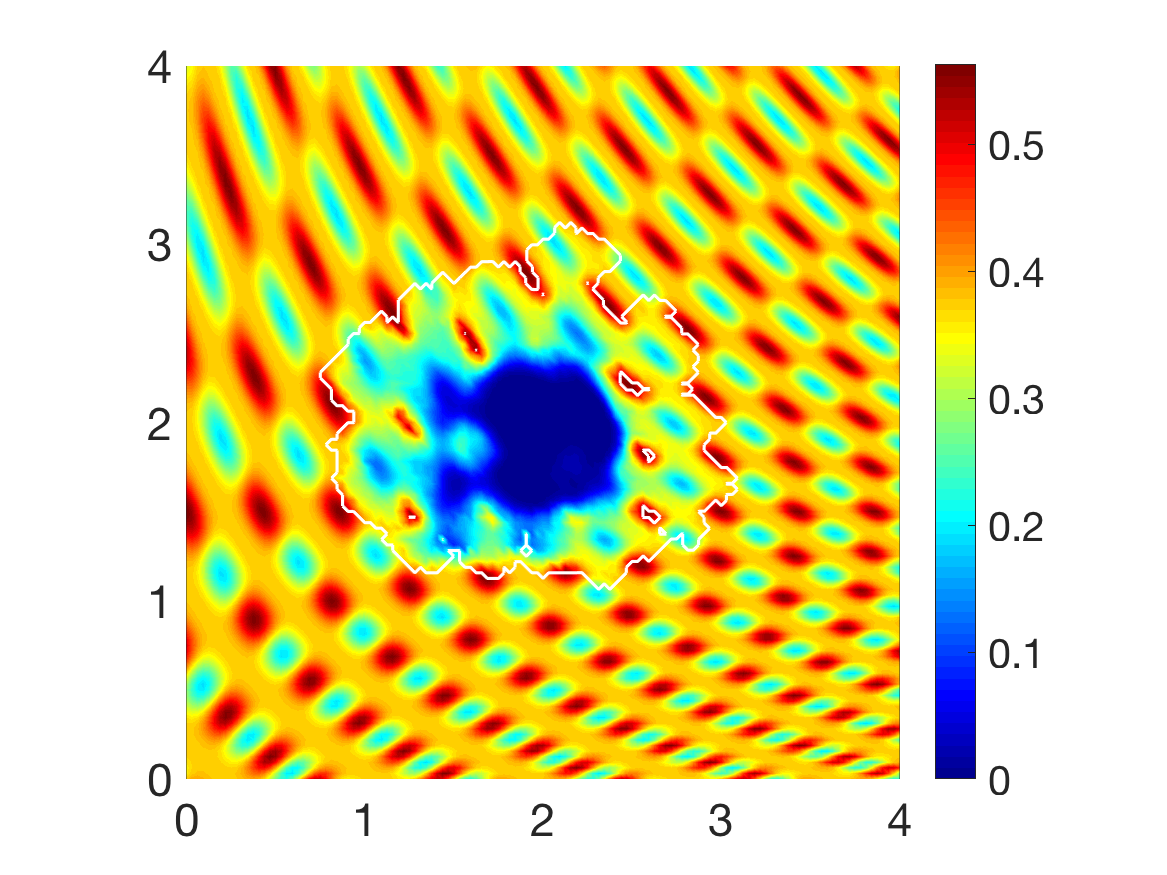}
  \caption{\emph{Non-fibres ECM phase}}
\end{subfigure}\hfil 
\begin{subfigure}{0.5\textwidth}
  \includegraphics[width=\linewidth]{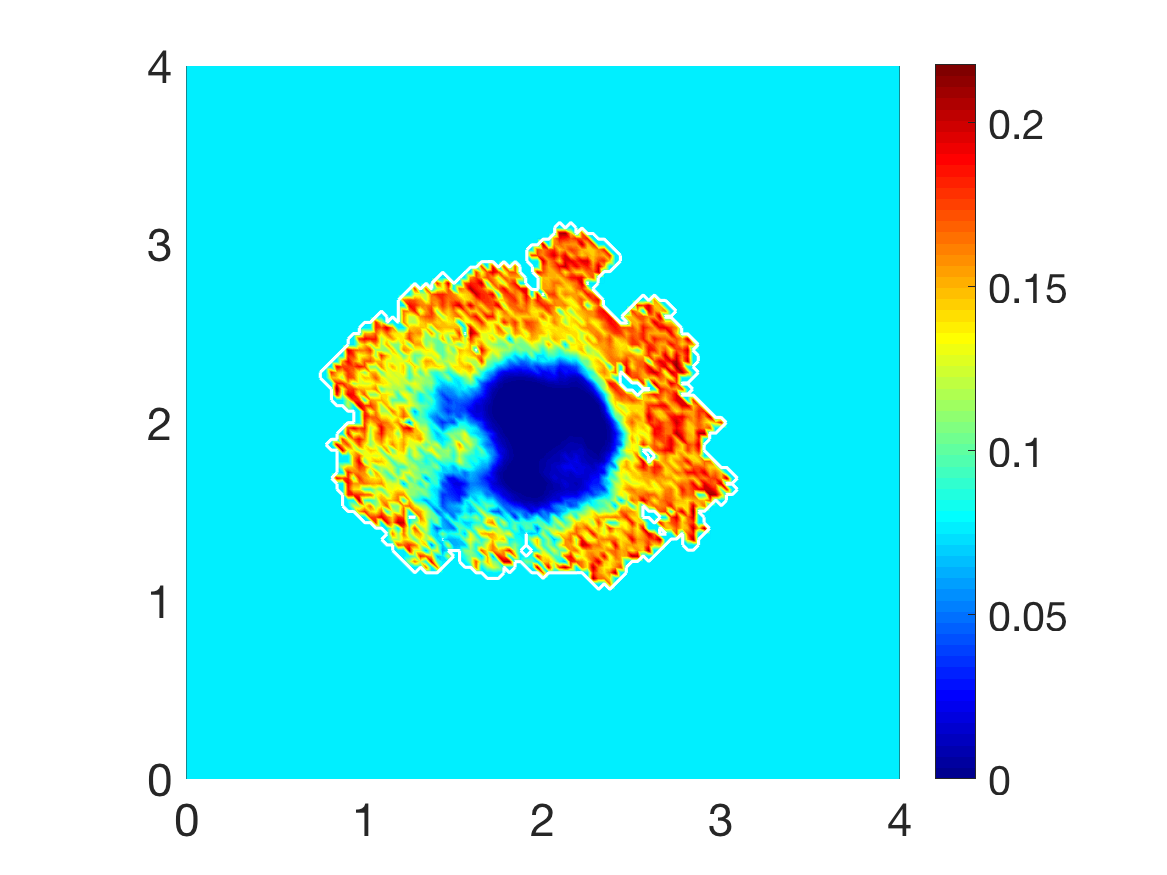}
  \caption{\emph{Macro-scale ECM fibres magnitude}}
\end{subfigure}\hfil 

\medskip
\begin{subfigure}{0.5\textwidth}
  \includegraphics[width=\linewidth]{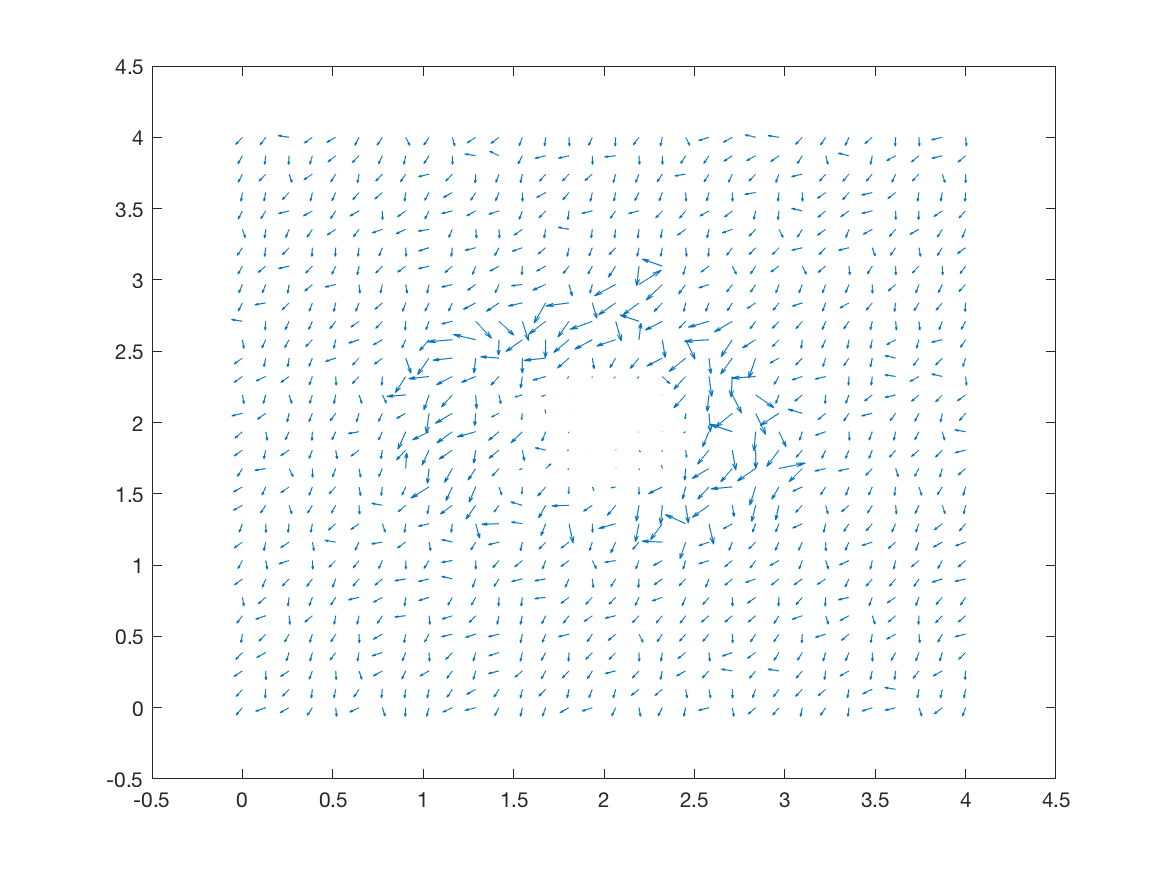}
  \caption{\emph{Oriented macro-scale ECM fibres}}
\end{subfigure}\hfil 
\begin{subfigure}{0.5\textwidth}
  \includegraphics[width=\linewidth]{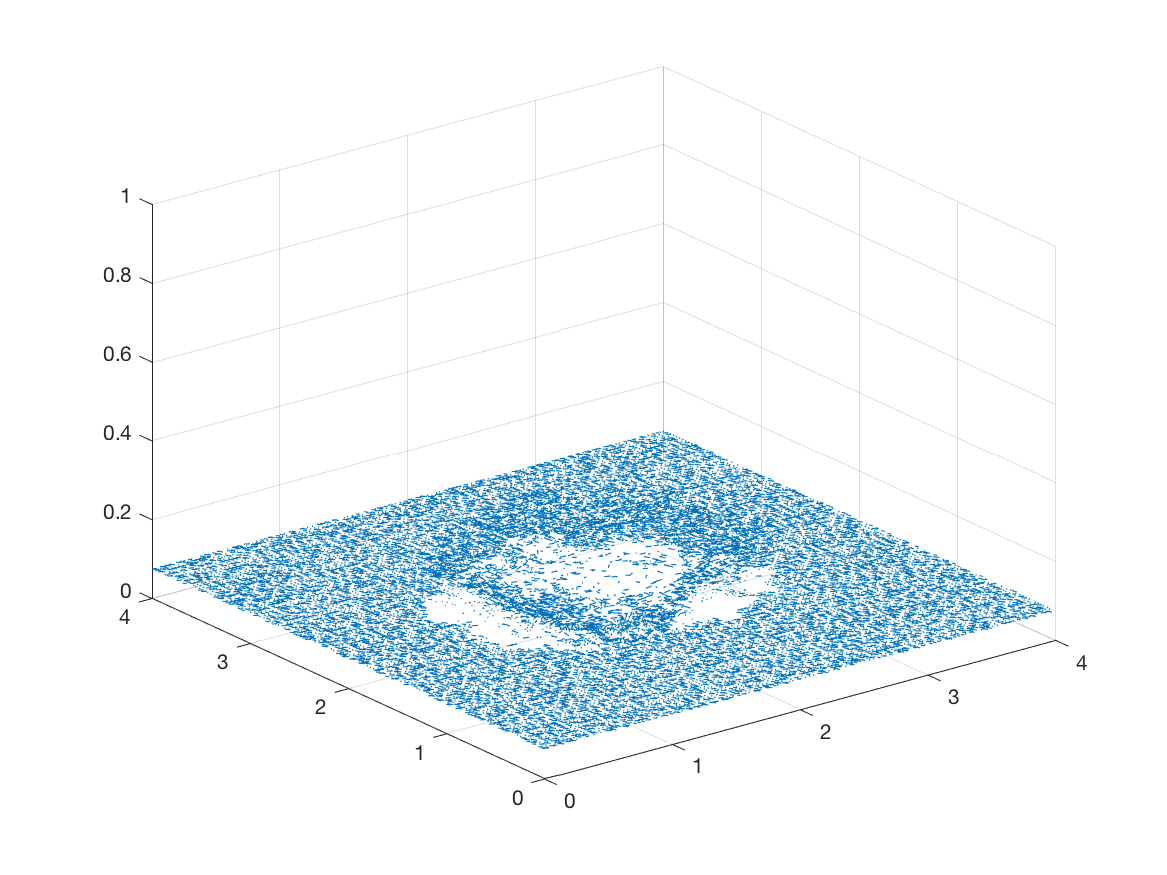}
  \caption{\emph{Oriented macro-scale ECM fibres in a 3D plot}}
\end{subfigure}\hfil 

\caption{Simulations at stage $75\Delta t$ with a heterogeneous distribution of non-fibres and a random initial $15\%$ homogeneous distribution of fibres.}
\label{fig:15homorandfib_75_cancer}
\end{figure}

Investigating the effects of different initial fibre distributions present in a homogeneous non-fibre ECM domain is effective in showing the influence of fibres on tumour invasion, however, is it crucial to examine these varying fibre distributions under a heterogeneous non-fibre ECM phase. We express heterogeneity of the non-fibre ECM phase using the initial condition \eqref{eq:matrix_IC} and we assume an initial $15\%$ homogeneous fibre distribution. Although the cell-non-fibre-ECM adhesion coefficients are kept low, we see the tumour of the boundary developing a lobular pattern, witnessed in subfigure \ref{fig:15homorandfib_75_cancer}(c), encroaching on the lower density regions of matrix. By first advancing on the low density regions of the matrix, the cancer cells proceed by engulfing the high density regions in their path and thus the tumour region becomes larger. The tumour continues to progress in this manner with small islands appearing over the regions of high matrix density. These islands arise when the cancer cells have failed to fully overrun areas within the ECM. This behaviour can occur when there are not yet enough cells to fill the area, or because the matrix density is simply too high and the cells must first degrade part of the matrix in order to make space available. Population $c_{2}$ is sparse in its spread \ref{fig:15homorandfib_75_cancer}(b), with only a few regions of cells visible, having gathered into pockets of the matrix where density of both fibre and non-fibre ECM is low, subfigures \ref{fig:15homorandfib_75_cancer}(c)-(d). Comparing to the simulations in Figure \ref{fig:homoland15randfib_75_cancer}, where initially the non-fibre ECM phase is homogeneous and the fibre phase is heterogeneous, we see a much larger overall spread of the tumour when the non-fibre ECM phase is taken as an initially heterogeneous distribution. This comparison implies that the non-fibre ECM phase plays a key role in the local invasion of cancer and is largely responsible for the progression of the tumour.

\begin{figure}
    \centering 
\begin{subfigure}{0.5\textwidth}
  \includegraphics[width=\linewidth]{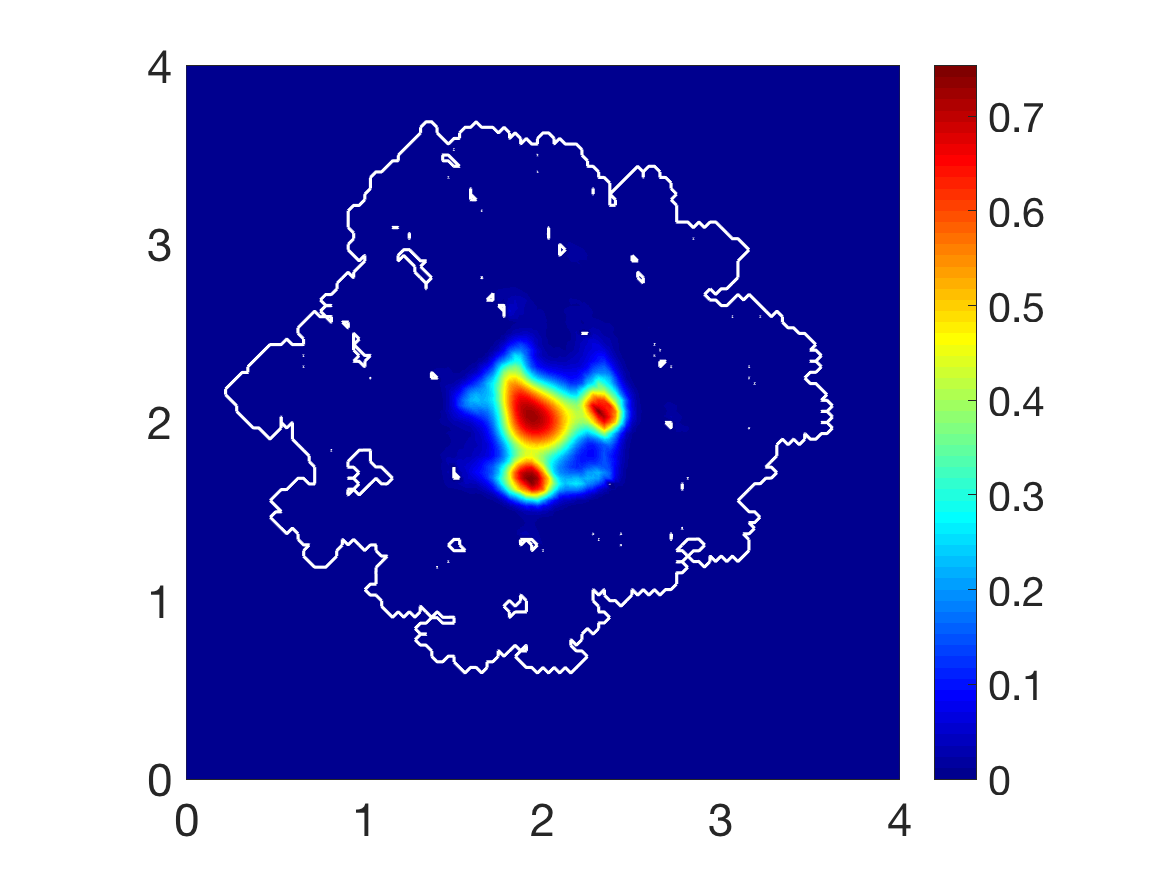}
  \caption{\emph{Cell population 1}}
\end{subfigure}\hfil 
\begin{subfigure}{0.5\textwidth}
  \includegraphics[width=\linewidth]{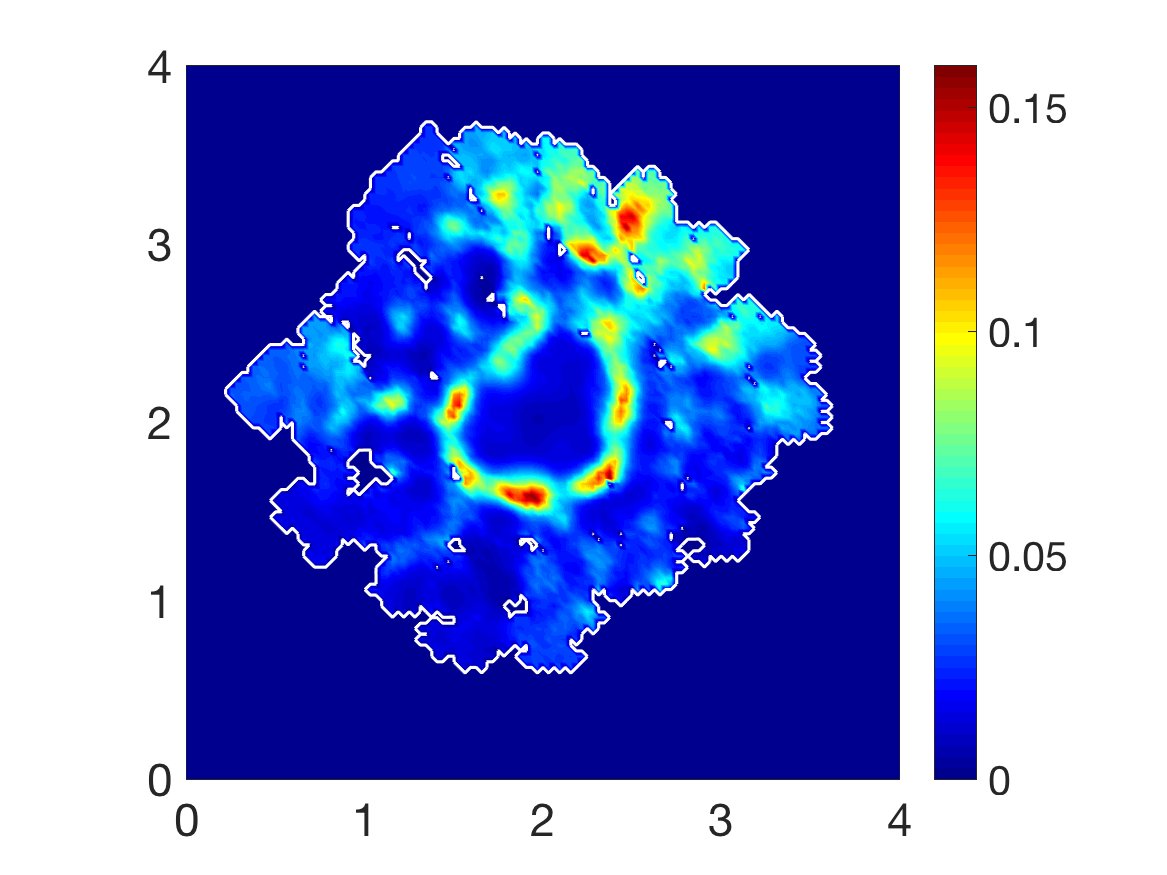}
  \caption{\emph{Cell population 2}}
\end{subfigure}\hfil 

\medskip
\begin{subfigure}{0.5\textwidth}
  \includegraphics[width=\linewidth]{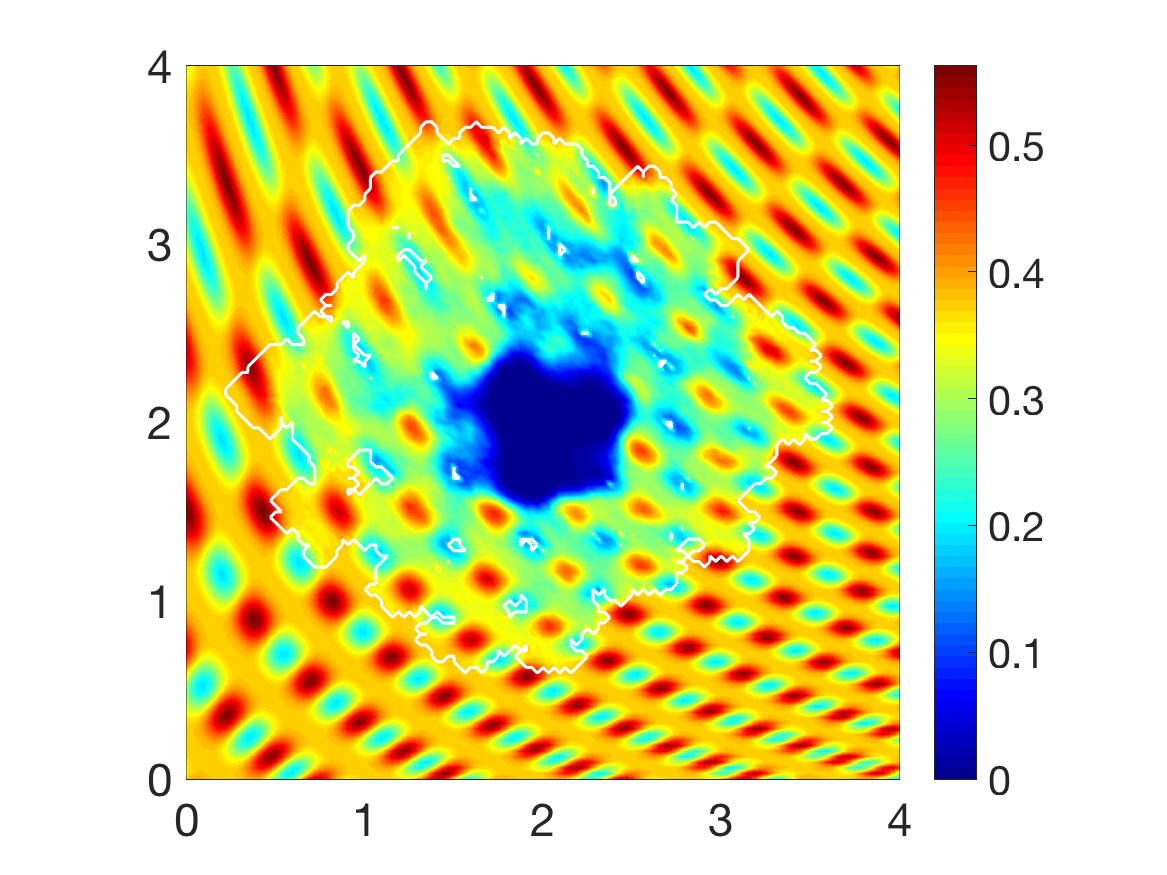}
  \caption{\emph{Non-fibres ECM phase}}
\end{subfigure}\hfil 
\begin{subfigure}{0.5\textwidth}
  \includegraphics[width=\linewidth]{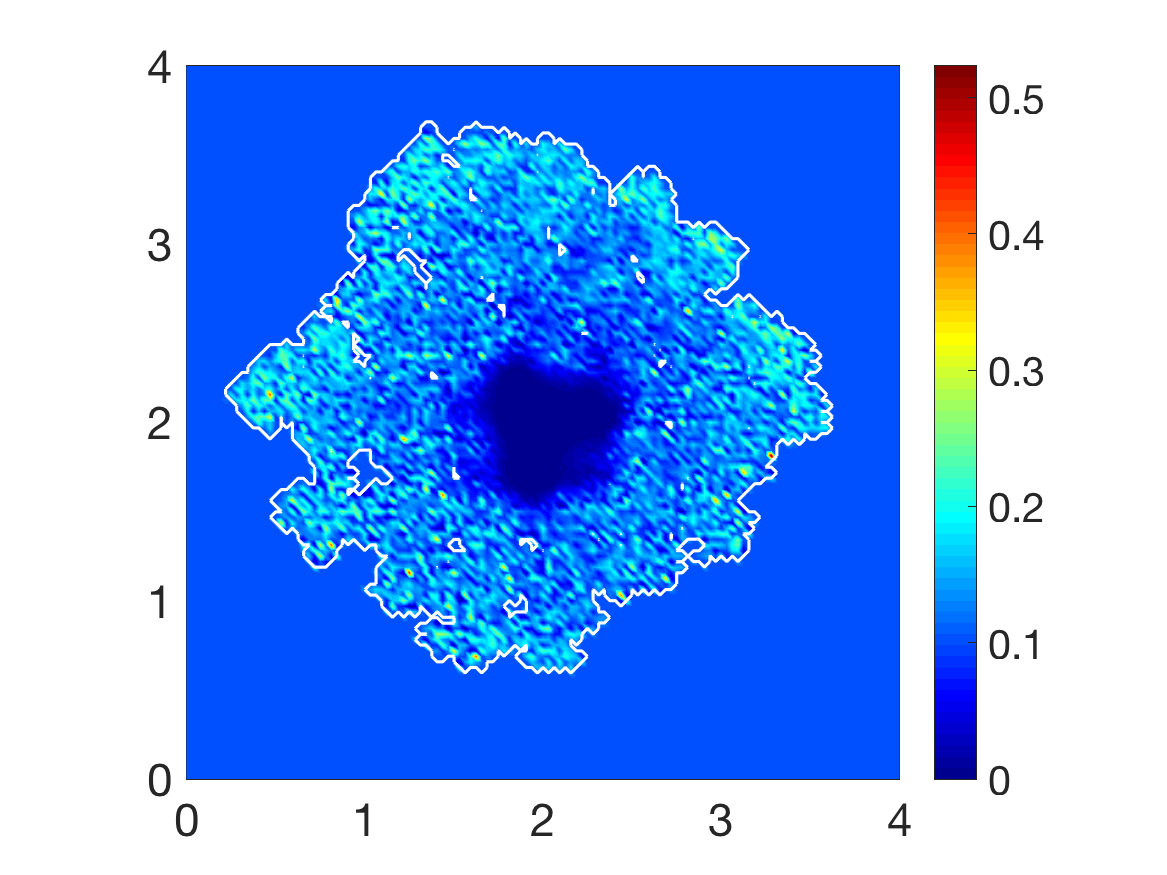}
  \caption{\emph{Macro-scale ECM fibres magnitude}}
\end{subfigure}\hfil 

\medskip
\begin{subfigure}{0.5\textwidth}
  \includegraphics[width=\linewidth]{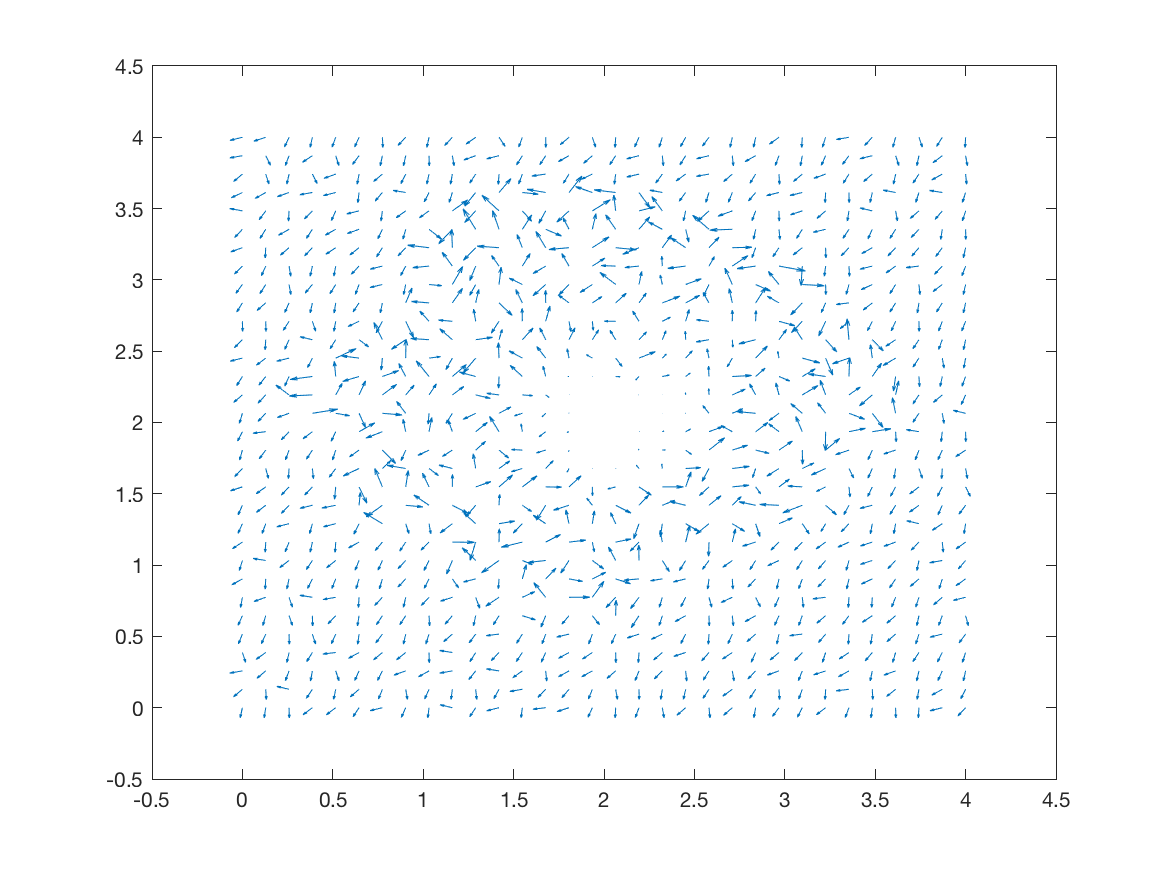}
  \caption{\emph{Oriented macro-scale ECM fibres}}
\end{subfigure}\hfil 
\begin{subfigure}{0.5\textwidth}
  \includegraphics[width=\linewidth]{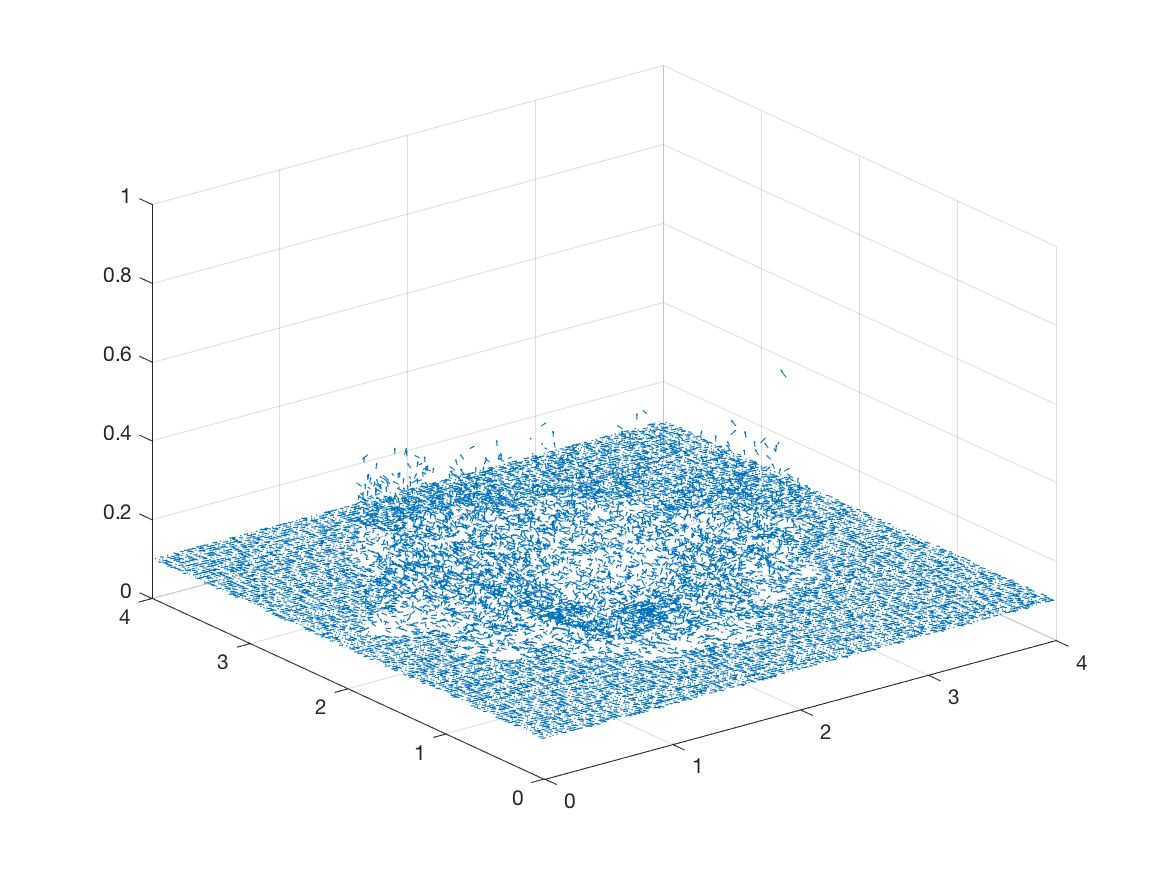}
  \caption{\emph{Oriented macro-scale ECM fibres in a 3D plot}}
\end{subfigure}\hfil 

\caption{Simulations at stage $75\Delta t$ with a heterogeneous distribution of non-fibres and a random initial $20\%$ homogeneous distribution of fibres.}
\label{fig:20homorandfib_75_cancer}
\end{figure}

Continuing to explore invasion in a heterogeneous non-fibre ECM phase, we increase the initial homogeneous fibre distribution to $20\%$ of the non-fibre phase. Figure \ref{fig:20homorandfib_75_cancer} give simulations at the final stage $75 \Delta t$. The tumour boundary is larger and more advanced when compared with the simulations in Figure \ref{fig:15homorandfib_75_cancer}, which have a lower initial fibre distribution. The increased fibres distributions within the ECM presents the cells with more opportunities for adhesion and as a result the cells spread at an increased rate into the surrounding matrix. Cell population $c_{2}$ has progressed outwards towards areas of the leading edge and formed small masses of cells in \ref{fig:20homorandfib_75_cancer}(b). The direction of migration is similar to that in Figure \ref{fig:homol_homof_20_70_cancer}, implying that even though they are subject to a rearrangement process, the initial fibre micro-domains influence the direction of migration of cells. It should also be noted that the density of cells in population $c_{2}$ is lower than that to previous simulations, this occurs as the cells have increased migration due to the increased fibre distributions, thus the cells do not build up in any one space, Figure \ref{fig:20homorandfib_75_cancer}(b). Cells of population $c_{1}$ are surrounded by the more aggressive cells of population $c_{2}$, hence their invasion of the tissue is restricted and they have created denser regions of cells through proliferation, Figure \ref{fig:20homorandfib_75_cancer}(a).

\begin{figure}
    \centering 
\begin{subfigure}{0.5\textwidth}
  \includegraphics[width=\linewidth]{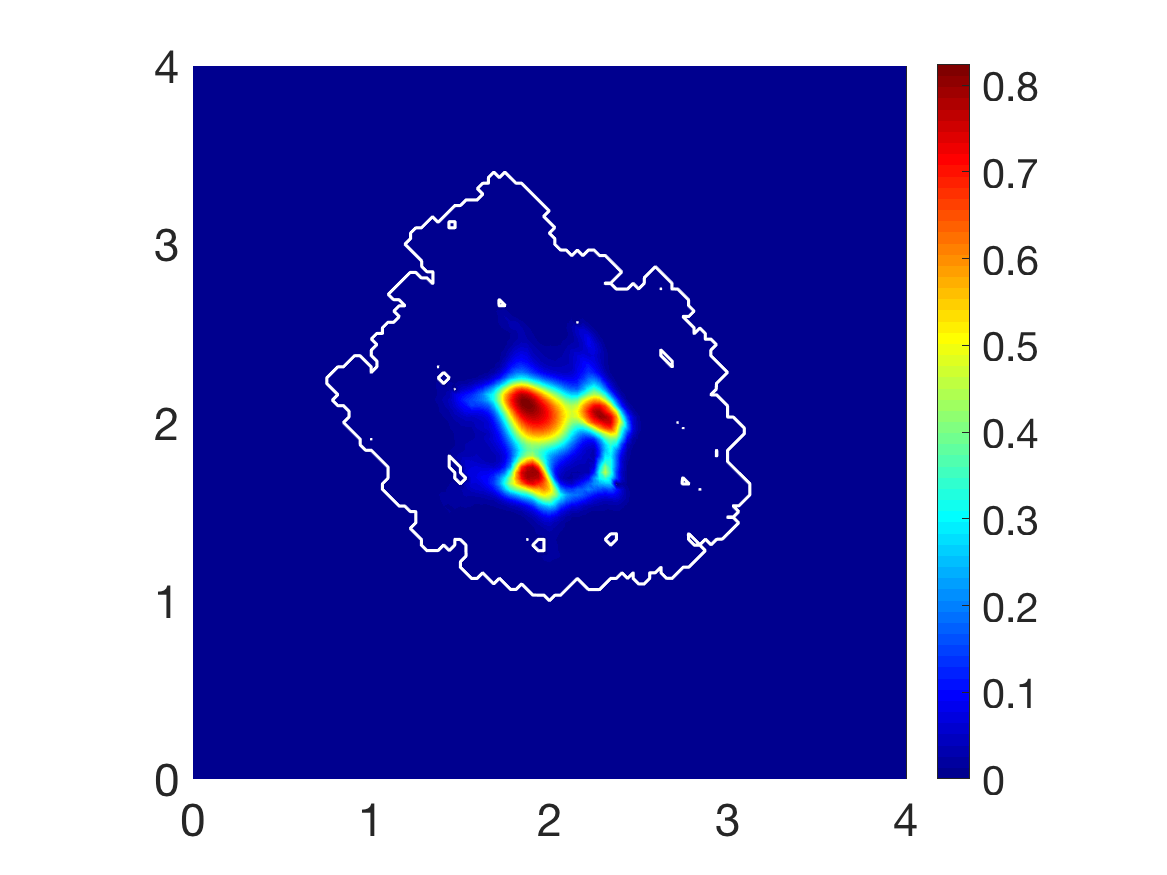}
  \caption{\emph{Cell population 1}}
\end{subfigure}\hfil 
\begin{subfigure}{0.5\textwidth}
  \includegraphics[width=\linewidth]{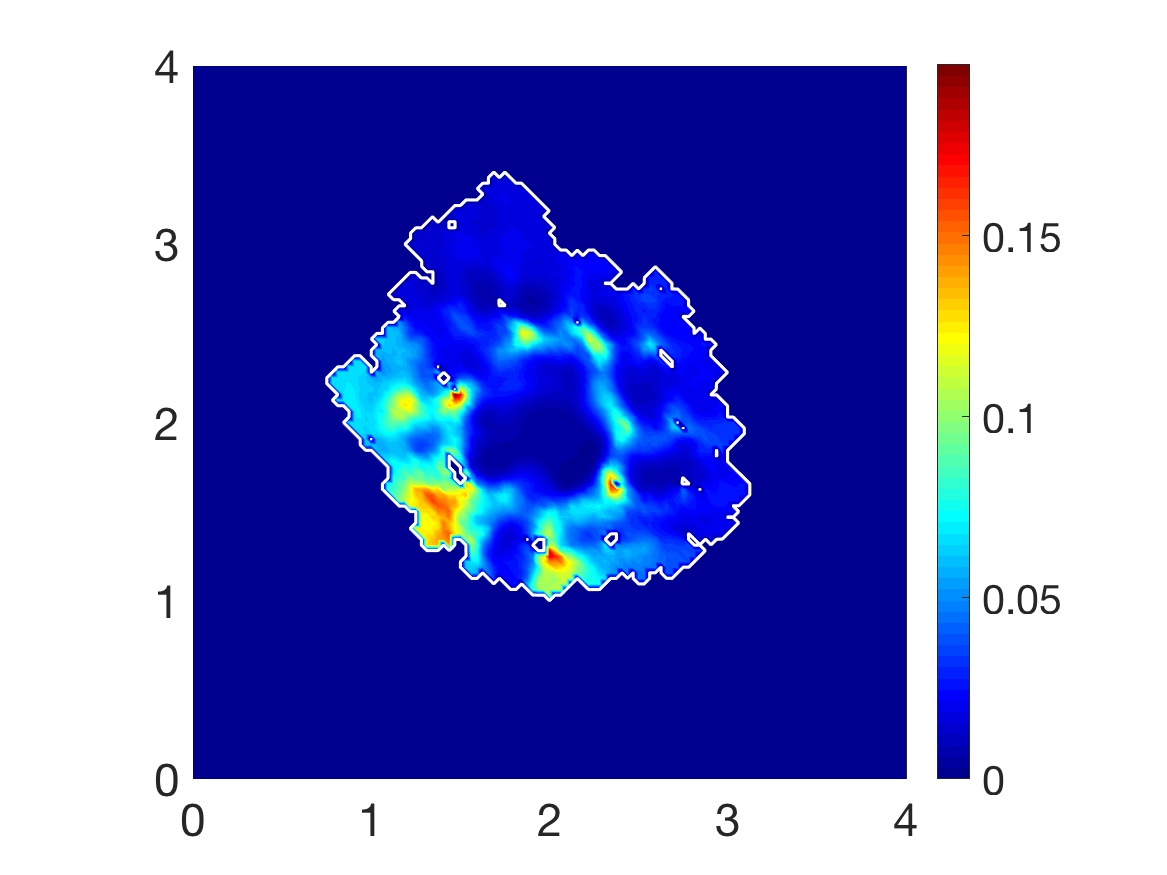}
  \caption{\emph{Cell population 2}}
\end{subfigure}\hfil 

\medskip
\begin{subfigure}{0.5\textwidth}
  \includegraphics[width=\linewidth]{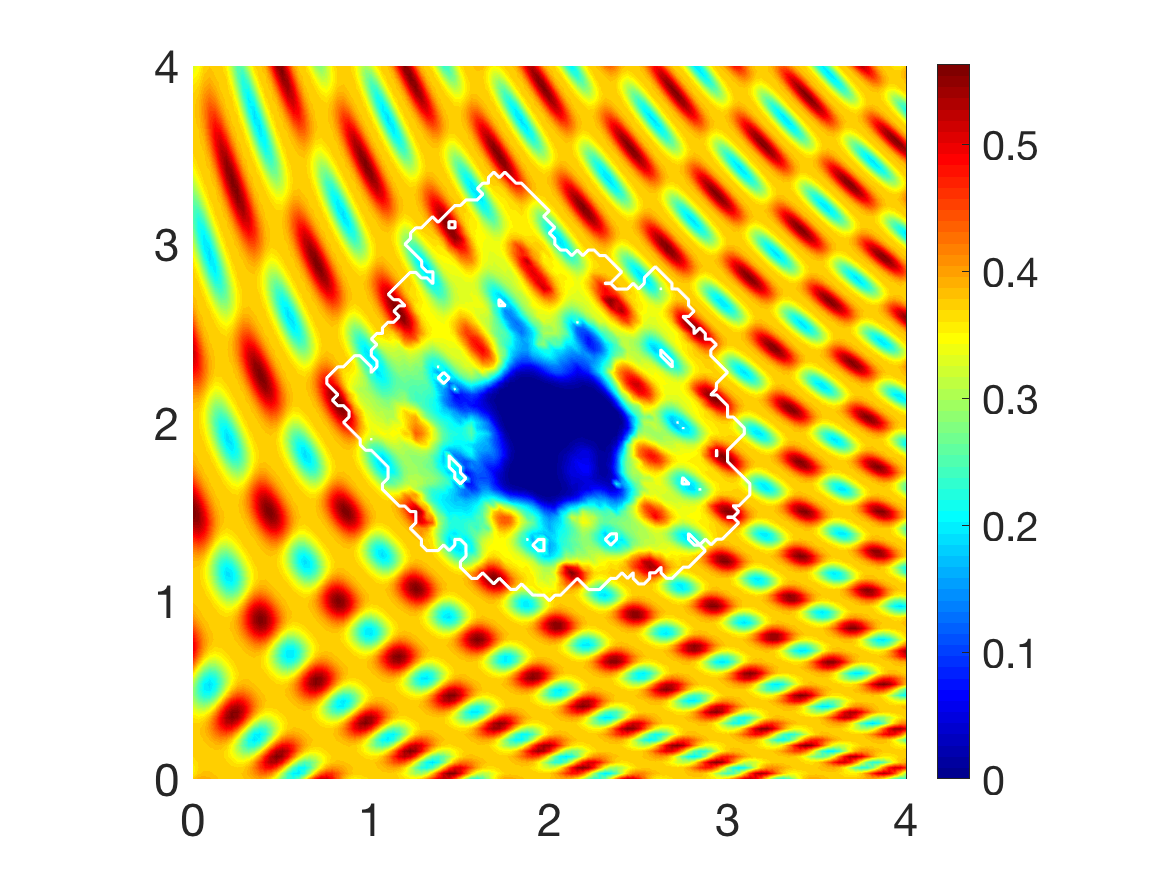}
  \caption{\emph{Non-fibres ECM phase}}
\end{subfigure}\hfil 
\begin{subfigure}{0.5\textwidth}
  \includegraphics[width=\linewidth]{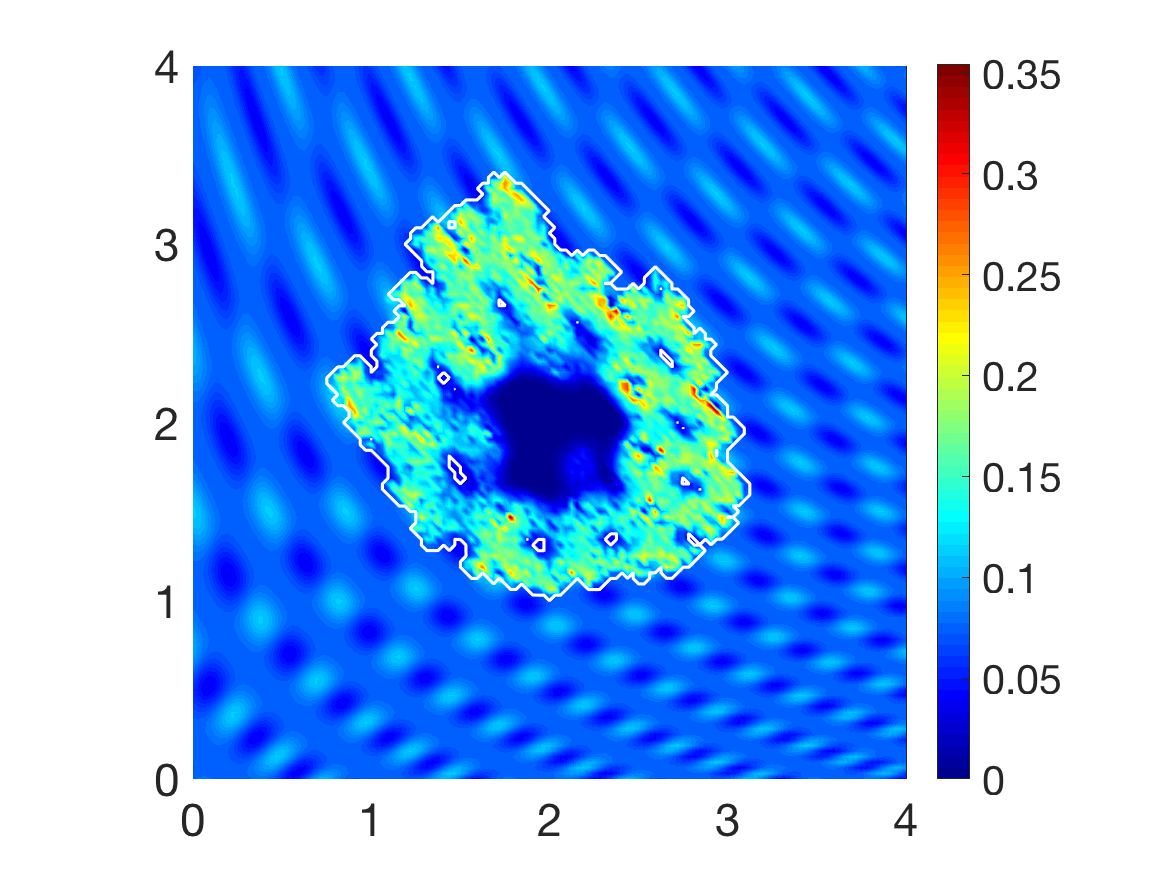}
  \caption{\emph{Macro-scale ECM fibres magnitude}}
\end{subfigure}\hfil 

\medskip
\begin{subfigure}{0.5\textwidth}
  \includegraphics[width=\linewidth]{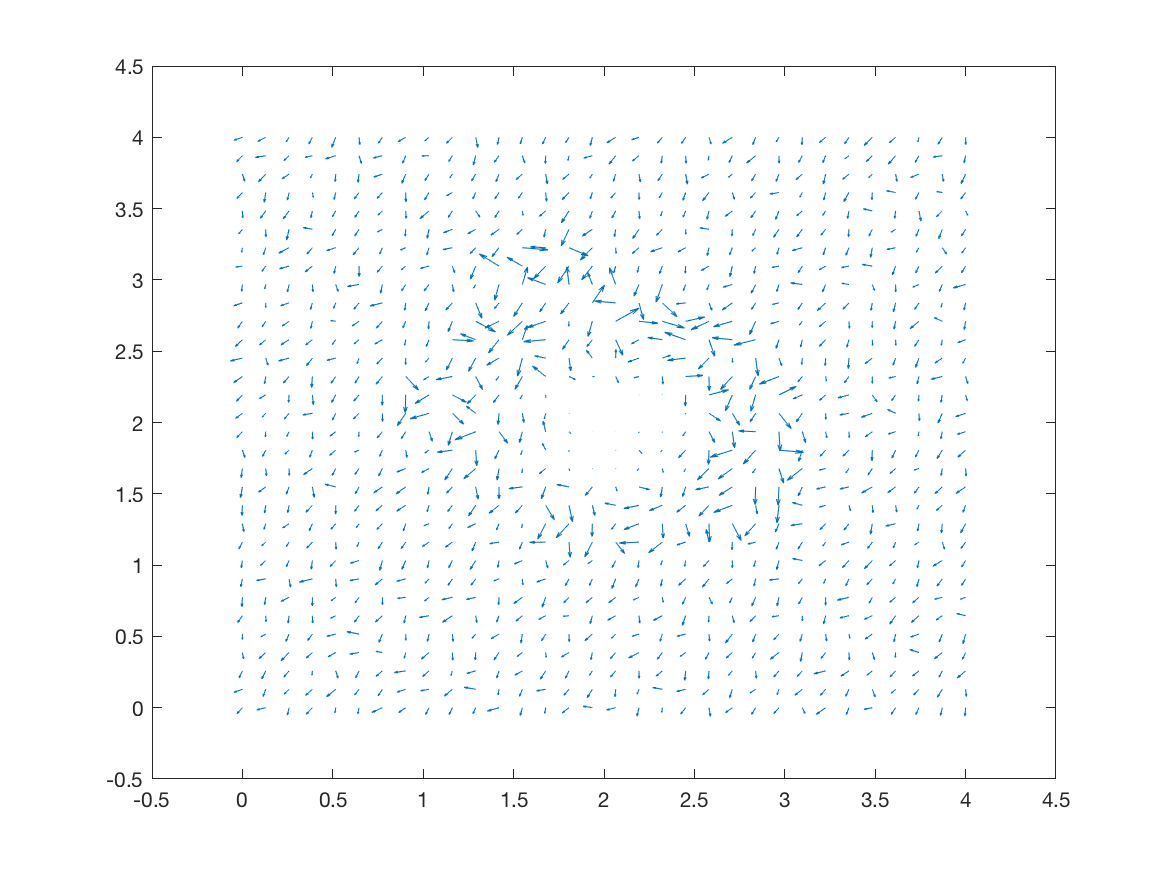}
  \caption{\emph{Oriented macro-scale ECM fibres}}
\end{subfigure}\hfil 
\begin{subfigure}{0.5\textwidth}
  \includegraphics[width=\linewidth]{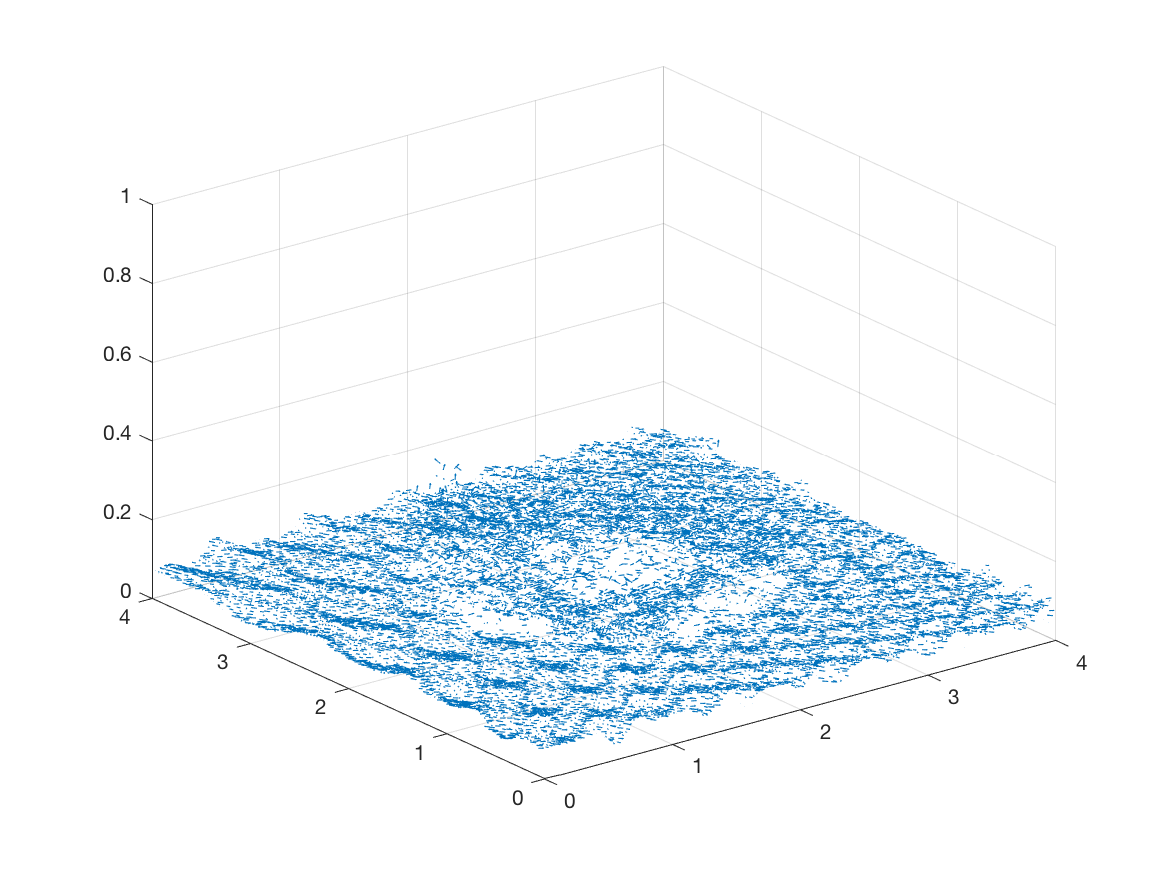}
  \caption{\emph{Oriented macro-scale ECM fibres in a 3D plot}}
\end{subfigure}\hfil 

\caption{Simulations at stage $70\Delta t$ with a heterogeneous distribution of non-fibres and a random initial $15\%$ heterogeneous distribution of fibres.}
\label{fig:hetero_15}
\end{figure}

We next consider here a fully heterogeneous ECM, with the fibre phase taking $15\%$ of the heterogeneous non-fibre phase, defined in \eqref{eq:matrix_IC}. Figure \ref{fig:hetero_15} shows simulations at final stage $70 \Delta t$. Cell population $c_{1}$ continues to be confined to the centre of the domain showing little outward movement whilst exhibiting the highest distribution of cells noted throughout all previous simulations. This increase in distribution can be described by the heterogeneous nature of the matrix; as the patches of low density ECM become more prominent, the cells have less areas to adhere and are therefore restricted to proliferation in these areas only, forming larger masses of cells. In contrast, cell population $c_{2}$ has established patches of high cell distributions following the general direction of the fibres, Figure \ref{fig:hetero_15}(e), and we again see small islands within the boundary forming over the lower density regions of ECM, Figure \ref{fig:hetero_15}(c), similar to the simulations in Figure \ref{fig:15homorandfib_75_cancer}. The tumour boundary is visibly more lobular and consistent in its pattern by following the structure of the ECM, attributable to the heterogeneity of the entire matrix. In line with previous simulations, the tumour is advancing on the higher density regions of matrix first, hence the fingering behaviour of the boundary towards these regions, Figure \ref{fig:hetero_15}(c)-(d).

\subsection{New family of pre-determined micro-fibre domains}\label{newmicdom}

\begin{figure}
    \centering 
\begin{subfigure}{0.5\textwidth}
  \includegraphics[width=\linewidth]{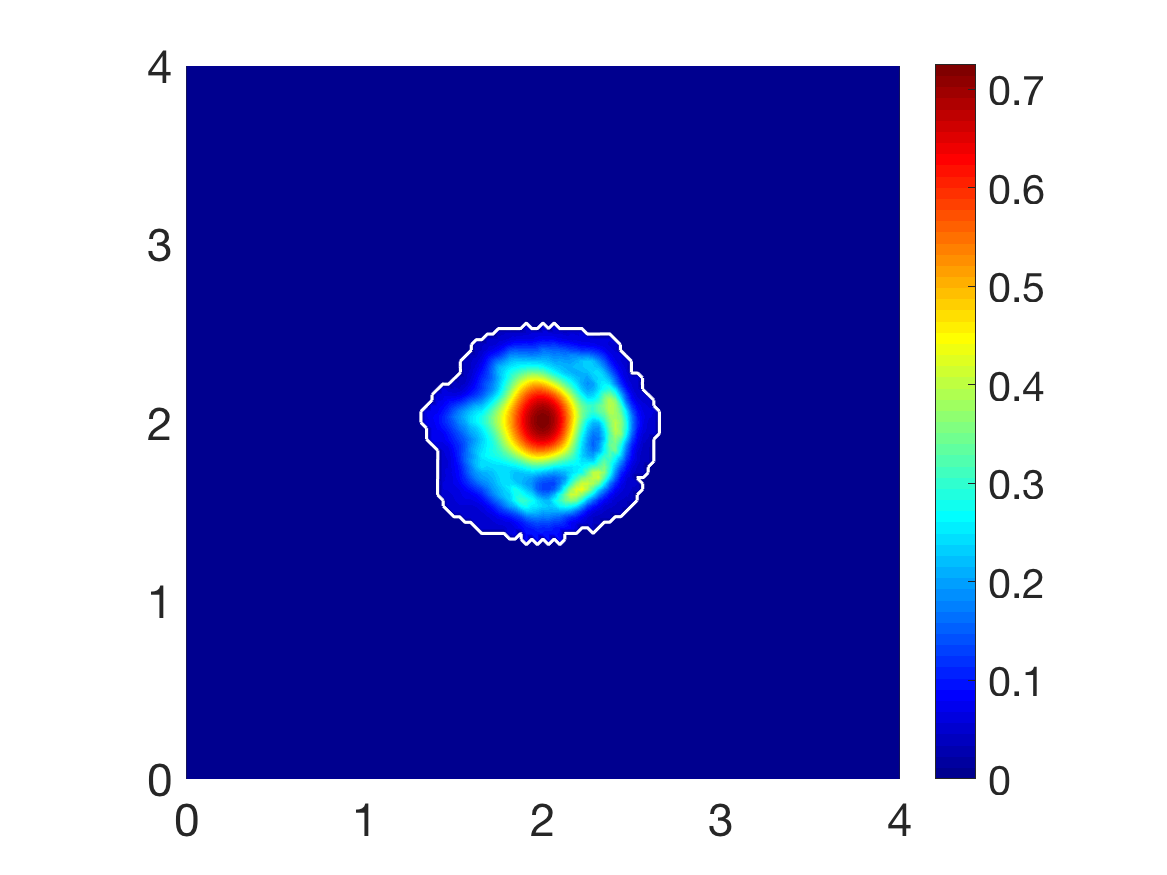}
  \caption{\emph{Cell population 1}}
\end{subfigure}\hfil 
\begin{subfigure}{0.5\textwidth}
  \includegraphics[width=\linewidth]{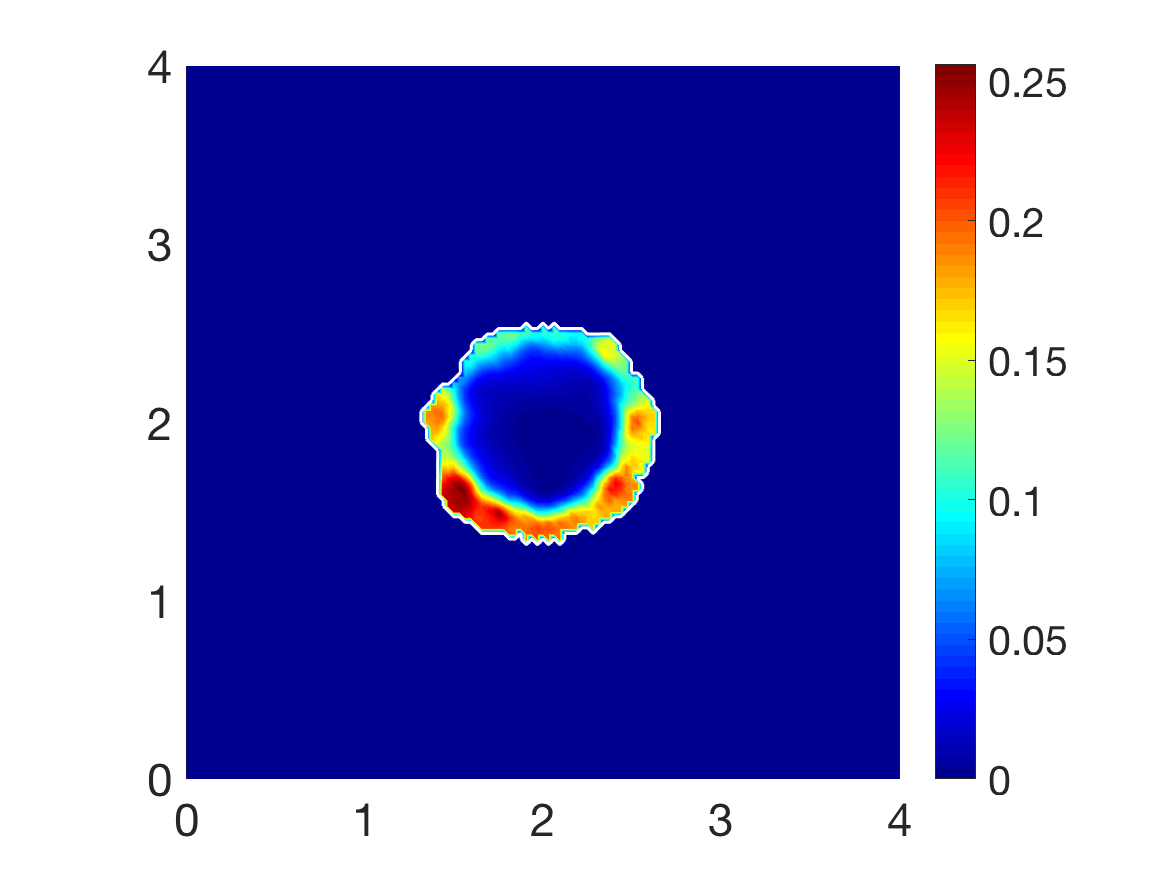}
  \caption{\emph{Cell population 2}}
\end{subfigure}\hfil 

\medskip
\begin{subfigure}{0.5\textwidth}
  \includegraphics[width=\linewidth]{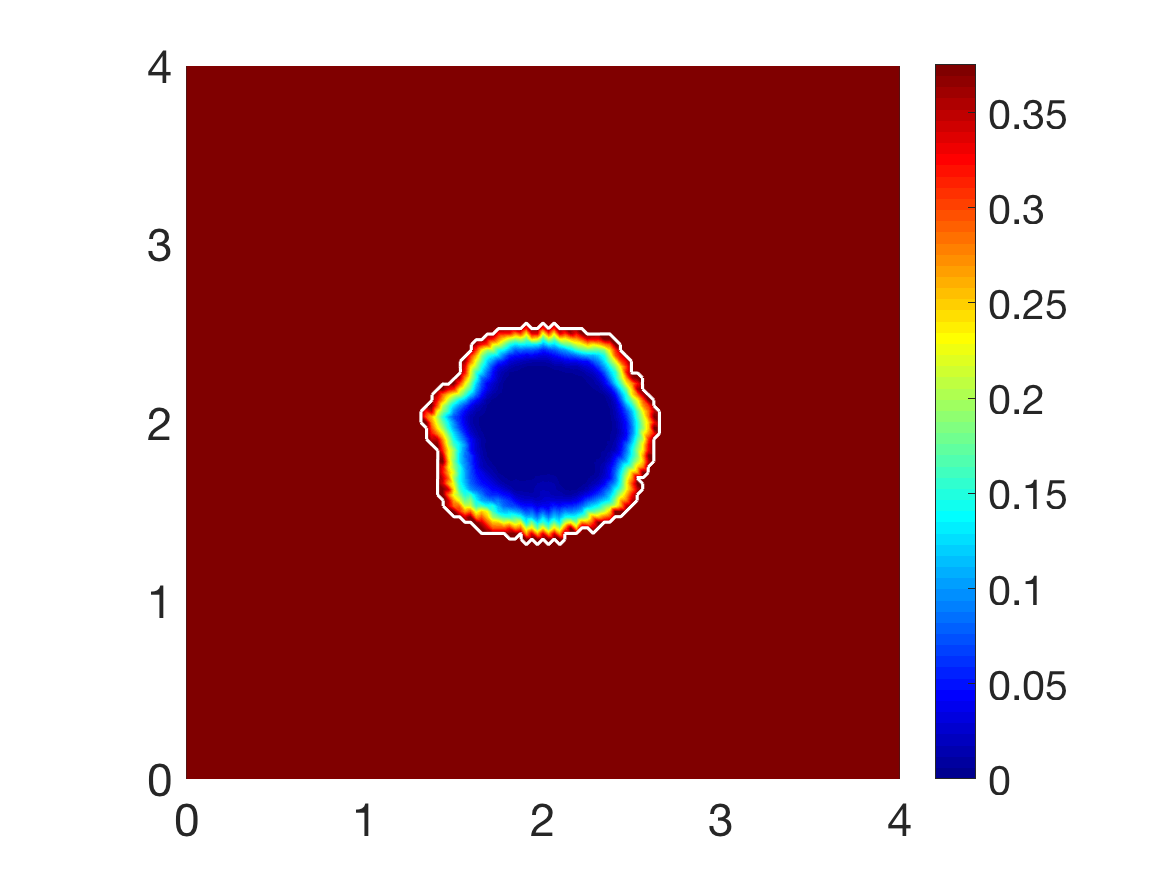}
  \caption{\emph{Non-fibres ECM phase}}
\end{subfigure}\hfil 
\begin{subfigure}{0.5\textwidth}
  \includegraphics[width=\linewidth]{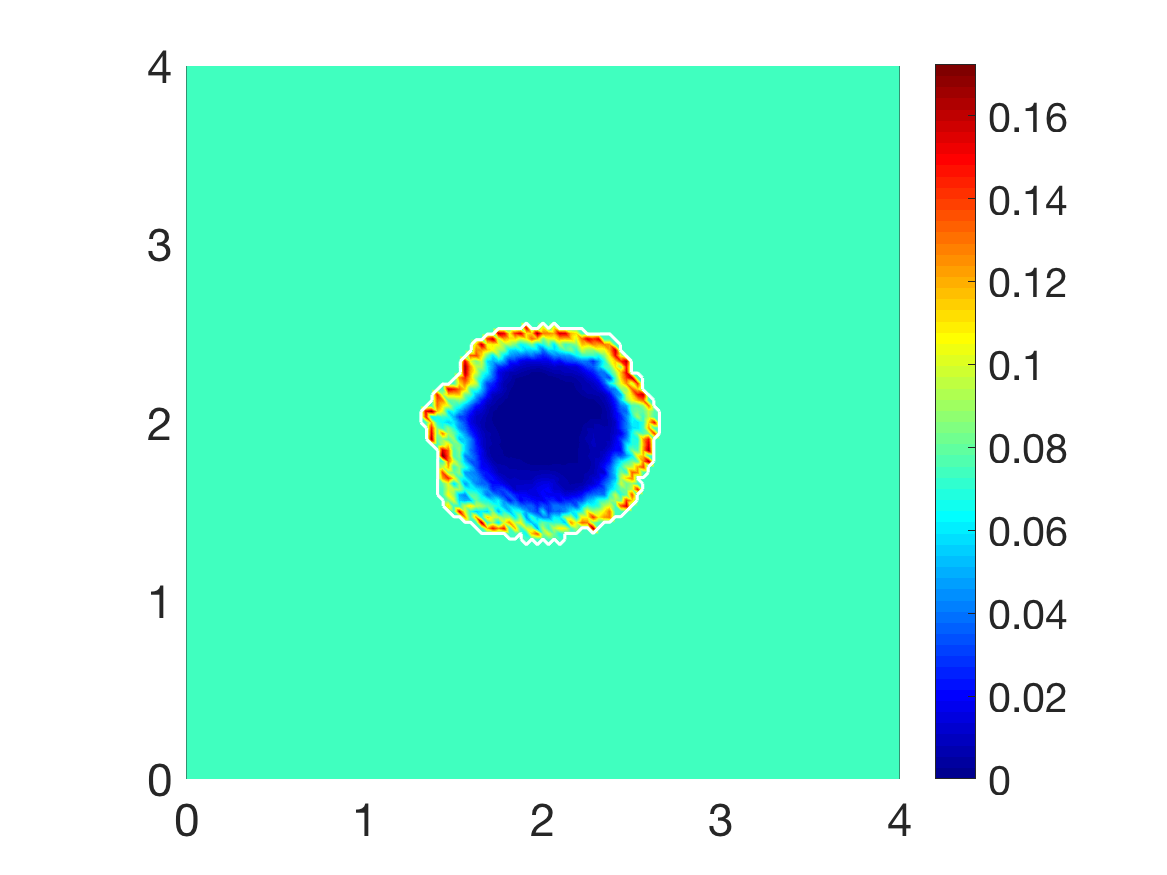}
  \caption{\emph{Macro-scale ECM fibres magnitude}}
\end{subfigure}\hfil 

\medskip
\begin{subfigure}{0.5\textwidth}
  \includegraphics[width=\linewidth]{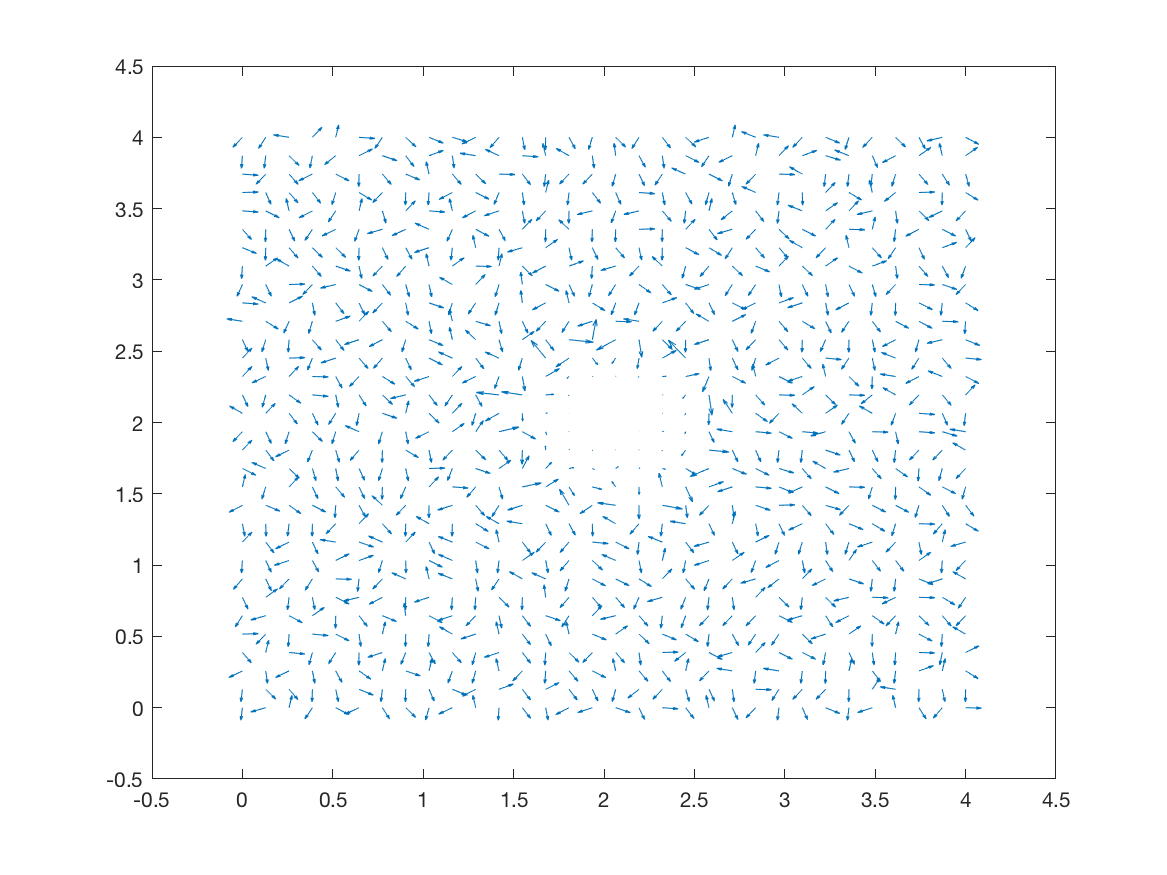}
  \caption{\emph{Oriented macro-scale ECM fibres}}
\end{subfigure}\hfil 
\begin{subfigure}{0.5\textwidth}
  \includegraphics[width=\linewidth]{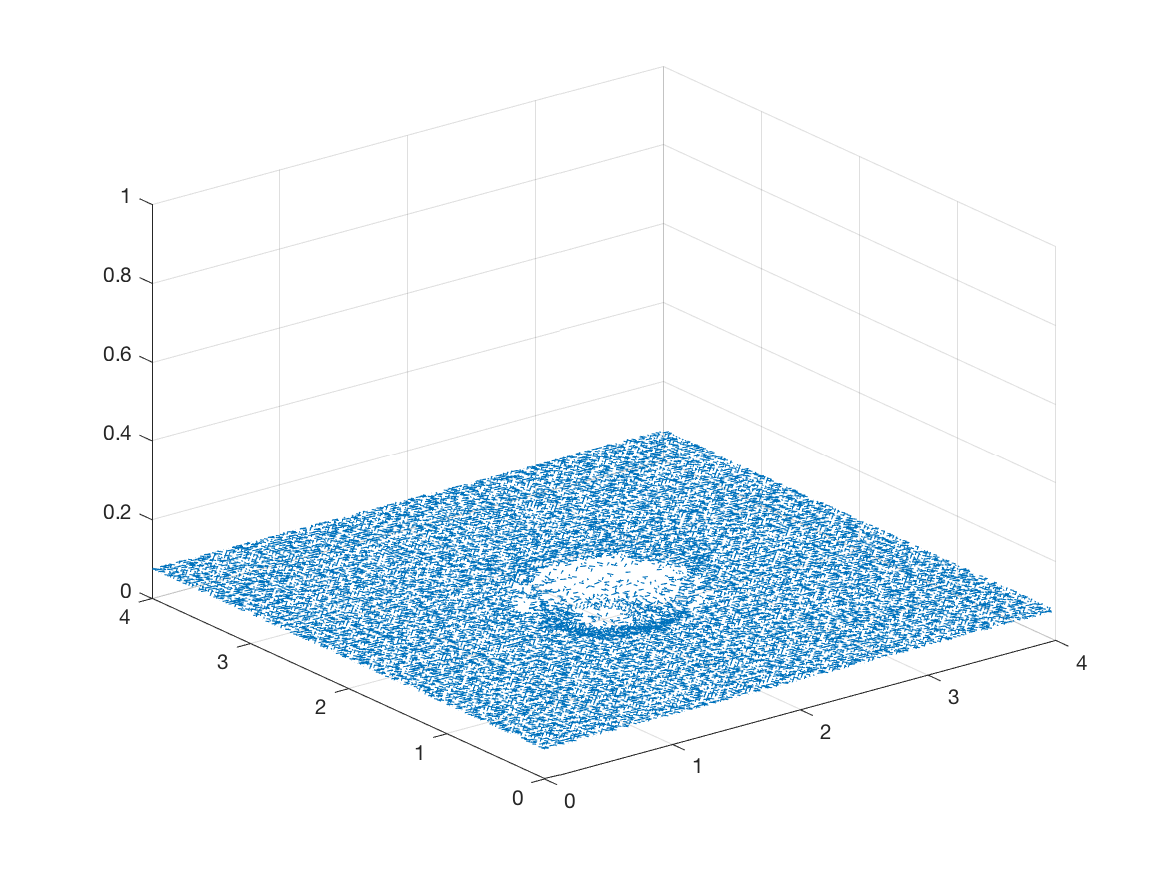}
  \caption{\emph{Oriented macro-scale ECM fibres in a 3D plot}}
\end{subfigure}\hfil 

\caption{Simulations at stage $75\Delta t$ with a homogeneous distribution of non-fibres and a different random initial $15\%$ homogeneous distribution of fibres.}
\label{fig:homol_homof_15_new_75_cancer}
\end{figure}

Currently, we have explored the local invasion of a tumour under the presence of randomly placed micro-domains, allocated from a family of five pre-determined micro-domains. Now we consider the impact changing two of these micro-domains to contain a different pattern of micro-fibres (\dt{i.e., } those given by the second family of paths $\{P^{2}_{i}\}_{i=1..5}$ defined in \ref{microfibres}) will have on the overall invasion of cancer. Using the same initial condition as before for the cell populations \eqref{cancer1ic}, \eqref{cancer2ic} and a homogeneous initial condition for the non-fibre ECM phase along with a $15 \%$ fibre phase, we obtain the computational results in Figure \ref{fig:homol_homof_15_new_75_cancer} at final stage $75 \Delta t$. Comparing with the results using the previous family of micro-domains, Figure \ref{fig:homol_homof_15_75_cancer}, the simulations appear similar, with only slight differences in the shape of the tumour boundary and the composition of the cells within each subpopulation. Cells in population $c_{2}$ have formed an increased number of high density regions in \ref{fig:homol_homof_15_new_75_cancer}(b), particularly at the lower part of the tumour boundary. The degradation of ECM is consistent to that of the previous results, showing regions of high degradation where the cells are of highest distribution, observed in subfigures \ref{fig:homol_homof_15_new_75_cancer}(c)-(d).

\begin{figure}
    \centering 
\begin{subfigure}{0.5\textwidth}
  \includegraphics[width=\linewidth]{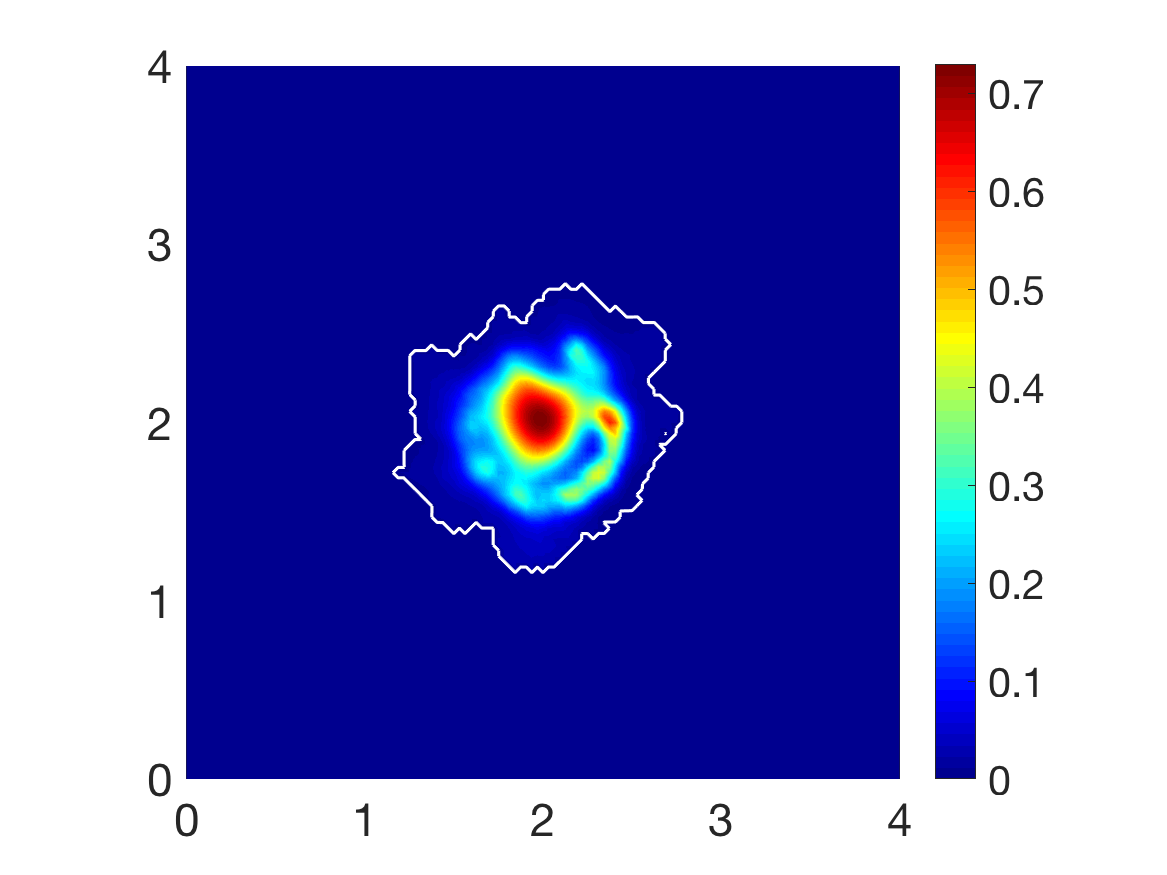}
  \caption{\emph{Cell population 1}}
\end{subfigure}\hfil 
\begin{subfigure}{0.5\textwidth}
  \includegraphics[width=\linewidth]{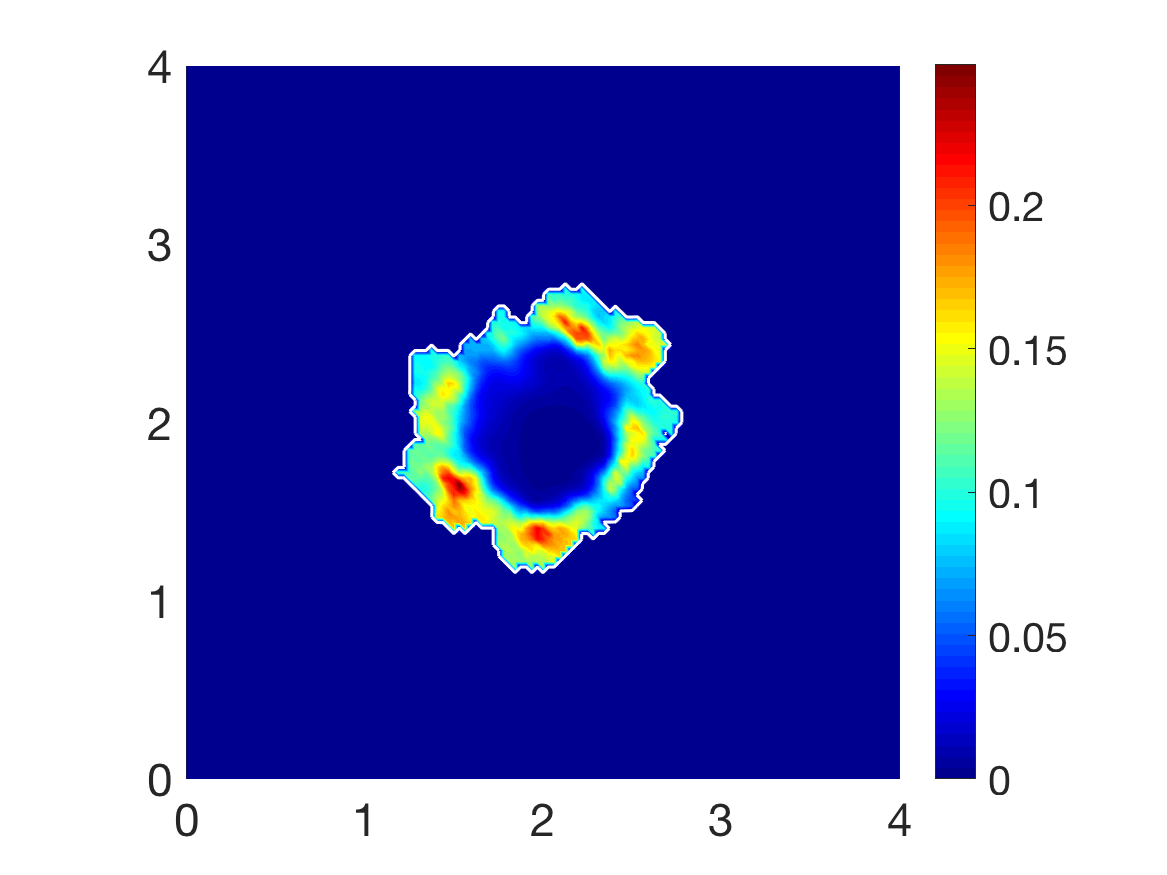}
  \caption{\emph{Cell population 2}}
\end{subfigure}\hfil 

\medskip
\begin{subfigure}{0.5\textwidth}
  \includegraphics[width=\linewidth]{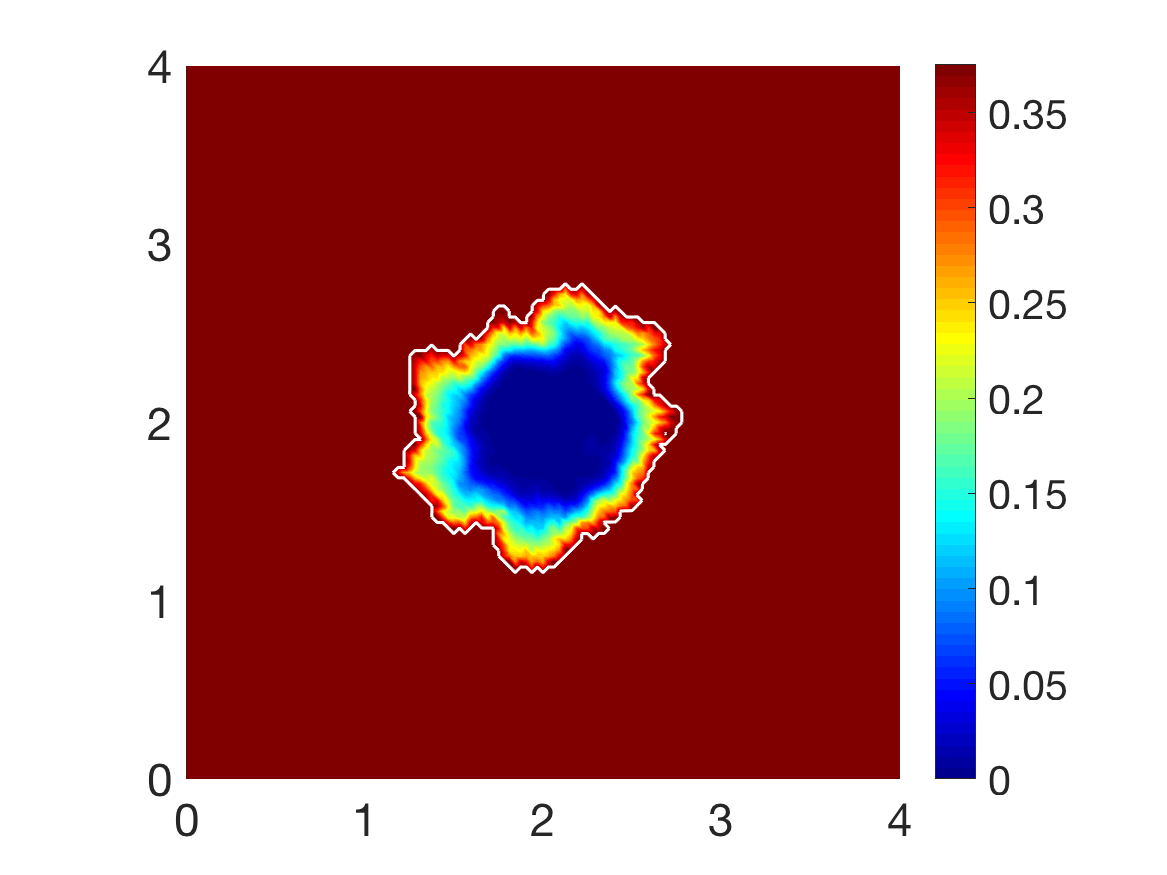}
  \caption{\emph{Non-fibres ECM phase}}
\end{subfigure}\hfil 
\begin{subfigure}{0.5\textwidth}
  \includegraphics[width=\linewidth]{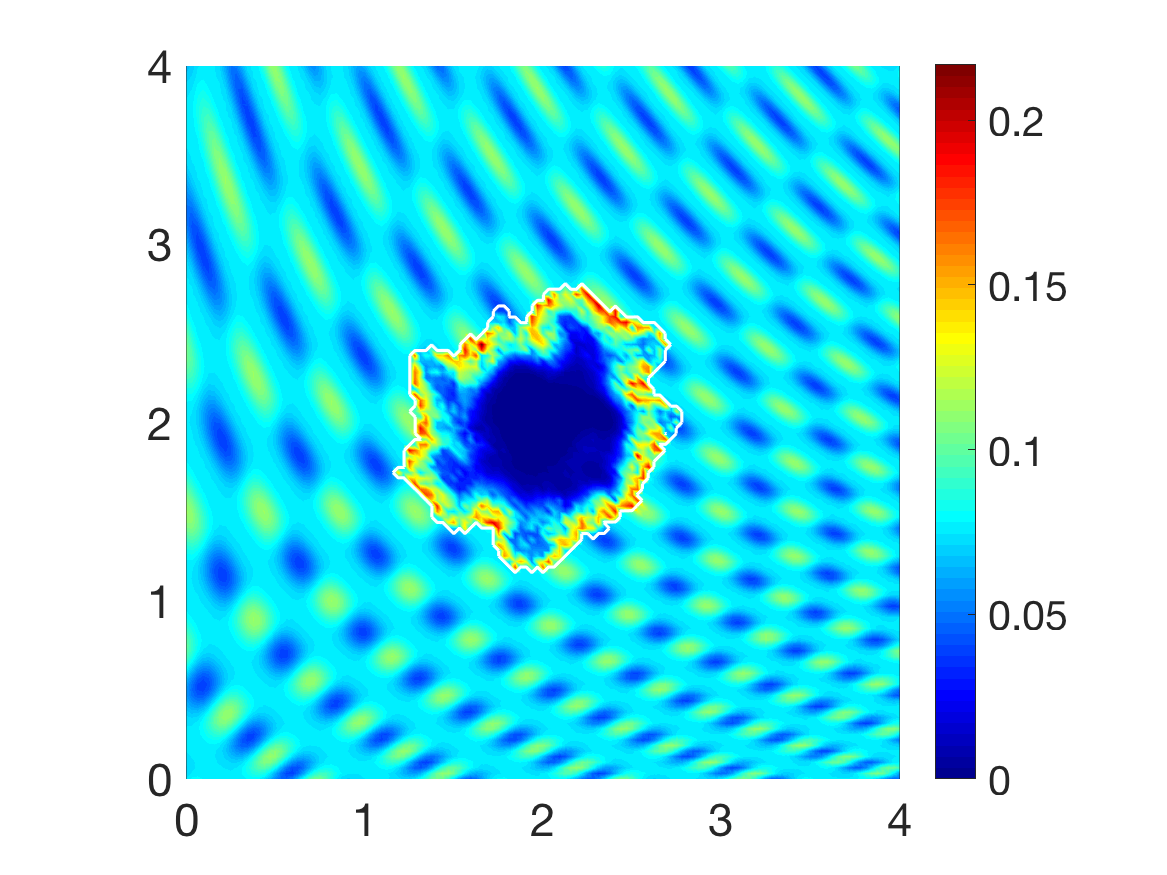}
  \caption{\emph{Macro-scale ECM fibres magnitude}}
\end{subfigure}\hfil 

\medskip
\begin{subfigure}{0.5\textwidth}
  \includegraphics[width=\linewidth]{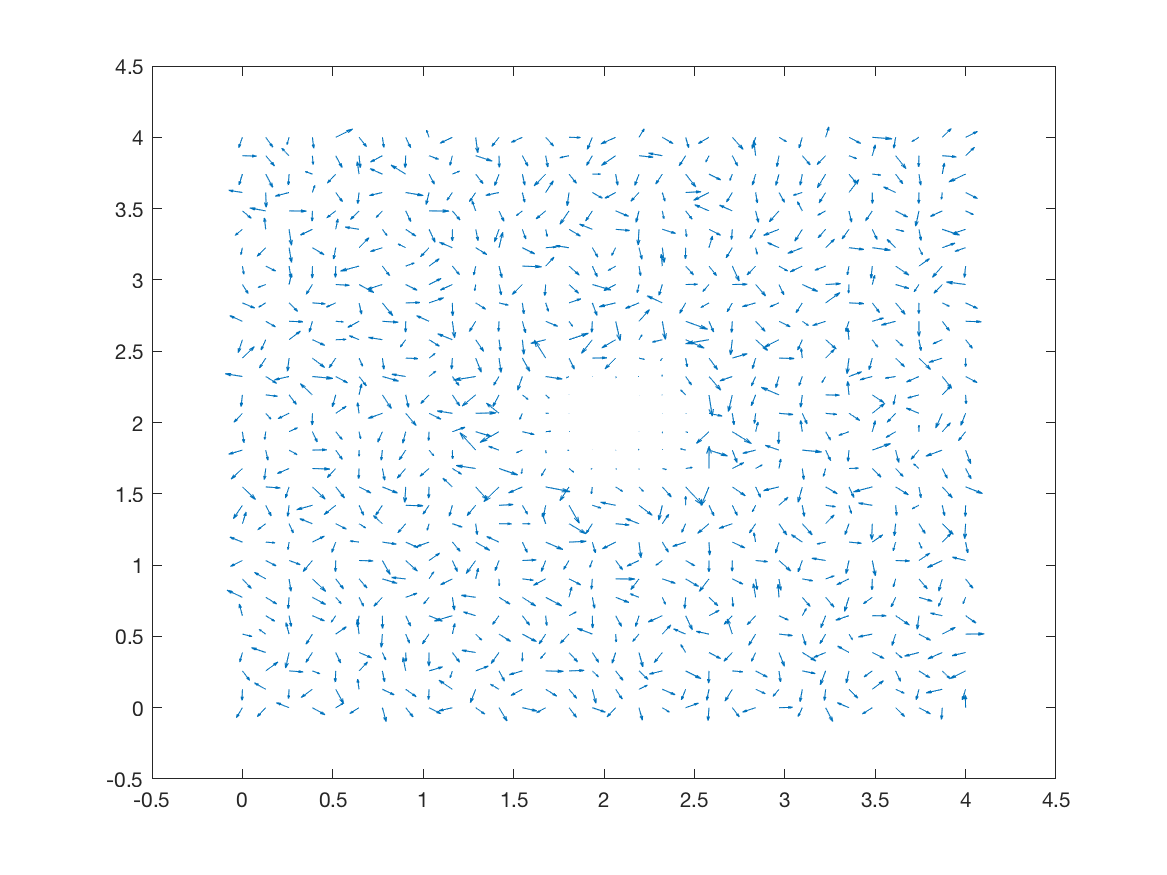}
  \caption{\emph{Oriented macro-scale ECM fibres}}
\end{subfigure}\hfil 
\begin{subfigure}{0.5\textwidth}
  \includegraphics[width=\linewidth]{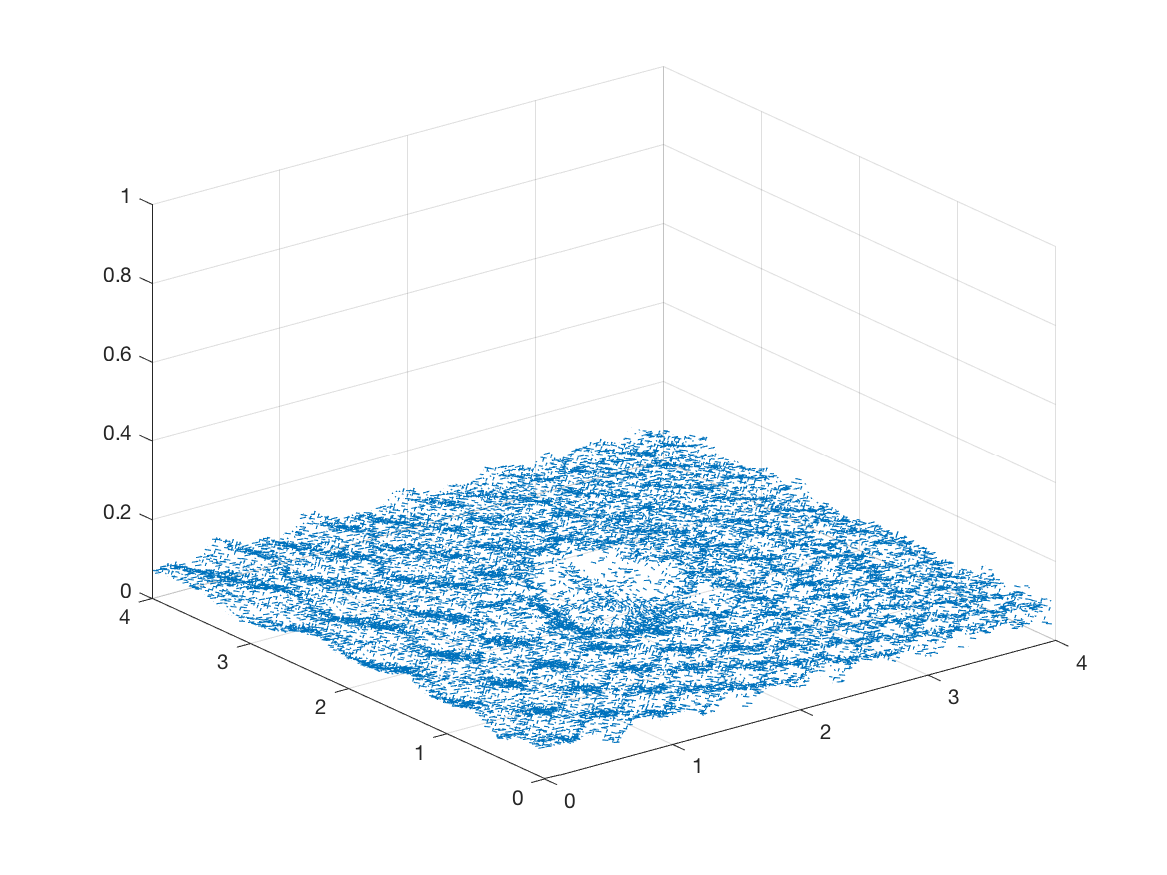}
  \caption{\emph{Oriented macro-scale ECM fibres in a 3D plot}}
\end{subfigure}\hfil 

\caption{Simulations at stage $75\Delta t$ with a homogeneous distribution of the non-fibres and a different random initial $15\%$ heterogeneous distribution of fibres.}
\label{fig:15newfib_homol_75_cancer}
\end{figure}

To complete our comparison of the pre-determined fibre micro-domains, we explore cancer invasion within a heterogeneous $15\%$ fibre component and a homogeneous non-fibre ECM component. Figure \ref{fig:15newfib_homol_75_cancer} displays simulations at final stage $75 \Delta t$, and when compared to Figure \ref{fig:homoland15randfib_75_cancer} there are definite changes in the migration of cells in population $c_{2}$, Figure \ref{fig:15newfib_homol_75_cancer}(b), with pools of high cell distributions forming in different regions to that of Figure \ref{fig:homoland15randfib_75_cancer}(b). Here, the tumour boundary is remaining closer to the main body of cells, with degradation of the ECM occurring explicitly within the boundary, subfigures \ref{fig:15newfib_homol_75_cancer}(c)-(d). The boundary of the tumour has expanded in a pattern consistent with the heterogeneity of the fibre phase \ref{fig:15newfib_homol_75_cancer}(d). Population $c_{1}$ has developed in a similar fashion to that in Figure \ref{fig:homoland15randfib_75_cancer}, however cells in population $c_{2}$ have formed regions of high cell distributions around the entire boundary \ref{fig:15newfib_homol_75_cancer}(b). From these simulations we cannot strictly conclude which family and arrangement of fibre micro-domains gives the more aggressive spread of cancer, however we can deduce that the initial placement of fibre micro-domains has a strong influence on the overall invasion of cancer, owing to the initial orientation of the fibres.

\section{Discussion}
We have presented a novel multiscale moving boundary model with dynamic fibre redistribution, first introduced in \citep{Shutt_2018}, which now includes a \dt{heterotypic} cancer cell population, \dt{with two sub-populations, $c_{1}$ and $c_{2}$}. This model was developed to explore both the random disposition of microscopic fibre distributions and the inclusion of a heterogeneous cancer cell population and their combined effects on the overall invasion pattern during tumour growth and local invasion of tissue. Paying attention in the first instance to the heterogeneous cell population, we introduced a secondary cell subpopulation, mutated from the primary cell population, which holds a lower cell-cell adhesion rate and a higher affinity for cell-fibre adhesion. Mutations from the primary to secondary subpopulation were irreversible, explored through a mutation rate dependent on the underlying ECM density. Focussing next on the fibre component of the ECM, we explored several scenarios; a homogeneous and heterogeneous macroscopic distribution of fibres, varying the initial ratio of macroscopic fibre distributions in relation to the non-fibre ECM phase, and finally the exploration of randomly allocated fibre micro-domains. The multiscale dynamics of the underlying fibre network were explored and modelled within a two-part multiscale model, where the micro-scale dynamics are connected to the macro-scale through a double feedback loop. We considered the macroscopic representations of the micro-fibres and their resulting effects on the macro-scale dynamics, in particular cellular adhesion, whilst at the same time allowing the distribution of both cancer cell subpopulations to cause macroscopic fibre degradation. In addition, through their spatial flux the cancer cells were able to influence the direction of microscopic fibre rearrangement. Finally, we include to this model the previously developed multiscale moving boundary framework developed in \citep{shutt_chapter} that considers the effects of a heterogeneous cancer cell population and its combined contribution to the leading edge proteolytic MDE dynamics. Thus, the model proposed here combines two multi-scale systems that contribute to the same tissue- (macro-) scale dynamics whilst having separate cell- (micro-) scale processes that are simultaneously connected via two double feedback loops. As in \citep{Shutt_2018}, the first multi-scale model gives the cell-scale fibre rearrangement process within the tumour region, and the second multi-scale system describes the MDE proteolytic activity within a cell-scale neighbourhood of the tumour boundary. 

To accommodate a heterogeneous cancer cell population, we adapt the macro-scale dynamics introduced in \citep{Shutt_2018} to include a similar set of dynamics for a second cell subpopulation which has been mutated from the primary cell population. In doing so, we adjust the term for cellular adhesion to include both cell subpopulations allowing for different parameters for each population \citep{Domschke_et_al_2014}, implementing a weaker self-adhesion coefficient and a higher rate of cell-fibre adhesion for the mutated population to promote migration and thus promote the overall invasive capabilities of the tumour. The heterogeneous cancer cell population is then incorporated within the micro-scale dynamics of both multi-scale systems. Firstly, it has influence in determining the source term for MDEs within each cell-scale neighbourhood, with the mutated population inducing a higher amount of MDEs than the primary cell population. Secondly, the distribution and spatial flux of cancer cells that determines the direction of micro-fibre redistribution is now taken as the combined cancer cell distributions and the addition of the spatial fluxes of both populations. Finally, we use the combined cell distributions to contribute equally to the degradation of both the fibre and non-fibre components of the ECM.

Comparing with results from the previous framework developed in \citep{Shutt_2018} that describes the local invasion of a \dt{single population of } tumour \dt{cells} in the presence of a macroscopically homogeneous fibre network, we witness some similarities with our simulations. In the presence of a heterogeneous component of the ECM (fibre or non-fibre phase) we observe a fingering, lobular pattern of invasion, where the cells first flood the low density cavities and proceed to engulf the higher density regions on their route of invasion. In the presence of a homogeneous non-fibre ECM phase and an initially homogeneous fibre phase, we see the boundary of the tumour stray from the symmetric invasion witnessed in \citep{shutt_chapter}, implying that the underlying fibre network indeed plays a role in tumour invasion. Increasing the initial macroscopic fibre density by only $5\%$ from $15\%$ to $20\%$ of the non-fibre ECM phase results in a very fast growing tumour that almost takes over the entire domain. This solidifies the reasoning that the fibre network plays a key role in the local invasion of cancer. Additionally, we have explored this model with two sets of randomly placed micro-fibre domains. Comparing the simulations of each family of micro-domains, particularly under the presence of both components of the ECM being initially homogeneous, the pattern of the subpopulations of cancer cells is affected. The initial orientation of the fibres has influenced the migration of the cells, most obviously observed within population $c_{2}$, where the cell-fibre adhesion rate is higher. As expected, this concludes that the orientation of the fibres is key during tumour growth and development, particularly at the initial stage of invasion.

Looking forward, this modelling framework allows for the opportunity to fully incorporate both multi-scale models; incorporating the micro-scale fibre network within the MDE micro-scale neighbourhoods by investigating the cellular effects of MDEs on the fibres, i.e., the degradation/slicing of the fibres by the matrix metallo-proteinases (MMPs) such as the freely moving MMP-2 and the membrane-bound MT1-MMP molecules. The exploration of TACS (tumour associated collagen signatures) could provide important information of the severity of the tumour. Biological experiments have shown that TACS-3, which is characterised by bundles of straightened and aligned collagen fibres that are orientated perpendicular to the tumour cells provide the poorest prognosis for patients \citep{Provenzano_2008}, whilst tumours in a TACS-2 environment, where the angle between cells and the fibres are between $0\degree-30\degree$, have been shown to provide an environment in which tumour progression is slower \citep{Conklin_2011}. Further work will be focussed on these areas, with the hope that as both the model and biological experiments develop, we can obtain a more realistic representation of fibres \textit{in vivo} and incorporate these into the model to gain a deeper understanding of the role of fibres within the ECM and during the local invasion of cancer.

\appendix
\section{Microscopic fibre domains}\label{microfibres}

For the fibres initial conditions, on the micro-domains $\delta Y(x)$, we consider two families of five distinctive micro-fibres patterns, \dt{$\{P^{1}_{i}\}_{i\in J}$ and $\{P^{2}_{i}\}_{i=1..5}$, which are defined by the union of paths $P^{l}_{i}=\bigcup\limits_{j=1..5} h^{l}_{i,j}$, $l=1,2$, which are given as follows.
\paragraph{For the first family of fibre paths $P^{1}$, we have:}
\bequd
h^{1}_{1,1}: z_{1}=z_{2};  \quad
h^{1}_{1,2}: z_{1}=\frac{1}{2}; \quad
h^{1}_{1,3}: z_{1}=\frac{1}{5}; \quad
h^{1}_{1,4}: z_{2}=\frac{2}{5};\quad
h^{1}_{1,5}: z_{2}=\frac{4}{5}.
\eequd
\vspace{-0.6cm}
\bequd
h^{1}_{2,1}: z_{1}=z_{2};  \quad
h^{1}_{2,2}: z_{1}=\frac{1}{2}; \quad
h^{1}_{2,3}: z_{1}=\frac{1}{5}; \quad
h^{1}_{2,4}: z_{2}=\frac{2}{5};\quad
h^{1}_{2,5}: z_{2}=\frac{1}{10}.
\eequd
\vspace{-0.6cm}
\bequd
h^{1}_{3,1}: z_{1}=z_{2};  \quad
h^{1}_{3,2}: z_{1}=\frac{1}{10}; \quad
h^{1}_{3,3}: z_{1}=\frac{9}{10}; \quad
h^{1}_{3,4}: z_{2}=\frac{2}{5};\quad
h^{1}_{3,5}: z_{2}=\frac{1}{10}.
\eequd
\vspace{-0.6cm}
\bequd
h^{1}_{4,1}: z_{1}=z_{2};  \quad
h^{1}_{4,2}: z_{1}=\frac{1}{10}; \quad
h^{1}_{4,3}: z_{1}=\frac{3}{10}; \quad
h^{1}_{4,4}: z_{2}=\frac{2}{5};\quad
h^{1}_{4,5}: z_{2}=\frac{1}{10}.
\eequd
\vspace{-0.6cm}
\bequd
h^{1}_{5,1}: z_{1}=z_{2};  \quad
h^{1}_{5,2}: z_{1}=\frac{4}{5}; \quad
h^{1}_{5,3}: z_{1}=\frac{3}{10}; \quad
h^{1}_{5,4}: z_{2}=\frac{1}{10};\quad
h^{1}_{5,5}: z_{2}=\frac{7}{10}.
\eequd
\paragraph{For the second family of fibre paths $P^{2}$, we have:}
\bequd
h^{2}_{1,1}: z_{1}=z_{2};  \quad
h^{2}_{1,2}: z_{1}=\frac{1}{2}; \quad
h^{2}_{1,3}: z_{1}=\frac{1}{5}; \quad
h^{2}_{1,4}: z_{2}=\frac{2}{5};\quad
h^{2}_{1,5}: z_{2}=\frac{4}{5}.
\eequd
\vspace{-0.6cm}
\bequd
h^{2}_{2,1}: z_{1}=z_{2};  \quad
h^{2}_{2,2}: z_{1}=\frac{1}{2}; \quad
h^{2}_{2,3}: z_{1}=\frac{1}{5}; \quad
h^{2}_{2,4}: z_{2}=\frac{2}{5};\quad
h^{2}_{2,5}: z_{2}=\frac{1}{10}.
\eequd
\vspace{-0.6cm}
\bequd
h^{2}_{3,1}: z_{1}=z_{2};  \quad
h^{2}_{3,2}: z_{1}=\frac{1}{10}; \quad
h^{2}_{3,3}: z_{1}=\frac{9}{10}; \quad
h^{2}_{3,4}: z_{2}=\frac{2}{5};\quad
h^{2}_{3,5}: z_{2}=\frac{1}{10}.
\eequd
\vspace{-0.6cm}
\bequd
h^{2}_{4,1}: z_{1}=z_{2};  \quad
h^{2}_{4,2}: z_{1}=\frac{4}{5}; \quad
h^{2}_{4,3}: z_{1}=\frac{3}{5}; \quad
h^{2}_{4,4}: z_{2}=\frac{4}{5};\quad
h^{2}_{4,5}: z_{2}=\frac{3}{5}.
\eequd
\vspace{-0.6cm}
\bequd
h^{2}_{5,1}: z_{1}=z_{2};  \quad
h^{2}_{5,2}: z_{1}=\frac{4}{5}; \quad
h^{2}_{5,3}: z_{1}=\frac{3}{5}; \quad
h^{2}_{5,4}: z_{2}=\frac{4}{5};\quad
h^{2}_{5,5}: z_{2}=\frac{1}{10}.
\eequd
}

\dt{
For each of these families of fibre paths $P^{l}$, $l=1,2$, as described in \cite{Shutt_2018}, the micro-scale fibres pattern within each micro-domain $\delta Y(x)$ is given as 
\bequ
f(z,t):=\sum\limits_{j=1}^{5}\psi_{h^{l}_{i,j}}(z)(\chi_{_{(\delta-2\gamma_{_{0}})Y(x)}}\ast \psi_{\gamma_{_{0}}})(z)
\eequ
where $\{\psi_{h^{l}_{i,j}}\}_{i,j=1..5}$ are smooth compact support functions of the form
\bequ
\begin{array}{l}
\hspace{3.0cm}\psi_{h_{j}}: \delta Y(x) \rightarrow \mathbb{R}\\[0.3cm]
\textrm{defined as follows:}\\[0.6cm]
\textrm{$Case\,1:$ if $h^{l}_{i,j}$ is not parallel to $z_{1}-$axis} \\
\textrm{(i.e., $h^{l}_{i,j}$ is identified as the graph of a function of $z_{2}$)}\\[0.2cm]
\textrm{we have:}\\[0.2cm]
\psi_{h^{l}_{i,j}}(z_1,z_2):=
\left\{
\begin{array}{ll}
C_{h^{l}_{i,j}} e^{-{\frac{1}{r^2-(h_{j}(z_2) - z_1)^2}}}, &\quad \text{if} \ z_1 \in [h^{l}_{i,j}(z_2)-r, h^{l}_{i,j}(z_2)+r], \\[0.6cm]
0, &\quad \text{if} \ z_1 \not\in [h^{l}_{i,j}(z_2)-r, h^{l}_{i,j}(z_2)+r];
\end{array}
\right.\\[0.8cm]
\textrm{$Case\,2:$ if $h^{l}_{i,j}$ is parallel to $z_{1}-$axis}\\ 
\textrm{(i.e., $h^{l}_{i,j}$ is identified as the graph of a constant function of $z_{1}$)}\\[0.2cm]
\textrm{we have:}\\[0.2cm]
\psi_{h^{l}_{i,j}}(z_1,z_2):=
\left\{
\begin{array}{ll}
C_{h^{l}_{i,j}} e^{-{\frac{1}{r^2-(h^{l}_{i,j}(z_1) - z_2)^2}}}, &\quad \text{if} \ z_2 \in [h^{l}_{i,j}(z_1)-r, h^{l}_{i,j}(z_1)+r], \\[0.6cm]
0, &\quad \text{if} \ z_2 \not\in [h^{l}_{i,j}(z_1)-r, h^{l}_{i,j}(z_1)+r].
\end{array}
\right.\\[0.6cm]
\end{array}
\label{eq:fib}
\eequ
Here $r>0$ is the width of the micro-fibres and $C_{h^{l}_{i,j}}$ are constants that determine the maximum height of $\psi_{h^{l}_{i,j}}$ along the smooth paths $\{h^{l}_{i,j}\}_{i,j=1..5}$ in $\delta Y(x)$. Finally, $\psi_{\gamma}$ is the standard mollifier defined in \ref{mollifier}, with $\gamma_{_{0}}= h/16 $.\\
}

\dt{Finally, the initial spatial configuration of the pattern of macroscopic ECM fibre phase is selected according to a randomly generated matrix of labels $A=(a_{i,j})_{i,j=1..n}$ corresponding to the entire $n\times n$ grid discretising $Y$, in which the entries $a_{i,j}$ are allocated values randomly selected from the set of configuration labels $\{1,2,3,4,5\}$ that will dictate the choice of micro-fibres pattern among those described above that will be assigned to the micro-domains $\delta Y(j\Delta x, i\Delta y)$, for all $i,j=1..n$.}

\section{Adhesion matrices}
The cell adhesion matrices use in the manuscript were considered as follows:
\bequ
\begin{array}{c}
\textbf{S}_{_{max}}=
\left( \begin{array}{cc}
S_{c_{1},c_{1}}& 0  \\
0 & S_{c_{2},c_{2}}  
\end{array} \right),\\[0.6cm]
 \textbf{S}_{cF}=
\left( \begin{array}{cc}
S_{_{c_{1},F}} & 0  \\
0 & S_{c_{2},F}  \end{array} \right)
\quad \text{and} \quad\textbf{S}_{cl}=
\left( \begin{array}{cc}
S_{_{c_{1},l}} & 0  \\
0 & S_{c_{2},l}  \end{array} \right).
\end{array}
\label{norm_matrices}
\eequ

\section{The mollifier $\psi_{\gamma}$} \label{mollifier}
The standard mollifier $\psi_{\gamma}:\R^{N}\to\R_{+}$ (which was used also in \citep{Dumitru_et_al_2013,Shutt_2018}) is defined as usual, namely
\[
\psi_{\gamma}(x):=\frac{1}{\gamma^{N}}\psi\big(\frac{x}{\gamma}\big),
\]
where $\psi$ is the smooth compact support function given by
\bequd
\psi(x):=
\left\{
\begin{array}{cll}
\frac{exp\frac{1}{\nor{x}^{2}_{_{2}}-1}}{\int\limits_{\Bila(0,1)}exp\frac{1}{\nor{z}^{2}_{_{2}}-1}dz},& \quad if & x\in \Bila(0,1),\\[0.3cm]
0, & \quad if & x\not\in \Bila(0,1).
\end{array}
\right.
\eequd

\section{Table for the parameter set $\Sigma$}      \label{paramSection}
Here we present a table for the parameter set $\Sigma$.
\begin{table}[htb!]
\begin{adjustwidth}{-2cm}{}
\centering
\begin{threeparttable}
\centering
\caption{The parameters in $\Sigma$}
\begin{tabular}{ c c c c}
  \hline		
  Parameter & Value & Description & Reference \\
  \hline
  $D_1$ & $3.5 \times10^{-4}$ & diffusion coeff. for cell population $c_{1}$ & \cite{Domschke_et_al_2014}\\
  $D_2$ & $7\times10^{-4}$ & diffusion coeff. for cell population $c_{2}$ & \cite{Domschke_et_al_2014}\\
  $D_m$ & $10^{-3}$ & diffusion coeff. for MDEs & Estimated \\
  $\mu_1$ & $0.25$ & proliferation coeff. for cell population $c_{1}$ & \cite{Domschke_et_al_2014} \\
  $\mu_2$ & $0.25$ & proliferation coeff. for cell population $c_{2}$ & \cite{Domschke_et_al_2014} \\
  $\gamma_{1}$ & 2 & non-fibrous ECM degradation coeff. & \cite{Shutt_2018} \\
  $\gamma_{2}$ & 1.5 & macroscopic fibre degradation coeff. & \cite{Peng2016} \\
  $\omega$ & 0-0.02 & non-fibrous ECM remodelling coeff. & \cite{Domschke_et_al_2014} \\
  $\alpha_{1}$ & $1$ & MDE secretion rate of $c_{1}$ & Estimated \\
  $\alpha_{2}$ & $1.5$ & MDE secretion rate of $c_{2}$ & Estimated \\
  $\textbf{S}_{max}$ & $\left( \begin{array}{cc}
0.5& 0  \\
0 & 0.3  \end{array} \right)$ & cell-cell adhesion coeff. matrix & \cite{Domschke_et_al_2014} \\
  $\textbf{S}_{cF}$ & $\left( \begin{array}{cc}
0.1& 0  \\
0 & 0.2  \end{array} \right)$ & cell-fibre adhesion coeff. & Estimated \\
  $\textbf{S}_{cl}$ & $\left( \begin{array}{cc}
0.05& 0  \\
0 & 0.05  \end{array} \right)$ & cell-matrix adhesion coeff. & Estimated \\
  $R$ & 0.15 & sensing radius & \cite{Shutt_2018} \\
  $r$ & 0.0016 & width of micro-fibres & \cite{Shutt_2018} \\
  $f_{\text{max}}$ & 0.6360 & max. micro-density of fibres & \cite{Shutt_2018} \\
  $p$ & 0.15-0.2 & percentage of non-fibrous ECM & Estimated \\
  $h$ & 0.03125 & macro-scale spatial discretisation size & \cite{Dumitru_et_al_2013}\\
  $\epsilon$ & 0.0625 & size of micro-domain $\epsilon Y$ & \cite{Dumitru_et_al_2013} \\
  $\delta$ & 0.03125 & size of micro-domain $\delta Y$ & \cite{Shutt_2018} \\
   \hline 
  \label{table:parameters} 
\end{tabular}
\end{threeparttable}
\end{adjustwidth}
\end{table}
\newpage

\section*{Acknowledgment} RS and DT would like to acknowledge the support received through the EPSRC DTA Grant EP/M508019/1 on the project: \emph{Multiscale modelling of cancer invasion: the role of matrix-degrading enzymes and cell-adhesion in tumour progression}.  

\section*{References}
\bibliography{ThesisReferences}

\end{document}